\documentclass[preprint,showpacs,preprintnumbers,amsmath,amssymb]{revtex4}

\usepackage[section]{placeins}
\usepackage{graphicx}
\usepackage{dcolumn}
\usepackage{bm}

\usepackage{epsfig}
\usepackage{amssymb}
\usepackage{amsmath}
\usepackage{flafter}
\usepackage{array}

\begin{document}
\title{ $B_{(s)},D_{(s)} \to \pi, K, \eta, \rho, K^*, \omega,
\phi$ Transition Form Factors and Decay Rates  with Extraction of
the CKM parameters $|V_{ub}|$, $|V_{cs}|$, $|V_{cd}|$ }
\author{Y. L. Wu, M. Zhong and Y.B. Zuo}
\affiliation{ Institute of Theoretical Physics, Chinese Academy of
Sciences, Beijing 100080, China}
\begin{abstract}
A systematic calculation for the transition form factors of heavy
to light mesons ( $B,B_s,D,D_s \to \pi, K, \eta, \rho, K^*,
\omega, \phi$ ) is carried out by using light-cone sum rules in
the framework of heavy quark effective field theory. The heavy
quark symmetry at the leading order of $1/m_Q$ expansion enables
us to reduce the independent wave functions and establish
interesting relations among form factors. Some relations hold for
the whole region of momentum transfer. The meson distribution
amplitudes up to twist-4 including the contributions from higher
conformal spin partial waves and light meson mass corrections are
considered. The CKM matrix elements $|V_{ub}|$, $|V_{cs}|$ and
$|V_{cd}|$ are extracted from some relatively well-measured decay
channels. A detailed prediction for the branching ratios of heavy
to light meson decays is then presented. The resulting predictions
for the semileptonic and radiative decay rates of heavy to light
mesons ( $B,B_s,D,D_s \to \pi, K, \eta, \rho, K^*, \omega, \phi$ )
are found to be compatible with the current experimental data and
can be tested by more precise experiments at B-factory, LHCb,
BEPCII and CLEOc.
\end{abstract}
 \pacs{13.20.-v, 13.20.Fc, 13.20.He}
\maketitle

\section{Introduction}

Heavy meson exclusive decays play an important role in extracting
the CKM matrix elements and probing new physics beyond the
standard model. They are clean in experiments but difficult in
theoretical calculations due to the requirement of the knowledge
of nonperturbative QCD. In a simple type of processes with a heavy
meson decaying into a light final meson, the nonperturbative
effects are generally parameterized by form factors. A systematic
evaluation for the form factors was first performed in
Ref.\cite{BSW} by using constituent quark model. Recently, the
form factors of $B$ and $B_s$ to light mesons were calculated via
light-cone sum rule in full QCD \cite{P1,P2}. Since the heavy
meson $B_{(s)}$ or $D_{(s)}$ contains one heavy quark and one
light quark, it is useful to adopt the heavy quark effective field
theory (HQEFT) based on a large component QCD
\cite{HQEFT1,HQEFT2,HQEFT3} with treating quark and anti-quark
fields on the same footing in a fully symmetric way. The leading
term of HQEFT coincides with the heavy quark effective theory
\cite{HG} in the infinite mass limit\cite{HQL1,HQL2} and possesses
the heavy quark spin-flavor symmetry\cite{HQS1,HQS2}, which
enables one to relate different heavy quarks and helps to improve
our understanding of heavy to light decays. The HQEFT can also
simplify significantly the evaluation of hadronic matrix elements
and provide a systematic and consistent approach to calculate the
$1/m_Q$ corrections. The HQEFT has well been applied to the heavy
to heavy meson decays\cite{W1,W2,W3,W4,W5} up to $1/m_Q^2$
corrections, and also extended to the heavy to light
decays\cite{WY2,WY3,WY,ZWW} including the $1/m_Q$ corrections for
$B\to \pi e \nu$ decay\cite{WWZ0} and $B \to \rho e \nu$
decay\cite{WWZN}. Some interesting relations have been shown to
hold for the whole region of momentum transfer at the leading
order of $1/m_Q$ expansion\cite{ZWW}. In this paper, we shall
provide a systematic analysis and calculation for all the
transition form factors of heavy to light mesons ( $B,B_s,D,D_s
\to \pi, K, \eta, \rho, K^*, \omega, \phi$ ) via the light-cone
sum rule in the framework of HQEFT. Within the framework of HQEFT,
the contributions from higher twist distribution amplitudes are
manifestly suppressed with a higher power of $1/m_M$ at small
momentum transfer and the much higher twist distribution
amplitudes can be neglected safely at small momentum transfer. We
consider the meson distribution amplitudes up to twist-4 including
the contributions from higher conformal spin partial waves and
light meson mass corrections. The CKM matrix elements $|V_{ub}|$,
$|V_{cs}|$ and $|V_{cd}|$ can then be extracted from the most
recent more precise experimental data. For a consistent check, we
also present a detailed prediction for the branching ratios of all
the heavy to light semileptonic and radiative exclusive meson
decays. As we consider in this paper only the leading
contributions of $1/m_Q$ expansion, the results should be
universal since the effective field theories of heavy quarks
resulting from different approaches are all the same in the
infinite mass limit\cite{Mainz}. The $1/m_Q$ corrections have been
shown to be small in the B-meson decays\cite{WWZ0,WWZN}. For the
D-meson decays, the $1/m_Q$ corrections are expected to be
sizable, but it is likely to have sizable effects only on some of
the form factors which are not dominant for the decay rates. This
is because, as we are going to show in this paper, the leading
contributions can result in consistent predictions for the
branching ratios.

The paper is organized as follows: in Sec.II, we begin with the
definitions of form factors via hadronic transition matrix elements
and then formulate them by using light-cone sum rules in HQEFT. Our
numerical analysis and results of heavy to light form factors are
presented in sec.III. In sec.IV, we provide a detailed calculation
for the branching ratios of heavy to light exclusive decays with
extraction of the CKM matrix elements $|V_{ub}|$, $|V_{cs}|$,
$|V_{cd}|$. A brief conclusion and remark is given in the last
section.

\section{Definitions of Form Factors and Formulation by using Light-cone Sum Rules}
\subsection{Definitions of Form Factors}

Following Refs.\cite{BSW,WY,ZWW}, we define the form factors via
transition matrix elements, which may be grouped into semileptonic
and penguin types. The semileptonic ones are

\begin{eqnarray}
& & \langle P(p)|\bar{q} \gamma^\mu Q |M(p+q) \rangle  =  \left [
(2p+q)^\mu - \frac{m_M^2-m_P^2}{q^2} q^\mu \right ] f_+ (q^2)
\nonumber \\
& & \hspace{5cm}+\left [\frac{m_M^2-m_P^2}{q^2} q^\mu \right ] f_0(q^2) ,\\
& & \langle V(p,\epsilon^*)|\bar{q} \gamma^\mu (1-\gamma^5)
Q|M(p+q)\rangle  =  -i (m_M
+ m_V) A_1(q^2) \epsilon^{* \mu} \nonumber \\
& & \hspace{2cm} +i \frac{A_2(q^2)}{m_M + m_V} [\epsilon^* \cdot
(p+q)](2p+q)^\mu
+ i \frac{A_3(q^2)}{m_M +m_V} [ \epsilon^* \cdot (p+q)] q^\mu     \nonumber \\
& & \hspace{2cm} +\frac{2 V(q^2)}{m_M + m_V} \varepsilon^{\mu \alpha
\beta \gamma} \epsilon^*_\alpha (p+q)_\beta p_\gamma .
\end{eqnarray}

Here the initial heavy mesons $B, B_s, D, D_s$ are denoted as $M$,
the final light pseudoscalar and vector mesons are labelled as $P$
and $V$ respectively. $Q$ denotes any heavy quark ($b$ or $c$) and
$q$ in the currents represents light quarks ($u, d $ or $s$). $m_M$
and $m_{P(V)}$ are the heavy and light pseudoscalar (vector) meson
masses respectively.

The penguin matrix elements can be written as
\begin{eqnarray}
& & \langle P(p)|\bar{q} \sigma^{\mu \nu} q_\nu (1+\gamma^5)
Q|M(p+q)\rangle
  =  i \frac{f_T(q^2)}{m_M + m_P} [ q^2 (2p+q)^\mu  \nonumber \\
 & &  \hspace{6cm} - ( m_M^2 - m_P^2) q^\mu ] , \\
& & \langle V(p,\epsilon^*)|\bar{q} \sigma^{\mu \nu} q_\nu (1 +
\gamma^5) Q|M(p+q)\rangle  =
-i \varepsilon^{\mu \alpha \beta \gamma } \epsilon^*_\alpha ( p+q)_\beta p_\gamma 2 T_1(q^2) \nonumber \\
& & \hspace{3cm}+T_2(q^2) \left \{ (m_M^2-m_V^2) \epsilon^{* \mu}
- [\epsilon^* \cdot (p+q) ](2p+q)^\mu \right \} \nonumber \\
& & \hspace{3cm }+T_3(q^2) [\epsilon^* \cdot (p+q) ]\left [q^\mu -
\frac{q^2}{m_M^2 - m_V^2} ( 2p+q)^\mu \right ].
\end{eqnarray}

In HQEFT, the matrix elements can be expanded into the powers of
$1/m_Q$ and be simply expressed by a set of heavy spin-flavor
independent universal wave functions\cite{W1}
\begin{eqnarray}
\frac{1}{\sqrt{m_M}}<P(V)|\bar{q} \Gamma Q|M> \frac{1}{\sqrt{\bar{\Lambda}_M}} [ <P(V)| \bar{q} \Gamma Q_v^+ |
M_v> +O(1/m_Q)],
\end{eqnarray}
where $\bar{\Lambda}_M = m_M - m_Q$ is the binding energy. From
heavy quark symmetry, the leading order matrix elements can
generally be written in the form\cite{WY2,WY3}
\begin{eqnarray}
& & <P(p)| \bar{q} \Gamma Q_v^+ | M_v> = - Tr [ \pi(v,p) \Gamma {\cal M}_v ] ,\\
& & <V(p,\epsilon^*)| \bar{q} \Gamma Q_v^+ |M_v> = - i Tr[ \rho
(v,p) \Gamma {\cal M}_v],
\end{eqnarray}
with
\begin{eqnarray}
\pi (v,p) & = & \gamma^5 [ A(v \cdot p) + \hat{p\hspace{-0.17cm}\slash} B (v \cdot p) ] ,\\
\rho(v,p) & = & L_1(v \cdot p) \epsilon^* \hspace{-0.28cm} \slash
+L_2(v \cdot p)
( v \cdot \epsilon^* ) + [L_3(v \cdot p) \epsilon^* \hspace{-0.28cm} \slash \nonumber \\
& & +L_4(v \cdot p) ( v \cdot \epsilon^*) ] \hat{p}
\hspace{-0.17cm} \slash ,
\end{eqnarray}
and
\begin{eqnarray}
\hat{p}^\mu = \frac{p^\mu}{v \cdot p}, \qquad {\cal M}_v = -
\sqrt{\bar{\Lambda}} \frac{1+v \hspace{-0.2cm} \slash}{2}
\gamma^5.
\end{eqnarray}
Here $A$, $B$ and $L_i(i=1,2,3,4)$ are the leading order wave
functions characterizing the heavy to light transition matrix
elements. ${\cal M}_v$ is the heavy pseudoscalar spin wave
function in HQEFT. $\bar{\Lambda} = \lim_{m_Q \rightarrow \infty}
\bar{\Lambda}_M$ is the heavy flavor independent binding energy.
The four-velocity of heavy meson $v^\mu$ satisfies $v^2 =1$.

The form factors defined in Eqs.(1-4) can then be expressed by the
universal wave functions $A,B$ and $L_i(i=1,2,3,4)$
\begin{eqnarray}
f_+(q^2) & = & \frac{1}{m_M} \sqrt{\frac{m_M
\bar{\Lambda}}{\bar{\Lambda}_M}}
[ A(v \cdot p) + B(v \cdot p) \frac{m_M}{v \cdot p} ]+ \cdot \cdot \cdot ,\\
f_0(q^2) & = & \frac{1}{m_M} \sqrt{\frac{m_M
\bar{\Lambda}}{\bar{\Lambda}_M}}
\left [ (1+\frac{q^2}{m_M^2- m_P^2}) A(v \cdot p)\right. \nonumber \\
& & \left. + (1-\frac{q^2}{m_M^2-m_P^2}) B(v \cdot p) \frac{m_M}{v \cdot p}  \right] + \cdot \cdot \cdot ,\\
f_T(q^2) & = & \frac{m_M+m_P}{m_M} \sqrt{\frac{m_M
\bar{\Lambda}}{\bar{\Lambda}_M}} \frac{B^\prime (v \cdot p)}{v \cdot
p} + \cdot \cdot \cdot ,\\
 A_1(q^2) & = & \frac{2}{m_M +
m_V}\sqrt{\frac{m_M \bar{\Lambda}}{\bar{\Lambda}_M}}
[ L_1(v \cdot p) + L_3(v \cdot p) ] + \cdot \cdot \cdot ,\\
A_2(q^2) & = & 2 (m_M + m_V) \sqrt{\frac{m_M
\bar{\Lambda}}{\bar{\Lambda}_M}}\left
 [ \frac{L_2(v \cdot p)}{2 m_M^2} + \frac{L_3(v \cdot p) - L_4(v \cdot p)}{2 m_M (v \cdot p)}\right ]
 + \cdot \cdot \cdot , \nonumber \\
 \\
A_3(q^2) & = & 2 (m_M + m_V) \sqrt{\frac{m_M
\bar{\Lambda}}{\bar{\Lambda}_M}}\left [ \frac{L_2(v \cdot p)}{2
m_M^2} - \frac{L_3(v \cdot p) - L_4(v \cdot p)}{2 m_M (v \cdot p)}
\right ]
 + \cdot \cdot \cdot , \nonumber \\
\\
V(q^2) & = & \sqrt{\frac{m_M
\bar{\Lambda}}{\bar{\Lambda}_M}}\frac{m_M + m_V}{m_M ( v \cdot p)}
L_3 ( v \cdot p) + \cdot \cdot \cdot ,\\
T_1(q^2) & = & \sqrt{\frac{m_M \bar{\Lambda}}{\bar{\Lambda}_M}}\left
[\frac{L_1^\prime (v \cdot p)}{m_M}
+ \frac{L_3^\prime ( v \cdot p)}{v \cdot p} \right] + \cdot \cdot \cdot ,\\
T_2(q^2) & = & 2 \sqrt{\frac{m_M
\bar{\Lambda}}{\bar{\Lambda}_M}}\frac{1}{m_M^2 - m_V^2}
\left [(m_M -v \cdot p) L_1^\prime (v \cdot p)\right. \nonumber \\
& & \left. + \frac{m_M v \cdot p - m_V^2}{v \cdot p} L_3^\prime (v \cdot p) \right] + \cdot \cdot \cdot ,\\
T_3(q^2) & = & \sqrt{\frac{m_M \bar{\Lambda}}{\bar{\Lambda}_M}}\left
[ - \frac{L_1^\prime ( v\cdot p)}{m_M} + \frac{L_3^\prime (v \cdot
p)}{v \cdot p}
- \frac{m_M^2 -m_V^2}{m_M^2 v \cdot p} L_4^\prime ( v \cdot p) \right ] + \cdot \cdot \cdot ,\nonumber \\
\end{eqnarray}
where $y \equiv v \cdot p = \frac{m_M^2+m_{P(V)}^2 -q^2}{2 m_M}$ is the energy of final light meson.

Evidently the universal wave functions $A, B$ and $L_i(i=1,2,3,4)$
are heavy flavor independent at the leading order of $1/m_Q$
expansion, which enables us to obtain the form factors for all heavy
to light transitions.

\subsection{Formulation by using Light-cone Sum Rules in HQEFT}

Now let us calculate the wave functions $A,B$ and $L_i(i=1,2,3,4)$
via light-cone sum rules in HQEFT. Using the same analysis given in
Refs.\cite{WY2,WY3,WY,ZWW} with simply generalizing the relevant
quantities to universal initial and final states, we have
\begin{eqnarray}
A(y) & = & - \frac{f_P}{4 F y} \int^{s_0}_0 d s e^{\frac{2 \bar{\Lambda}_M - s}{T}}
 \left[ \frac{1}{y} \frac{\partial}{\partial u} g_2(u) - \mu_P \phi_p (u) \right. \nonumber \\
& & - \left. \frac{\mu_P}{6} \frac{\partial}{\partial u }\phi_\sigma (u) \right]
 \left|_{u = 1- \frac{s}{2 y}}\right. ,\\
B(y) & = & - \frac{f_P}{4 F} \int^{s_0}_0 d s e^{\frac{2 \bar{\Lambda}_M -s}{T}}\left [
- \phi_P (u) + \frac{1}{y^2} \frac{\partial^2}{\partial u^2} g_1 (u) \right.\nonumber \\
& & - \left. \frac{1}{y^2} \frac{\partial}{\partial u } g_2(u) +
\frac{\mu_P}{6 y} \frac{\partial}{\partial u} \phi_\sigma (u)
\right] \left|_{u= 1-\frac{s}{2 y}}\right. ,
\end{eqnarray}
for $M \rightarrow P$ decays and
\begin{eqnarray}
L_1(y) & = & \frac{1}{4 F } e^{\frac{2 \bar{\Lambda}_M}{T}} \int^{s_0}_0 d s e^{-\frac{s}{T}}
\frac{1}{y}\left [ f_V m_V g^{(v)}_\perp (u) \right. \nonumber \\
& & \left. +\frac{1}{4} ( f_V - f^T_V \frac{m_{q_1} + m_{q_2}}{m_V}
) m_V
\frac{\partial}{\partial u} g^{(a)}_\perp (u) \right. \nonumber \\
& & \left. + \frac{f^T_V m_V^2}{2 y} C_T(u)\right ]\left |_{u = 1- \frac{s}{2 y}}\right. ,\\
L_2(y) & = & \frac{1}{4 F} e^{\frac{2 \bar{\Lambda}_M }{T}}
\int^{s_0}_0 d s e^{-\frac{s}{T}}
\left  \{ \frac{f^T_V m_V^2}{y^2} [ \frac{1}{2} C_T(u) \right. \nonumber \\
& & \left.+ B_T(u) + \frac{1}{2} \frac{\partial}{\partial u}
h^{(s)}_\parallel (u) ]
+f_V m_V^2 [ \frac{m_V}{2 y^3} C(u) \right. \nonumber \\
& & \left. - \frac{m_{q_1}+m_{q_2}}{2 y^2 m_V}
\frac{\partial}{\partial u} h^{(s)}_\parallel (u)  ] \right \}
\left|_{u=1-\frac{s}{2 y}}\right.,\\
L_3(u) & = & \frac{1}{4 F} e^{\frac{2 \bar{\Lambda}_M }{T}} \int^{s_0}_0 d s e^{-\frac{s}{T}}\left
\{ - \frac{1}{4 y} [ f_V\right. \nonumber \\
& & \left. - f^T_V \frac{m_{q_1} + m_{q_2}}{m_V} ]
m_V [ \frac{\partial}{\partial u} g^{(a)}_\perp (u) ] + f^T_V [ \phi_\perp(u) \right. \nonumber \\
& & \left. - \frac{m_V^2}{16 y^2} \frac{\partial^2}{\partial u^2}
A_T(u)] \right\}
\left|_{u = 1- \frac{s}{2y}}\right. ,\\
L_4(y) & = & \frac{1}{4 F} e^{\frac{2 \bar{\Lambda}_M}{T}} \int^{s_0}_0 d s e^{- \frac{s}{T}}
\frac{1}{y}\left \{ f_V m_V [ \phi_\parallel (u) - g^{(v)}_\perp (u) \right. \nonumber \\
& & \left. - \frac{1}{4} \frac{\partial}{\partial u} g^{(a)}_\perp -
\frac{m_V^2}{16 y^2}
\frac{\partial^2}{\partial u^2} A(u) ]+\frac{f^T_V m_V^2}{y} B_T (u) \right. \nonumber \\
& & \left. + \frac{1}{4} f^T_V (m_{q_1} + m_{q_2})
\frac{\partial}{\partial u} g^{(a)}_\perp (u) \right\}\left|_{u=1-
\frac{s}{2y}}\right. ,
\end{eqnarray}
for $M \rightarrow V$ decays. $s_0$ and $T$ are the heavy meson
threshold energy and Borel transformation parameter respectively.
$m_{q_1}$ and $m_{q_2}$ are the quark masses in light final state
meson, for which we choose $m_u = m_d =0$ and $m_s =0.15$GeV at
present. We consider all the meson distribution amplitudes up to
twist-4, their definitions are presented in Appendix A.

As has been proved in Ref.\cite{ZWW}, the wave functions
$B^\prime(y)$ and $L_i^\prime (y)$ for penguin type form factors
are exactly the same as $B(y)$ and $L_i(y)$ at the leading order
of $1/m_Q$ expansion. So there are four exact relations that
relate the penguin type form factors with the semileptonic type
ones
\begin{eqnarray}
f_T (q^2) & = & \frac{m_M + m_P}{2 m_M} \left[ (1+\frac{m_M^2
-m_P^2}{q^2})
f_+ (q^2) -\frac{m_M^2 - m_P^2}{q^2} f_0(q^2) \right] ,\\
T_1(q^2) & = & \frac{m_M^2 - m_V^2 +q^2}{2 m_M} \frac{V(q^2)}{m_M + m_V} + \frac{m_M + m_V}{2 m_M} A_1(q^2) ,\\
T_2(q^2) & = & \frac{2}{m_M^2-m_V^2} \left [ \frac{(m_M - y) (m_M +m_V)}{2} A_1(q^2) \right. \nonumber \\
& & \left. + \frac{m_M (y^2 - m_V^2)}{m_M +m_V} V(q^2) \right] ,\\
T_3(q^2) & = & -\frac{m_M + m_V}{2 m_M} A_1(q^2) +\frac{m_M - m_V}{2 m_M}[ A_2(q^2) -A_3(q^2) ]\nonumber \\
& & +\frac{m_M^2 + 3 m_V^2 -q^2}{2 m_M ( m_M + m_V)} V(q^2).
\end{eqnarray}
which hold for the whole region of momentum transfer. The second
relation was also noticed in ref.\cite{ABS} by using QCD sum rule
approach. In Refs. \cite{pvesrb,pab}, it was shown that the
Isgur-Wise relations are satisfied very well at $q^2\rightarrow 0$
(large recoil) and hold with about $80\%$ accuracy at large $q^2$.
In Ref.\cite{cfss}, it was found that the Isgur-Wise relations are
valid up to $70\%$ in the whole $q^2$ region by applying for the
three point QCD sum rules method. In the quark model, the authors
of Refs.\cite{stech,soares} concluded that Isgur-Wise relations
also hold at large recoil. It was also shown in \cite{ZWW} that
the large energy effective theory (LEET) relations \cite{cyopr}
hold within 80\% accuracy at large recoil point on the whole, and
most of them even hold better than 90\% accuracy.

It is seen that within the framework of HQEFT, we have only six
independent form factors at the leading order of $1/m_Q$
expansion, two for $M \rightarrow P$ and four for $ M \rightarrow
V$ decays, namely $f_+, f_0, A_i(i=1,2,3)$ and $V$. They can be
represented by the universal wave functions $A, B$ and $L_i$
($i=1,2,3,4$), which have been formulated by using light-cone sum
rules in Eqs.(21-26).

\section{Numerical Analysis and Results of the Form Factors}

Given the formulae above, we are now in the stage to evaluate the
form factors. For that, it needs to know the light meson
distribution amplitudes ( DAs ) which have been studied by several
groups. We shall use the results given in Refs.\cite{PB1, AK1, AK2,
BF1} for pseudoscalar mesons and the ones in Refs.\cite{PB2,PB3} for
vector mesons.

The leading twist ( twist-2 ) meson distribution amplitudes are
given by
\begin{eqnarray}
\phi_P (u, \mu) & = & 6 u (1-u) \left [ 1+ \sum^4_{n=1} a^P_n (\mu) C^{3/2}_n (2u-1) \right] ,\\
\phi_{\parallel (\perp)} (u, \mu) & = & 6 u (1-u) \left\{ 1
+ 3 a^{\parallel (\perp)}_1 ( \mu) (2u-1)\right.  \nonumber \\
& & \left.+ a^{\parallel (\perp)}_2 ( \mu) \frac{3}{2}[ 5 (2u-1)^2
- 1 ] \right\} ,
\end{eqnarray}
Distribution amplitudes of higher twist are listed in Appendix B. We
choose $ \mu_b = \sqrt{m^2_B - m^2_b} \simeq \sqrt{2\bar{\Lambda}_B
m_B} \simeq 2.4 $GeV and $ \mu_c = \sqrt{m^2_D -m^2_c} \simeq
\sqrt{2\bar{\Lambda}_D m_D} \simeq 1.3 $GeV as the scales of
$B_{(s)}$ and $D_{(s)}$ decays respectively, which are the typical
virtualities of heavy quarks.
%
%
From Eqs.(11, 12, 14-17, 21-26) and the relations given by
Eqs.(27-30), we are able to calculate systematically all the form
factors of heavy to light meson decays at the leading order of
$1/m_Q$ expansion.

The parameters relevant to specific light mesons are collected in
Tabs.1-2. Other parameters are listed in the following:

\begin{eqnarray}
& & B_2(\mu_b) =0.29  \hspace{0.5cm} B_4(\mu_b) = 0.58  \hspace{0.5cm} C_2(\mu_b)
= 0.059  \hspace{0.5cm} C_4(\mu_b) = 0.034 \nonumber \\
& & B_2(\mu_c) =0.41  \hspace{0.5cm} B_4(\mu_c) = 0.925  \hspace{0.3cm} C_2(\mu_c)
= 0.087  \hspace{0.5cm} C_4(\mu_c) = 0.054 \nonumber \\
& & \delta^2(\mu_b) =0.17 {\rm GeV}^2  \hspace{0.5cm} \epsilon(\mu_b) =0.36 \nonumber \\
& & \delta^2(\mu_c) =0.19 {\rm GeV}^2  \hspace{0.5cm} \epsilon(\mu_c) =0.45 \nonumber \\
& & \mu_P(\mu_b) = 2.02 {\rm GeV}  \hspace{0.5cm}  \mu_P(\mu_c) 1.76 {\rm GeV}
\end{eqnarray}
for the pseudoscalar mesons
\begin{eqnarray}
& & \zeta_3(\mu_c)=0.032 \hspace{0.5cm} \zeta_4(\mu_c)=0.15 \hspace{0.5cm}
\zeta^T_4(\mu_c)=0.10 \hspace{0.5cm} \tilde{\zeta}^T_4(\mu_c)=-0.10\nonumber \\
& & \zeta_3(\mu_b)=0.018 \hspace{0.5cm} \zeta^T_4(\mu_b)=0.06 \hspace{0.4cm}
 \tilde{\zeta}^T_4(\mu_b)=-0.06\nonumber \\
& & \omega^V_3(\mu_c)=3.8 \qquad \omega^A_3(\mu_c)=-2.1 \hspace{0.3cm}
\omega^T_3(\mu_c)=7.0 \hspace{0.6cm} \omega^A_4(\mu_c)=0.8 \nonumber \\
& & \omega^V_3(\mu_b)=3.6 \qquad \omega^A_3(\mu_b)=-1.7 \hspace{0.3cm} \omega^T_3(\mu_b)=7.2  \nonumber \\
& & \ll Q^{(1)} \gg(\mu_c) =-0.15 \qquad \ll Q^{(3)} \gg (\mu_c) =0 \nonumber \\
& & \ll Q^{(1)} \gg(\mu_b) =-0.07 \qquad \ll Q^{(3)} \gg (\mu_b) =0
\end{eqnarray}
for the vector mesons and
\begin{eqnarray}
& & m_B = 5.28 {\rm GeV} \qquad \bar{\Lambda}_B = 0.53 {\rm GeV}
\qquad \bar{\Lambda} = 0.53 {\rm GeV} \nonumber \\
& & m_{B_s} = 5.37 {\rm GeV} \qquad
 \bar{\Lambda}_{B_s} = 0.62 {\rm GeV}
\qquad F = 0.30 {\rm GeV}^{3/2} \mbox{\cite{W3}} \nonumber \\
 & & m_D = 1.87 {\rm GeV} \qquad \bar{\Lambda}_D = 0.53 {\rm GeV} \nonumber \\
 & & m_{D_s} =1.97 {\rm GeV} \qquad
 \bar{\Lambda}_{D_s} = 0.63 {\rm GeV}
\end{eqnarray}
for the heavy mesons in initial states. Note that the twist-4
distribution amplitudes $g_3(u,\mu)$ and ${\mathbb
A}_\parallel(u,\mu)$ have been neglected for $B_{(s)}$ decays,
because their contributions are negligible and $\omega^A_4(\mu_b)$
is not known\cite{P2,PB2}. We choose the region of threshold energy
$s_0$ and Borel parameter $T$ so that the curves of form factors
become most stable. In the evaluation, we adjust $s_0$ and $T$ for
all decays of $B \rightarrow P$ consistently. The same procedures
are performed for $B_s \rightarrow P$, $B_{(s)} \rightarrow V$,
$D_{(s)} \rightarrow P$ and $D_{(s)} \rightarrow V$ decays. Our
interesting regions for the Borel parameter $T$ are around
$T=2.0$GeV and $1.5$GeV for $B_{(s)}$ and $D_{(s)}$ decays
respectively. As illustrations, we show how the form factors vary as
functions of $T$ for different $s_0$ in Fig.1 for $B \rightarrow
K^*$ decay and in Fig.2 for $B \rightarrow \rho$ decay.

It is well known that the light-cone sum rules may be broken down at
large momentum transfer, i.e. $q^2 \sim m_Q^2$. To get reasonable
behavior of the form factors in the whole kinematically accessible
region, we use the following parametrization
\begin{eqnarray}
F(q^2) = \frac{F(0)}{1-a_F q^2/m_M^2 + b_F ( q^2/m_M^2)^2} ,
\end{eqnarray}
where $F(q^2)$ can be any of the form factors $f_+$, $f_0$,
$A_i(i=1,2,3)$ and $V$. For $B \rightarrow \pi, B_s \rightarrow K$
and $D \rightarrow \pi (K)$ decays, we may use the single pole
approximations for the form factor $f_+$ in the large $q^2$ region.
\begin{eqnarray}
f_+ (q^2) = \frac{f_{M^*} g_{M^* M \pi}}{2 m_{M^*}(1-q^2/m_{M^*}^2)}
\end{eqnarray}
with $f_{B^*}=0.16 \pm 0.03$ GeV, $g_{B^* B \pi}=29 \pm
3$\cite{AK1}, $f_{D^*}g_{D^* D \pi} = 2.7 \pm 0.8$ GeV,
$f_{D^*_s}g_{D^*_s D K } = 3.1 \pm 0.6$ GeV\cite{AK2} and
$f_{B^*}g_{B^*B_sK} = 3.88 \pm 0.31$ GeV\cite{ZH}. For other decays,
we use only the light-cone sum rule predictions to fit the
parameters in Eq.(36). With the above considerations, we obtain the
form factors in the whole kinematically accessible region shown in
Figs.3-32. Numerical results are presented in Tabs.4-7. To see the
contributions from higher twist distribution amplitudes and meson
mass corrections, we also present here the form factors obtained
with only considering the leading twist meson DAs for $B_{(s)},
D_{(s)} \rightarrow V$ transitions. The comparison of our results
with other groups are given in Tabs.8-10, where the form factor
ratios are defined as $R_V \equiv V(0)/A_1(0)$, $R_2 \equiv
A_2(0)/A_1(0)$. The penguin type form factors at $q^2=0$ given by
Eqs.(27-30) are also listed explicitly in Tab.11 for a complete
analysis.

Roughly speaking, the form factor in this work are consistent with
the ones obtained via light-cone sum rules in full QCD\cite{P1,P2}
but smaller than the quark model predictions\cite{BSW}. Note that
our present results are also lower than the previous
ones\cite{WY,ZWW}. This is mainly because of a different choice for
the threshold energy $s_0$ and Borel parameter $T$ when consistently
considering all the relevant decay channels. The form factor ratio
$R_V$ agrees with the measurements of FOCUS\cite{ML} and
BEATRICE\cite{MA} roughly, while $R_2$ is in the low side of the
experimental data when including the contributions from higher twist
light meson DAs and mass corrections, especially for the $D$ meson
transitions. It is seen via comparing with the results obtained from
only considering the leading twist meson DAs. This may indicate that
when including the contributions from higher twist distribution
amplitudes, the $1/m_Q$ corrections in heavy quark expansion may
need to be considered. On the other hand, it still needs to improve
the experimental measurements for a consistent check. In our present
considerations, the uncertainties of form factors are mainly due to
the variations of parameters $s_0$ and $T$. As illustrated in
Figs.1-2, the form factors become larger when increasing $s_0$ and
decreasing $T$, but the allowed regions of $s_0$ and $T$ are
constrained by the stability. In general, it leads to about $(5-10)
\%$ uncertainties. The possible uncertainties of form factors in the
whole kinematically accessible region are shown in Figs.3-32.

\section{Branching Ratios of Heavy to Light Exclusive Decays with extraction of the CKM matrix elements $|V_{ub}|$,
$|V_{cs}|$, $|V_{cd}|$}

In this section,  we shall apply the above obtained form factors to
calculate the branching ratios for heavy to light exclusive
semileptonic and radiative decays. As some of the decay rates have
been well measured, they can be used to extract the important CKM
matrix elements $|V_{ub}|$, $|V_{cs}|$ and $|V_{cd}|$. Then other
branching ratios are predicted, which can be tested by further more
precise experiments.

The relevant decay width formulae of rare decays have the
following forms\cite{AP}
\begin{eqnarray}
\frac{d \Gamma}{d \hat{s}} & = & \frac{G^2_F \alpha^2 m_M^2}{2^{10} \pi^5}\left[(
{|A^\prime|}^2 + {|C^\prime|}^2 ) ( \lambda - \frac{{\hat{u} ( \hat{s})}^2}{3} )\right. \nonumber \\
& & \left. + {|C^\prime|}^2 4 \hat{m}^2_l ( 2 + 2 \hat{m}^2_P - \hat{s} ) + Re(C^\prime D^{\prime *} )
 \right. \nonumber \\
& & \left. \times 8 \hat{m}^2_l ( 1- \hat{m}^2_P) + {|D^\prime|}^2 4 \hat{m}^2_l \hat{s} \right]
\end{eqnarray}
for the heavy to pseudoscalar meson decays $M \rightarrow P \ell^+
\ell^-$,

\begin{eqnarray}
\frac{d \Gamma}{d \hat{s}} & = & \frac{G^2_F \alpha^2 m_M^2}{2^{10}
\pi^5} \left \{ \frac{{|A|}^2}{3} \hat{s} \lambda ( 1 + 2
\frac{\hat{m}^2_l}{\hat{s}} ) + {|E|}^2 \hat{s}
\frac{{\hat{u}(\hat{s})}^2}{3} + \frac{1}{4 \hat{m}^2_V} \left[
|B|^2 ( \lambda  -\frac{{\hat{u}(\hat{s})}^2}{3} \right. \right. \nonumber \\
& & \left.+ 8 \hat{m}^2_V ( \hat{s} + 2 \hat{m}^2_l ) )  + |F|^2 (
\lambda - \frac{{\hat{u}(\hat{s})}^2}{3} + 8 \hat{m}^2_V ( \hat{s}
- 4 \hat{m}^2_l ) ) \right ] \nonumber \\
& & + \frac{\lambda}{4 \hat{m}^2_V} \left[ |C|^2 ( \lambda -
\frac{{\hat{u}(\hat{s})}^2}{3} ) + |G|^2 (  \lambda  -
\frac{{\hat{u}(\hat{s})}^2}{3} + 4 \hat{m}^2_l ( 2 + 2 \hat{m}^2_V -
\hat{s} ) )
\right] \nonumber \\
& & - \frac{1}{2 \hat{m}^2_V} \left [ Re(B C^*) ( \lambda  -
\frac{{\hat{u}(\hat{s})}^2}{3} )(
1- \hat{m}^2_V - \hat{s} ) + Re(F G^*) ( \lambda - \frac{{\hat{u}(\hat{s})}^2}{3} ) \right. \nonumber \\
& & \left. \times ( 1- \hat{m}^2_V  - \hat{s} ) + 4 \hat{m}^2_l
\lambda )\right ]
- 2 \frac{\hat{m}^2_l}{\hat{m}^2_V} \lambda  \left [ Re(F H^*) - Re(G H^*) ( 1- \hat{m}^2_V ) \right]\nonumber \\
& & \left. + \frac{\hat{m}^2_l}{\hat{m}^2_V} \hat{s} \lambda |H|^2
\right\}
\end{eqnarray}
for the heavy to vector meson decays $M \rightarrow V \ell^+
\ell^-$, and
\begin{eqnarray}
\Gamma = \frac{G^2_F \alpha }{32 \pi^4} m^2_Q m^3_M ( 1- \frac{m^2_V}{m^2_M})^3 |C^{eff}_7|^2 |T_1(0)|^2
\end{eqnarray}
for the $M \rightarrow V \gamma$ decays.

The auxiliary functions are defined as
\begin{eqnarray}
A^\prime(\hat{s}) & = & C^{eff}_9 (\hat{s}) f_+ (\hat{s}) + \frac{2 \hat{m}_Q}{1 + \hat{m}_P} C^{eff}_7 f_T(\hat{s}) ,\\
C^\prime (\hat{s})& = & C_{10} f_+ (\hat{s}) ,\\
D^\prime (\hat{s})& = & C_{10} \frac{1-\hat{m}_P}{\hat{s}} (f_0 (\hat{s}) - f_+(\hat{s})) ,\\
A(\hat{s}) & = & \frac{2}{1 + \hat{m}_V} C^{eff}_9 (\hat{s})
V(\hat{s}) + \frac{4 \hat{m}_Q}{\hat{s}}
C^{eff}_7 T_1 (\hat{s}) ,\\
B(\hat{s}) & = & (1+\hat{m}_V)\left [ C^{eff}_9 ( \hat{s} ) A_1
(\hat{s} ) + \frac{2 \hat{m}_Q}{\hat{s}}
( 1- \hat{m}_V ) C^{eff}_7 T_2 (\hat{s}) \right] ,\\
C(\hat{s}) & = & \frac{1}{1-\hat{m}^2_V}\left [ (1-\hat{m}_V )
C^{eff}_9 (\hat{s}) A_2 (\hat{s})
+ 2 \hat{m}_Q C^{eff}_7 ( T_3 (\hat{s}) \right. \nonumber \\
& & \left. + \frac{1- \hat{m}^2_V}{\hat{s}} T_2(\hat{s}) ) \right]
\\
E(\hat{s}) & = & \frac{2}{1+\hat{m}_V} C_{10} V(\hat{s}) ,\\
F (\hat{s}) & = & (1+\hat{m}_V) C_{10} A_1 (\hat{s}) ,\\
G(\hat{s}) & = & \frac{1}{1+\hat{m}_V} C_{10} A_2 (\hat{s}) ,\\
H (\hat{s} ) & = & \frac{1}{1 + \hat{m}_V} C_{10}A_3 (\hat{s}) \\
\lambda (\hat{s}) & = & 1 + \hat{m}^4_{P(V)} + \hat{s}^2 - 2 \hat{s} - 2 \hat{m}^2_{P(V)} ( 1 + \hat{s}) ,\\
\hat{u} ( \hat{s}) & = & \sqrt{\lambda(\hat{s}) ( 1- 4
\frac{\hat{m}^2_l}{\hat{s}} ) } ,
\end{eqnarray}
where $\hat{s} = \frac{q^2}{m^2_M}$, $\hat{m}_{P(V)} = \frac{m_{P(V)}}{m_M}$ and $ \hat{m}_l = \frac{m_l}{m_M}$.

For $B_{(s)}$ decays, the Wilson coefficients $C_i$ are calculated
in the naive dimensional regularization (NDR) scheme in \cite{AB}.
At the scale $\mu_b$, their values are listed in Tab.3.

The effective Wilson coefficient is defined as \cite{AG,AJB}
\begin{eqnarray}
C^{eff}_9 (\hat{s}) & = & C_9 + g(\hat{m}_c, \hat{s}) ( 3 C_1 + C_2 + 3 C_3 + C_4 + 3 C_5 + C_6 ) \nonumber \\
& & - \frac{1}{2} g(1, \hat{s}) ( 4 C_3 + 4 C_4 + 3 C_5 + C_6 ) - \frac{1}{2} g(0, \hat{s}) (C_3 + 3 C_4) \nonumber \\
& & + \frac{2}{9} ( 3 C_3 + C_4 + 3 C_5 + C_6 ) - \frac{V^*_{u q} V_{u b}}{V^*_{t q} V_{t b}} ( 3 C_1 +C_2) \nonumber \\
& & \times [ g(0, \hat{s}) - g(\hat{m}_c, \hat{s}) ]
\end{eqnarray}
with
\begin{eqnarray}
g(0, \hat{s}) & = & \frac{8}{27} - \frac{8}{9} \ln \frac{m_b}{\mu} - \frac{4}{9} \ln \hat{s} + \frac{4}{9} i \pi ,\\
g(z, \hat{s}) & = & \frac{8}{27} - \frac{8}{9} \ln \frac{m_b}{\mu} -
\frac{8}{9} \ln z + \frac{4}{9} x
 -\frac{2}{9} ( 2+ x) \sqrt{|1-x|}\nonumber \\
& & \times \left\{
\begin{array}{ll}
(\ln \left| \frac{\sqrt{1-x} +1}{\sqrt{1-x} - 1}\right| - i \pi ),
\hspace{1.05 cm} {\rm for}
\hspace{0.2cm} x \equiv \frac{4 z^2}{\hat{s}} \leq 1\\
2 \arctan \frac{1}{\sqrt{x-1}}, \hspace{1.7cm} {\rm for}
\hspace{0.2cm} x \equiv \frac{4z^2}{\hat{s}} > 1
\end{array}
\right. .
\end{eqnarray}
Note that we have dropped the CKM factor $|V_{tq}^*V_{tb}|$ which
have to be multiplied by $C_i$ in calculations. We have also
neglected the $c \bar{c}$ resonance contributions from $J/\Psi,
\Psi^\prime, \cdot \cdot \cdot, \Psi^{(v)}$.

For the relevant Wilson coefficients in $D_{(s)}$ decays, we take
the formulae given in Refs.\cite{BG,CG,FP}
\begin{eqnarray}
C^{eff}_7 (\mu_c) & = & \sum^8_{i =1}(\frac{\alpha_s
(m_b)}{\alpha_s(\mu_c)})^{c_i}
[ a_i C_1(m_b) + b_i C_2(m_b) ] ,\\
C^{eff}_9 & = & C_9 + \sum_{i = d, s} \lambda_i \left [ -
\frac{2}{9} \ln x_i + \frac{8}{9} \frac{z^2_i}{\hat{s}} -
\frac{1}{9} ( 2 + \frac{4 z^2_i}{\hat{s}})
\sqrt{\left|1-\frac{4 z^2_i}{\hat{s}}\right|}\right.\nonumber \\
& & \left. \times {\cal T} (z_i) \right] + \frac{3 \pi}{\alpha^2}
\sum_{i=\phi,\rho,\omega} \kappa_i \frac{m_{V_i} \Gamma_{V_i
\rightarrow
\ell^+ \ell^-}}{m^2_{V_i} - s - i m_{V_i} \Gamma_{V_i}} ,\\
C_{10} & = & - \sum_{i = d, s} \lambda_i \frac{(\bar{C}_i^{box}(x_i)
+ \bar{C}_i^Z(x_i))}{2 \sin^2\theta_W} ,
\end{eqnarray}
with
\begin{eqnarray}
C_{1,2}(m_b)&  = & \frac{1}{2}
(\frac{\alpha_s(M_W)}{\alpha_s(m_b)})^{\frac{6}{23}}
\mp (\frac{\alpha_s(M_W)}{\alpha_s(m_b)})^{-\frac{12}{23}} ,\\
C_9 & = & \sum_{i=d,s} \lambda_i \left[ - (F^i_1 (x_i) + 2
\bar{C}^Z_i (x_i))
+ \frac{(\bar{C}_i^{box}(x_i) + \bar{C}_i^Z(x_i))}{2 \sin^2\theta_W} \right] ,\\
{\cal T} (z) & = & \left\{
\begin{array}{ll}
2 \arctan \left [\frac{1}{\sqrt{\frac{4 z^2}{\hat{s}} - 1}}\right ]
 \hspace{0.5cm} {\rm for} \hspace{0.2cm} \hat{s} \leq 4 z^2 \\
\ln \left | \frac{1+\sqrt{1- \frac{4 z^2}{\hat{s}}}}{1- \sqrt{1-
\frac{4 z^2} {\hat{s}}}} \right | - i \pi \hspace{0.5cm} {\rm for}
\hspace{0.2cm} \hat{s} > 4 z^2
\end{array}
\right. ,
\end{eqnarray}
and
\begin{eqnarray}
& & a_i = ( -\frac{65710}{18413}, \hspace{0.3cm}\frac{22173}{8590},
\hspace{0.2cm}\frac{2}{5}, \hspace{0.5cm}0, \hspace{0.4cm}0.6524,
\hspace{0.3cm}-0.0532, \hspace{0.3cm}-0.0034,
\hspace{0.3cm}-0.0084) \nonumber \\
& & b_i = ( -\frac{675158}{165717},
\hspace{0.15cm}\frac{23903}{8590}, \hspace{0.2cm}\frac{2}{5},
\hspace{0.5cm}0, \hspace{0.4cm}0.8461, \hspace{0.4cm}0.0444,
\hspace{0.6cm}0.0068,
\hspace{0.4cm}-0.0059) \nonumber \\
& & c_i = (\hspace{0.5cm}\frac{14}{25}, \hspace{1.1cm}\frac{16}{25},
\hspace{0.4cm}\frac{6}{25}, \hspace{0.1cm}-\frac{12}{25},
\hspace{0.1cm}0.3469, \hspace{0.3cm}-0.4201, \hspace{0.3cm}-0.8451,
\hspace{0.3cm}0.1317)
\end{eqnarray}

Here we have defined $x_i = m_i^2/M^2_W$, $z_i = m_i/m_M$ and
$\lambda_i = V^*_{c i} V_{u i}$. The functions $\bar{C}^{box}_i (
x_i)$, $\bar{C}^Z_i ( x_i)$ and $F^i_1 (x_i)$ are given in
Appendix C. It is noted that the $\phi$, $\rho$ and $\omega$
resonance effects have been included in $D_{(s)}$ decays, which
comprise the largest long distance contributions.

According to Eqs.(56, 59, 62), the effective Wilson coefficient is
given by $C^{eff}_7 = 0.094$, $0.086$ and $0.079$ for $\mu_c 1.24$,
$1.34$ and $1.44 {\rm Gev}$ respectively. In this paper, we shall
use $C^{eff}_7 = 0.086$ corresponding to $\mu_c = 1.34 {\rm Gev}$
with the CKM factor $V^*_{cb}V_{ub}$ multiplied.

For semileptonic decays, we take the formulae in Ref.\cite{KW},
which have the following forms
\begin{eqnarray}
\frac{d \Gamma}{d \hat{s}}(M \rightarrow P  \ell \nu_\ell ) &  & \frac{G_F^2 |V_{Q q}|^2 m_M^5}{32 \pi } \lambda^{1/2} (1, \hat{m}_P^2,
 \hat{s}) \left [ H^2_T (\hat{s}) \rho_T (\hat{s}) \right.\nonumber \\
& & \left.+ H_L^2 (\hat{s}) \rho_L (\hat{s}) \right] , \\
\frac{d \Gamma}{d \hat{s}} (M \rightarrow V \ell \nu_\ell ) & & \frac{G_F^2 |V_{Q q}|^2 m_M^5}{32 \pi  } \lambda^{1/2} (1, \hat{m}_V^2, \hat{s}) \hat{s} \left \{ [ H^2_+
(\hat{s})+H_-^2 (\hat{s}) \right.\nonumber \\
& & \left. + H_{0 T}^2 (\hat{s}) ]\rho_T (\hat{s})+ H_{0 L}^2
(\hat{s}) \rho_L (\hat{s}) \right \} ,
\end{eqnarray}
with
\begin{eqnarray}
H_T (\hat{s}) & = & \lambda^{1/2} ( 1, \hat{m}_P^2, \hat{s}) f_+ (\hat{s}) ,\\
H_L (\hat{s}) & = & \lambda^{1/2} ( 1, \hat{m}_P^2, 0) f_0 (\hat{s}) ,\\
H_\pm (\hat{s}) & = & (1+\hat{m}_V) A_1(\hat{s}) \mp
\frac{\lambda^{1/2}
( 1, \hat{m}^2_V, \hat{s})}{1 + \hat{m}_V} V(\hat{s}) ,\\
H_{0 T} (\hat{s})&  = & \frac{1+ \hat{m}_V}{2 \hat{m}_V
\sqrt{\hat{s}}}\left [ ( 1 - \hat{m}_V^2
- \hat{s} ) A_1(\hat{s}) - \frac{\lambda^{1/2} (1, \hat{m}_V, \hat{s} ) }{ (1+\hat{m}_V)^2}\right ] A_2(\hat{s}) ,\\
H_{0 L} (\hat{s})&  = & \frac{1}{\sqrt{\hat{s}}} \lambda^{1/2} ( 1,
\hat{m}_V^2, \hat{s} )
 \frac{1}{2 \hat{m}_V} \left [ (1+\hat{m}_V ) A_1 (\hat{s})\right. \nonumber \\
& & \left. - ( 1- \hat{m}_V ) A_2(\hat{s}) -
\frac{\hat{s}}{1+\hat{m}_V} A_3 (\hat{s}) \right ] ,
\\
\rho_T(\hat{s}) & = & \frac{1}{3 (2 \pi )^2 \hat{s}^3} \lambda^{1/2}
( \hat{m}_1^2, \hat{m}_2^2, \hat{s})
[ 2 \hat{s}^2 - (\hat{m}_1^2 + \hat{m}_2^2 ) \hat{s}\nonumber \\
& &  - ( \hat{m}_1^2 - \hat{m}_2^2 )^2 ] ,\\
\rho_L (\hat{s}) & = & \frac{1}{(2 \pi)^2 \hat{s}^3} \lambda^{1/2} (
\hat{m}_1^2, \hat{m}_2^2, \hat{s} ) [
( \hat{m}_1^2 + \hat{m}_2^2 ) \hat{s} - ( \hat{m}_1^2 - \hat{m}_2^2 )^2 ] ,\\
& & \lambda ( x, y, z) = ( x - y - z)^2 - 4 y z .
\end{eqnarray}

With the above analysis, we can now extract the CKM matrix
elements $|V_{ub}|$, $|V_{cs}|$ and $|V_{cd}|$ from some well
measured decay channels. The numerical results are found to be
\begin{eqnarray}
|V_{ub}| & = & ( 3.53^{+0.22+0.28}_{-0.20-0.30}) \times 10^{-3}
\hspace{0.8cm} (B^0 \rightarrow \pi^- e^+ \nu_e , {\rm \hspace{0.2cm} up \hspace{0.2cm}
to \hspace{0.2cm} twist} \hspace{0.2cm} 4)\\
|V_{ub}| & = & ( 3.41^{+0.19+0.43}_{-0.17-0.49}) \times 10^{-3}
\hspace{0.8cm} (B^0 \rightarrow \rho^- e^+ \nu_e, {\rm \hspace{0.2cm} up \hspace{0.2cm}
to \hspace{0.2cm} twist} \hspace{0.2cm} 4)\\
|V_{ub}| & = & ( 3.47\pm 0.56) \times 10^{-3} \hspace{1.1cm} ( {\rm
\hspace{0.2cm} average
\hspace{0.2cm} result } )\\
|V_{ub}| & = & ( 3.72^{+0.21+0.42}_{-0.18-0.52}) \times 10^{-3}
\hspace{0.8cm} (B^+ \rightarrow \rho^0 e^+ \nu_e, {\rm
\hspace{0.2cm} up \hspace{0.2cm}
to \hspace{0.2cm} twist} \hspace{0.2cm} 2) \nonumber \\
|V_{cs}| & =& 1.003^{+0.074+0.027}_{-0.066-0.027} \hspace{2cm}
(D^0\rightarrow K^- \mu^+ \nu_\mu , {\rm \hspace{0.2cm} up \hspace{0.2cm}
to \hspace{0.2cm} twist} \hspace{0.2cm}4 ) \\
|V_{cs}| & =& 1.008^{+0.022+0.062}_{-0.022-0.066} \hspace{2cm} ( D^+
\rightarrow \bar{K}^{* 0} e^+ \nu_e, {\rm \hspace{0.2cm} up \hspace{0.2cm}
to \hspace{0.2cm} twist} \hspace{0.2cm} 4 )\\
|V_{cs}| & =& 1.006^{+0.045+0.062}_{-0.040-0.066} \hspace{2cm} (
D^+\rightarrow \bar{K}^{* 0} e^+ \nu_e, {\rm \hspace{0.2cm} up \hspace{0.2cm}
to \hspace{0.2cm} twist} \hspace{0.2cm}2 ) \nonumber \\
|V_{cd}| & = & 0.234^{+0.009+0.043}_{-0.011-0.053} \hspace{2cm} (D^+
\rightarrow \rho^0 e^+ \nu_e, {\rm \hspace{0.2cm} up \hspace{0.2cm}
to \hspace{0.2cm} twist} \hspace{0.2cm} 4 ) \\
|V_{cd}| & = & 0.248^{+0.010+0.046}_{-0.009-0.056} \hspace{2cm} (
D^+ \rightarrow \rho^0 e^+ \nu_e, {\rm \hspace{0.2cm} up
\hspace{0.2cm} to \hspace{0.2cm} twist} \hspace{0.2cm} 2)
\nonumber
\end{eqnarray}
The first uncertainties come from theoretical calculations and the
second ones from experimental measurements. Among these three CKM
matrix elements, $|V_{ub}|$ is the most important one and has been
studied by many groups either theoretically or experimentally. The
above theoretical values for $|V_{ub}|$ obtained from considering
the meson DAs up to twist-4 are consistent with the recent results
given in Ref.\cite{WWZN} and also the early results summarized in
Ref.\cite{WWZ}. A larger value $|V_{ub}| = (
3.72^{+0.21+0.42}_{-0.18-0.52}) \times 10^{-3}$ is resulted if only
considering the leading twist DAs. Using the formula in Ref\cite{W4}
for the inclusive decay ${\cal B} (B \rightarrow X_u e \bar{\nu} )$
and taking the recent updated measurement \cite{BAB}, we have
$|V_{ub}| = (3.93 \pm 0.82) \times 10^{-3}$. Thus our average value
from the exclusive and inclusive decays is compatible with the world
average value\cite{PDG}: $|V_{ub}| = (3.67 \pm 0.47) \times
10^{-3}$. For the CKM matrix elements $|V_{cs}|$ and $|V_{cd}|$, the
extracted values are also consistent with the world average
results\cite{PDG}.

We shall use the above extracted result for $|V_{ub}|$ to predict
the branching ratios for other decays. Considering the light meson
DAs to twist-4, the numerical results are given in Tabs.12, 13a, 14,
15a, 16, 17a, 18, 19a by taking the central value
$|V_{ub}|=3.47\times 10^{-3}$. The branching ratios for decays with
electron in the final state are almost equal to the corresponding
decays with muon as a whole (for $m_e \simeq m_\mu \simeq 0$), while
the decays with tau are smaller. Our predictions for the branching
ratios are all compatible with present experiments. For example, the
branching ratios of $B^+ \rightarrow K^{*+} e^+ e^-$, $B^0
\rightarrow \pi^- \ell \nu_\ell$ and $B^0 \rightarrow \rho^- \ell^+
\nu_\ell$ agree well with the experimental values ( see Tabs.13a,
14, 15a ) and the same for $D^0 \rightarrow K^- \mu^+ \nu_\mu$, $D^+
\rightarrow \rho^0 e^+ \nu_e$ and $D^0 \rightarrow K^{* -} e^+
\nu_e$ decays ( see Tabs.18, 19a ). From all the predicted heavy to
light meson decays, some channels such as $B_s \rightarrow \phi
\mu^+ \mu^-$ in Tab.13a, $B^+ \rightarrow \omega \ell^+ \nu_\ell$ in
Tab.15a are very close to the upper limits of experiments, they can
be tested in near future by more precise experiments. For
comparison, we also list the results obtained with only considering
the leading twist meson DAs by taking $|V_{ub}|=3.72\times 10^{-3}$
in Tabs.13b, 15b, 17b, 19b for the $B_{(s)}, D_{(s)} \rightarrow V$
decays.

\section{Conclusion}

In this paper, we have calculated almost all the heavy to light
form factors via light-cone sum rules at the leading order of
$1/m_Q$ expansion in HQEFT. The vector meson distribution
amplitudes are considered up to twist-4 with the contributions
from higher conformal spin partial waves and the light meson mass
corrections are also included. As shown in Ref.\cite{ZWW} due to a
set of exact relations at the leading order of $1/m_Q$ expansion,
the penguin type form factors can be read off from the
semileptonic ones and thus there are only seven independent form
factors at the leading order of $1/m_Q$ expansion, which has been
chosen to be $f_+$, $f_0$, $A_i ( i = 1, 2, 3 )$ and $V$ in our
present paper. This reduces the uncertainties resulting from
parameterizing the form factors in the large momentum transfer
region. In obtaining the form factors presented in Tabs.4-7, we
have chosen the free parameters $s_0$ and $T$ mainly according to
two criterions: One is that the curves of form factors should be
as stable as possible with the variations of these two parameters;
another is that the curves of form factors should be well behaved
at the large $q^2$ region, i.e. the curves with different $s_0$
and $T$ should be parallel with each other as possible. The second
criterion leads to a strong constraint on the parameters $s_0$ and
$T$. As a consequence, only a small variation is allowed for the
two free parameters $s_0$ and $T$, which then results in the small
uncertainties for the form factors. In general, it has been seen
that the form factors in the present work are consistent with the
ones obtained via light-cone sum rules in full QCD\cite{P1,P2} but
smaller than the ones in the quark model\cite{BSW}, which
indicates that the extracting values for $|V_{ub}|$ should be
larger than the previous ones. Another observation is that the
form factor ratio $R_2$ becomes smaller when including the
contributions from higher twist meson DAs and mass corrections,
especially for the $D$ meson decays. As the branching ratios are
not sensitive to the ratio $R_2$ (i.e., form factor $A_2$), the
CKM matrix elements $|V_{ub}|$, $|V_{cs}|$ and $|V_{cd}|$ have
consistently been extracted from some relatively well measured
decay rates. A detailed calculation for the branching ratios of
heavy to light semileptonic and radiative exclusive meson decays
has been performed. Our theoretical predictions are consistent
with the existing experimental data within the uncertainties, some
predictions can be tested by more precise experiments at
colliders. As our present calculations have been carried out at
the leading order of $1/m_Q$ expansion, the results should be
universal for all the frameworks of heavy quark effective field
theory resulting from different approaches. The effects of $1/m_Q$
corrections in the $B \rightarrow \pi e \bar{\nu} $\cite{WWZ} and
$B \rightarrow \rho e \bar{\nu} $\cite{WWZN} decays have been
evaluated and found to be insignificant as it is expected for the
bottom-quark hadronic systems. The corrections in the charm-quark
hadronic systems could become sizable, but it is likely that only
some of form factors (such as $A_2$) which are not dominant for
the branching ratios may receive large corrections. This is seen
from the case that the leading order contributions have already
provided a consistent explanation for some relatively well
measured decay rates. It is expected that the inclusion of $1/m_Q$
corrections within the framework of HQEFT will further improve the
predictions, which is under investigation. Furthermore, measuring
the individual form factors by more precise experiments, such as
B-factories and LHCb, BEPCII and CLEO-C, will be very helpful for
understanding the hadronic structure and low energy dynamics of
QCD.

\acknowledgments

\label{ACK} We would like to thank Drs. P.Ball and R.Zwicky for
their useful comments. This work was supported in part by the
National Science Foundation of China (NSFC) under the grant
10475105, 10491306, and the Project of Knowledge Innovation
Program (PKIP) of Chinese Academy of Sciences.

\appendix
\section*{Appendix}
\setcounter{equation}{0}
\renewcommand{\theequation}{A.\arabic{equation}}
\section{Definitions of Light Meson Distribution Amplitudes}
The light meson distribution amplitudes are defined as
\cite{P2,WY}
\begin{eqnarray}
& & <P(p)|\bar{q}_1 (x) \gamma^\mu \gamma^5 q_2(0) |0> = -i p^\mu
f_P \int^1_0 d u e^{i u p \cdot x}
[ \phi_P(u) + x^2 g_1(u) ] \nonumber \\
& & + f_P ( x^\mu - \frac{x^2 p^\mu}{x \cdot p}) \int^1_0 d u e^{i u p \cdot x} g_2 (u) \\
& & <P(p)|\bar{q}_1 (x) i \gamma^5 q_2 (0) |0> = \frac{f_P
m_P^2}{m_{q_1} + m_{q_2}} \int^1_0
d u e^{i u p \cdot x} \phi_p(u) \\
& & <P(p) | \bar{q}_1(x) \sigma^{\mu \nu} \gamma^5 q_2(0) |0>\nonumber \\
& & = i (p^\mu x^\nu - p^\nu x^\mu) \frac{f_P m_P^2}{6 (m_{q_1}
+m_{q_2})}\int^1_0 d u e^{i u p \cdot x}
\phi_\sigma (u) \\
& & <V(p,\epsilon^*)|\bar{q}_1 (x) \gamma^\mu q_2(0) |0> = f_V m_V
\left \{
\frac{\epsilon^* \cdot x}{p \cdot x} p^\mu \int^1_0 d u e^{i u p \cdot x} [ \phi_\parallel (u) \right. \nonumber \\
& & + \frac{m_V^2 x^2}{16}  A (u) ]+ ( \epsilon^{* \mu} - p^\mu
\frac{\epsilon^* \cdot x}{p \cdot x}) \int^1_0 d u e^{i u p \cdot x} g^{(v)}_\perp (u) \nonumber \\
& & \left. -\frac{1}{2} x^\mu \frac{\epsilon^* \cdot x}{(p \cdot
x)^2} m_V^2  \int^1_0 d u e^{i u p \cdot x}
C(u)\right \}\\
& & <V(p,\epsilon^*)|\bar{q}_1(x)\gamma^\mu \gamma^5 q_2(0)|0> \nonumber \\
& & = - \frac{1}{4} (f_V - f^T_V \frac{m_{q_1} + m_{q_2}}{m_V}) m_V
\varepsilon^\mu_{\nu \alpha \beta}
\epsilon^{* \nu} p^\alpha x^\beta \int^1_0 d u e^{i u p \cdot x} g^{(a)}_\perp (u) \\
& & <V(p,\epsilon^*)| \bar{q}_1 (x) \sigma^{\mu \nu} q_2(0)|0> \nonumber \\
& & = -i f^T_V \left \{ (\epsilon^{* \mu} p^\nu - \epsilon^{* \nu}
p^\mu) \int^1_0 d u e^{i u p \cdot x}
[ \phi_\perp (u) + \frac{m_V^2 x^2}{16}  A_T (u) ]\right. \nonumber \\
& & +(p^\mu x^\nu - p^\nu x^\mu) \frac{\epsilon^* \cdot x}{(p \cdot
x)^2} m_V^2 \int^1_0 d u e^{i u p \cdot x}
B_T(u)\nonumber \\
& & \left. + \frac{1}{2} ( \epsilon^{* \mu} x^\nu - \epsilon^{* \nu
} x^\mu) \frac{m_V^2}{p \cdot x} \int^1_0
d u e^{i u p \cdot x} C_T(u) \right \} \\
& & <V(p, \epsilon^*)| \bar{q}_1(x) q_2(0)|0> \nonumber \\
& & = - \frac{i}{2} ( f^T_V - f_V \frac{m_{q_1} + m_{q_2}}{m_V} )
(\epsilon^* \cdot x) m_V^2 \int^1_0 d u e^{ i u p \cdot x}
h^{(s)}_\parallel (u)
\end{eqnarray}
where $\phi_P(u), \phi_\parallel(u), \phi_\perp(u)$ are the leading
twist (twist-2) distribution amplitudes. $\phi_p(u), \phi_\sigma(u),
g^{(v)}_\perp(u), g^{(a)}_\perp(u), h^{(t)}_\parallel(u),
h^{(s)}_\parallel(u)$
 and $g_1(u), g_2(u),  A(u),  A_T(u), B_T(u),C(u), C_T(u)$ are twist-3 and twist-4
 respectively.

\renewcommand{\theequation}{B.\arabic{equation}}
\section{ Higher Twist Meson Distribution Amplitudes  }

The light meson distribution amplitudes of twist-3 and twist-4
including the contributions from higher conformal spin partial waves
and light meson mass corrections are presented in this
appendix\cite{PB1, AK1, AK2, BF1, PB2, PB3}.

For pseudoscalar mesons:
\begin{eqnarray}
& & \phi_p (u, \mu) = 1+ \left [ B_2 (\mu) - \frac{5}{2} \rho^2_P (\mu) \right ] C^{1/2}_2 (2u-1)\nonumber \\
& & + \left [ B_4 (\mu) - \frac{27}{10} \rho^2_P (\mu) - \frac{81}{10} \rho^2_P (\mu) a^P_2 ( \mu) \right ]
 C^{1/2}_4 ( 2u-1) \\
& & \phi_\sigma (u, \mu) = 6 u (1-u) \left \{ 1 + [ C_2(\mu) - \frac{7}{20} \rho^2_P (\mu) - \frac{5}{3}
\rho^2_P ( \mu) a^P_2(\mu) ] C^{3/2}_2 (2u-1)\right. \nonumber \\
& & \left. + C_4 (\mu) C^{3/2}_4 (2u-1) \right \} \\
& & g_1(u, \mu) = \frac{5}{2} \delta^2 (\mu) (1-u)^2 u^2 + \frac{1}{2} \varepsilon (\mu) \delta^2 (\mu)
\left \{ (1-u) u [ 2 + 13 (1-u) u ]\right. \nonumber \\
& & +10 u^3 (\ln u) ( 2-3u+\frac{6}{5} u^2 ) + 10 (1-u)^3 [ \ln (1-u) ] [ 2 - 3 (1-u) \nonumber \\
& & \left. +\frac{5}{6} (1-u)^2 ] \right \}+a_2 ( \mu) \rho^2_P (\mu) \left \{ \frac{9}{16} u^2
+ \frac{9}{4} u^3- \frac{153}{16} u^4+ \frac{81}{8} u^5- \frac{27}{8} u^6 \right. \nonumber \\
& & + \frac{9}{80} [ 4 (-1+u)^3 (1+3u +6 u^2 ) \ln (1-u) + u (-4+3u-88u^2+269 u^3\nonumber \\
& & \left. -270 u^4+90 u^5-4 u^2 ( 10-15 u+6 u^2) \ln u ) ] \right \} \\
& & g_2(u, \mu) = \frac{10}{3} \delta^2(\mu) (1-u) u ( 2u -1) \nonumber \\
& & + a_2(\mu) \rho^2_P(\mu) \left [ \frac{81}{4} u^5- \frac{405}{8} u^4+\frac{153}{4} u^3
-\frac{27}{4} u^2 - \frac{9}{8} u \right ]
\end{eqnarray}

For vector mesons:
\begin{eqnarray}
& & g^{(a)}_\perp (u, \mu) = 6 u (1-u) \left \{ 1+ a^\parallel_1 (\mu) (2u-1) + [ \frac{1}{4}
a^\parallel_2 (\mu) + \frac{5}{3} \zeta_3(\mu) ( 1\right. \nonumber \\
& &\left. - \frac{3}{16} \omega^A_3 (\mu) + \frac{9}{16} \omega^V_3 (\mu) ) ] (5 (2u-1)^2 -1)\right \}
+6 \tilde{\delta}_+ (\mu) \left [ 3 u (1-u) \right.\nonumber \\
& & \left. + (1-u) \ln (1-u) + u \ln u \right ] + 6 \tilde{\delta}_- (\mu)\left [ (1-u) \ln (1-u)
 - u \ln u \right ] \\
& & g^{(v)}_\perp (u, \mu) = \frac{3}{4} \left (1+ (2u-1)^2 \right ) + a^\parallel_1
(\mu) \frac{3}{2}(2u-1)^3 + \left ( \frac{3}{7} a^\parallel_2(\mu)\right. \nonumber \\
& & \left. + 5 \zeta_3 (\mu) \right ) \left (3 (2u-1)^2 -1\right ) +\left [ \frac{9}{112}
 a^\parallel_2 (\mu) + \frac{15}{64} \zeta_3 (\mu) ( 3 \omega^V_3 (\mu) - \omega^A_3) \right ]\nonumber \\
& & \times \left (3-30 (2u-1)^2 + 35 (2u-1)^4 \right ) + \frac{3}{2} \tilde{\delta}_+ (\mu)
 \left [ 2 + \ln u + \ln (1-u) \right ] \nonumber \\
& & + \frac{3}{2} \tilde{\delta}_- (\mu) \left [ 2 (2u-1) + \ln (1-u) - \ln u \right ]
\end{eqnarray}

\begin{eqnarray}
& & h^{(s)}_\parallel ( u, \mu) = 6 u (1-u) \left [ 1+ a^\perp_1(\mu) (2u-1)
+ ( \frac{1}{4} a^\perp_2 (\mu)\right. \nonumber \\
& & \left.+ \frac{5}{8} \zeta_3 (\mu) \omega^T_3 (\mu) ) ( 5 (2u-1)^2 -1) \right ]+ 3 \delta_+ (\mu)
\left [ 3 u (1-u) \right. \nonumber \\
& & \left. + (1-u) \ln (1-u) + u \ln u \right ] + 3 \delta_- (\mu) \left [ (1-u) \ln (1-u) - u \ln u \right ] \\
& & h^{(t)}_\parallel (u, \mu) = 3 (2u-1)^2 + \frac{3}{2} a^\perp_1(\mu) (2u-1) [3 (2u-1)^2 -1]\nonumber \\
& & + \frac{3}{2} a^\perp_2(\mu) (2u-1)^2 [ 5 (2u-1)^2 -3 ] + \frac{15}{16} \zeta_3(\mu)
\omega^T_3(\mu) [ 3 - 30 (2u-1)^2 \nonumber \\
& & + 35 (2u-1)^4 ] + \frac{3}{2} \delta_+ (\mu) [ 1+ (2u-1) \ln \frac{1-u}{u} ] \nonumber \\
& & + \frac{3}{2} \delta_- (\mu) (2u-1) [ 2 + \ln u + \ln (1-u)]
\end{eqnarray}
and
\begin{eqnarray}
& & A ( u, \mu) = 30 u^2 (1-u)^2 \left \{ \frac{4}{5} [1+
\frac{1}{21} a^\parallel_2 (\mu)
+ \frac{10}{9} \zeta_3 (\mu) + \frac{25}{9} \zeta_4 (\mu) ] \right. \nonumber \\
& & \left. + \frac{1}{5} [ \frac{9}{14} a^\parallel_2 (\mu) + \frac{1}{18} \zeta_3 (\mu)
+ \frac{3}{8} \zeta_3(\mu) ( \frac{7}{3}\omega^V_3 (\mu) - \omega^A_3(\mu) ) ] C^{5/2}_2 ( 2u-1) \right \} \nonumber \\
& & +10 \left [ -2 a^\parallel_2 (\mu) - \frac{14}{3} \zeta_3 (\mu) + \frac{9}{2} \zeta_3(\mu) \omega^V_3 (\mu)
- 42 \zeta_4 (\mu) \omega^A_4 (\mu) \right ] \nonumber \\
& & \times \left [ \frac{1}{10} u (1-u) ( 2 + 13 u (1-u) ) + \frac{1}{5} u^3 ( 10 - 15 u +6 u^2 )
 \ln u \right. \nonumber \\
& & \left. + \frac{1}{5} (1-u)^3 ( 10 - 15 (1-u) + 6 (1-u)^2 ) \ln (1-u) \right ]\\
& & A_T (u, \mu) = 30 u^2 (1-u)^2 \left \{ \frac{2}{5} [ 1+
\frac{2}{7} a^\perp_2 (\mu)
+ \frac{10}{3} \zeta^T_4(\mu) - \frac{20}{3} \tilde{\zeta}^T_4 ] \right. \nonumber \\
& & \left. + [\frac{3}{35} a^\perp_2 (\mu) + \frac{1}{40} \zeta_3(\mu) \omega^T_3 (\mu) ] C^{5/2}_2 (2u-1)
 \right \}\nonumber \\
& & - \left [ \frac{18}{11} a^\perp_2 (\mu) - \frac{3}{2} \zeta_3(\mu) \omega^T_3 (\mu)
+ \frac{126}{55} \ll Q^{(1)} \gg (\mu)+ \frac{70}{11} \ll Q^{(3)} \gg (\mu) \right ] \nonumber \\
& & \times [u (1-u) ( 2 + 13 u (1-u)) + 2 u^3 ( 10 - 15 u + 6 u^2 ) \ln u \nonumber \\
& & + 2 (1-u)^3 ( 10 - 15 (1-u) + 6 (1-u)^2 ) \ln (1-u) ] \\
& & g_3(u, \mu) = 1 + \left [-1 - \frac{2}{7} a^\parallel_2 (\mu) + \frac{40}{3} \zeta_3(\mu)
- \frac{20}{3} \zeta_4 (\mu) \right ] C^{1/2}_2 (2u-1) \nonumber \\
& & + \left [ - \frac{27}{28} a^\parallel_2 (\mu) + \frac{5}{4}
\zeta_3 (\mu) - \frac{15}{16} \zeta_3 (\mu) ( \omega^A_3 (\mu) + 3
\omega^V_3 (\mu) ) \right ] C^{1/2}_4 (2u-1) \\
& & h_3 (u, \mu) = 1+ \left [ -1 + \frac{3}{7} a^\perp_2 (\mu) - 10 ( \zeta^T_4 (\mu) + \tilde{\zeta}^T_4 )
\right ] C^{1/2}_2 (2u-1) \nonumber \\
& & + \left [ - \frac{3}{7} a^\perp_2 (\mu) - \frac{15}{8} \zeta_3 (\mu) \omega^T_3 (\mu0 \right ] C^{1/2}_4 (2u-1)
\end{eqnarray}
with
\begin{eqnarray}
& & C(u) = g_3(u)+\phi_{\|}-2g^{(v)}_\bot \\
& & B_T(u)=h^{(t)}_\parallel(u)-\frac{1}{2} \phi_\bot-\frac{1}{2}
h_3(u) \\
& & C_T(u)=h_3(u)-\phi_\bot(u)
\end{eqnarray}

\renewcommand{\theequation}{C.\arabic{equation}}
\section{ The Functions $\bar{C}^{box}_i (x_i)$, $\bar{C}^Z_i ( x_i )$ and
$F^i_1 ( x_i)$ in Eqs.(58),(60).}

The functions $\bar{C}^{box}_i (x_i)$, $\bar{C}^Z_i ( x_i )$ and
$F^i_1 ( x_i)$ appearing in Eqs.(58),(60) read\cite{BG,FP}:

\begin{eqnarray}
& & \bar{C}_i^{box}(x_i)={3\over 8} \biggl[-{1\over x_i-1}+{x_i\ln
x_i \over (x_i-1)^2}\biggr]-\gamma (\xi,x_i)
\\
& & \bar{C}_i^Z(x_i)={x_i\over 4}-{3\over 8}{1\over x_i-1}+{3\over
8}{2x_i^2-x_i\over
(x_i-1)^2}\ln x_i +\gamma (\xi,x_i)\nonumber\\
\end{eqnarray}
\begin{eqnarray}
F_1^i(x_i)&=&Q\Biggl(\biggl[{1\over 12}{1\over x_i-1}+{13\over
12}{1\over
(x_i-1)^2}-{1\over 2}{1\over (x_i-1)^3}\biggr]x_i\nonumber\\
&+&\biggl[{2\over 3}{1\over x_i-1}+\biggl({2\over 3}{1\over
(x_i-1)^2}-{5\over 6}{1\over (x_i-1)^3}+{1\over 2}{1\over
(x_i-1)^4}\biggr)x_i\biggr]\ln
x_i\Biggr)\nonumber\\
&-&\biggl[{7\over 3}{1\over x_i-1}+{13\over 12}{1\over
(x_i-1)^2}-{1\over
2}{1\over (x_i-1)^3}\biggr]x_i\nonumber\\
&-&\biggl[{1\over 6}{1\over x_i-1}-{35\over 12}{1\over
(x_i-1)^2}-{5\over 6}{1\over (x_i-1)^3}+{1\over 2}{1\over
(x_i-1)^4}\biggr]x_i\ln x_i-2\gamma (\xi,x_i)
\end{eqnarray}
where $Q=-1/3$ is the corresponding charge of the intermediate
quarks $d, s, b$. The gauge dependent term $\gamma (\xi,x_i)$
cancels out in the combinations of $F^i_i$, $\bar{C}^{box}_i$, and
$\bar{C}^Z_i$.

\newpage
\begin{figure}[h]
\begin{center}
\includegraphics[width=2.2in]{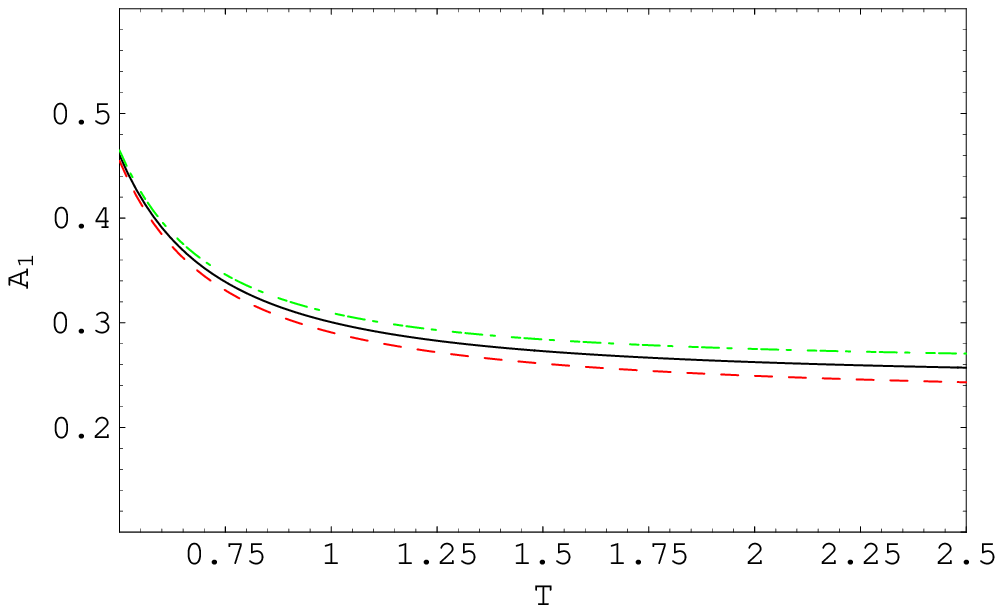}
\hspace{0.2in}
\includegraphics[width=2.2in]{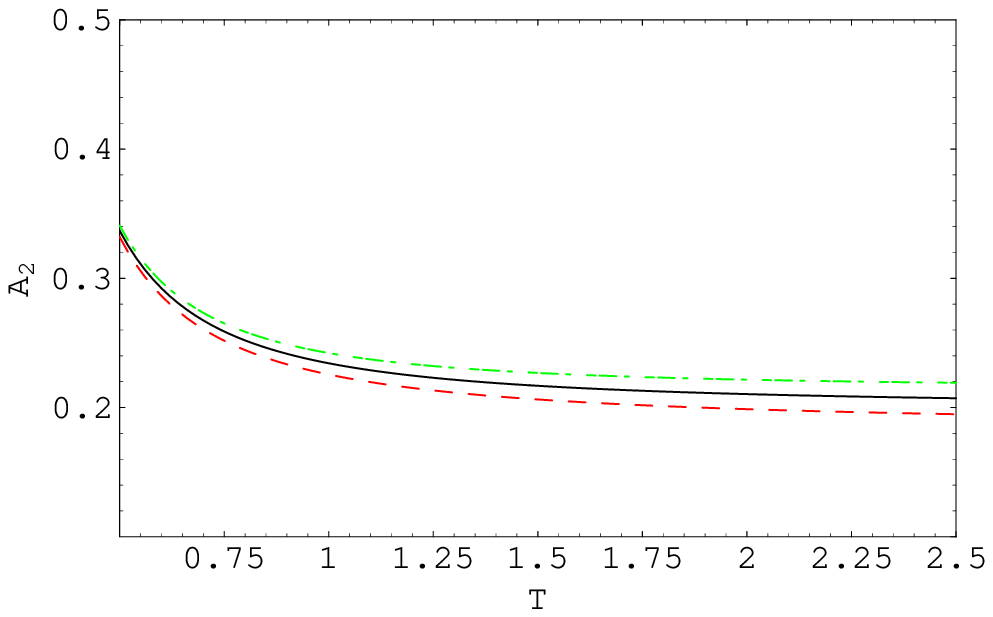}
\hspace{0.2in}
\includegraphics[width=2.2in]{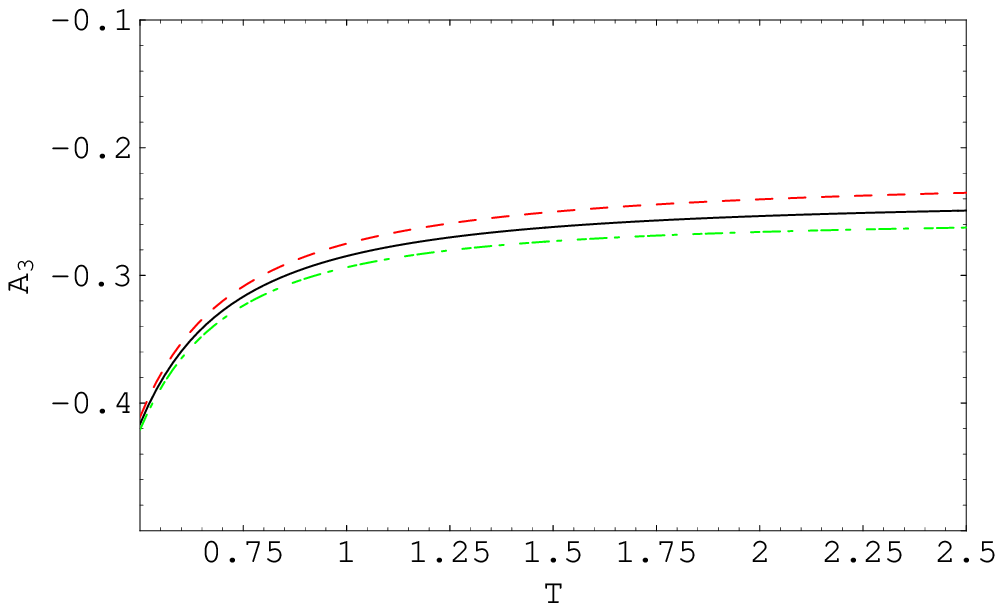}
\hspace{0.2in}
\includegraphics[width=2.2in]{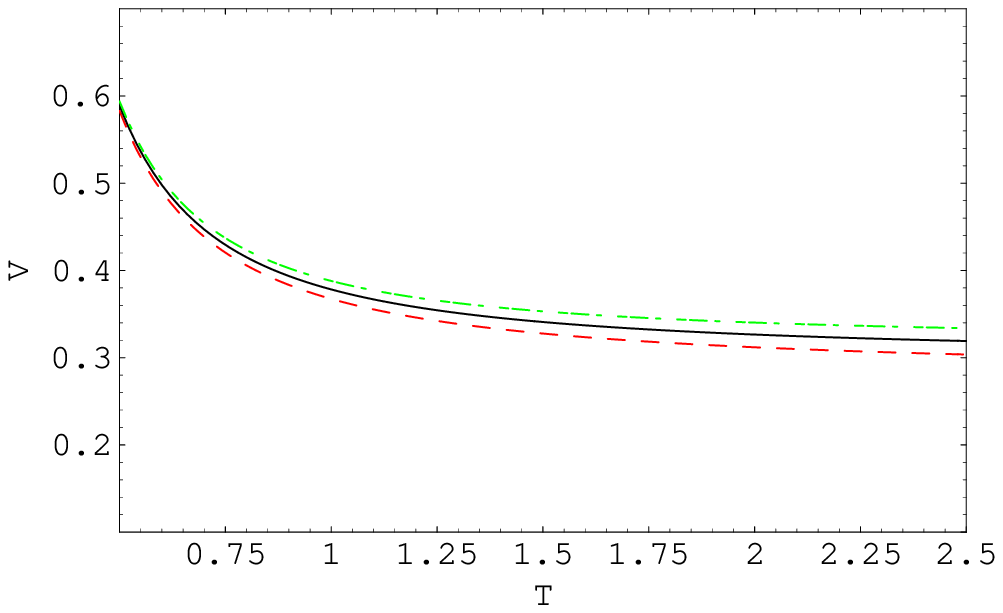}
\end{center}
\begin{flushleft}
{\bf Fig.1}: Form factors of $B \rightarrow K^*$ decays as functions
of $T$ for different $s_0$ at momentum transfer $q^2 = 0 {\rm Gev}^2
$, which are obtained with considering the meson DAs up to twist-4.
The dashed solid and dot dashed lines correspond to $s_0 = 1.6, 1.7$
and $1.8$GeV respectively.
\end{flushleft}
\end{figure}

\begin{figure}[h]
\centering
\includegraphics[width=2.2in]{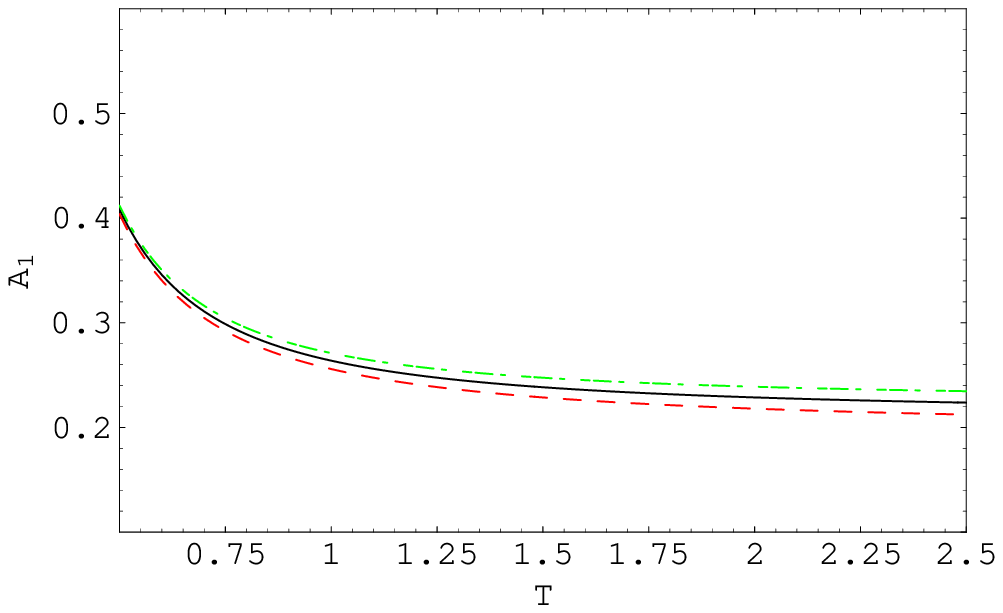}%
\hspace{0.2in}
\includegraphics[width=2.2in]{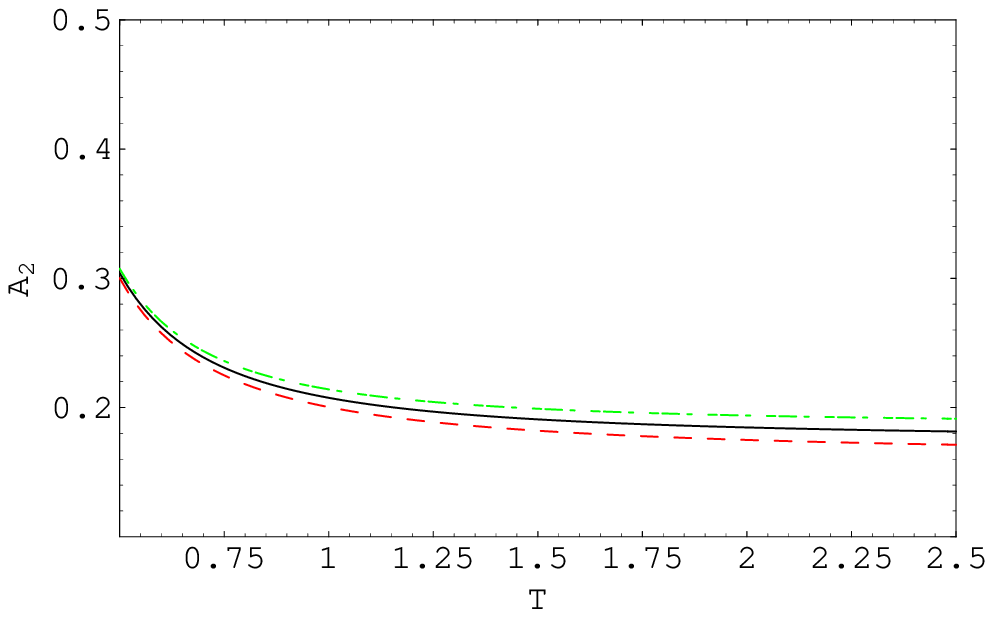}
\hspace{0.2in}
\includegraphics[width=2.2in]{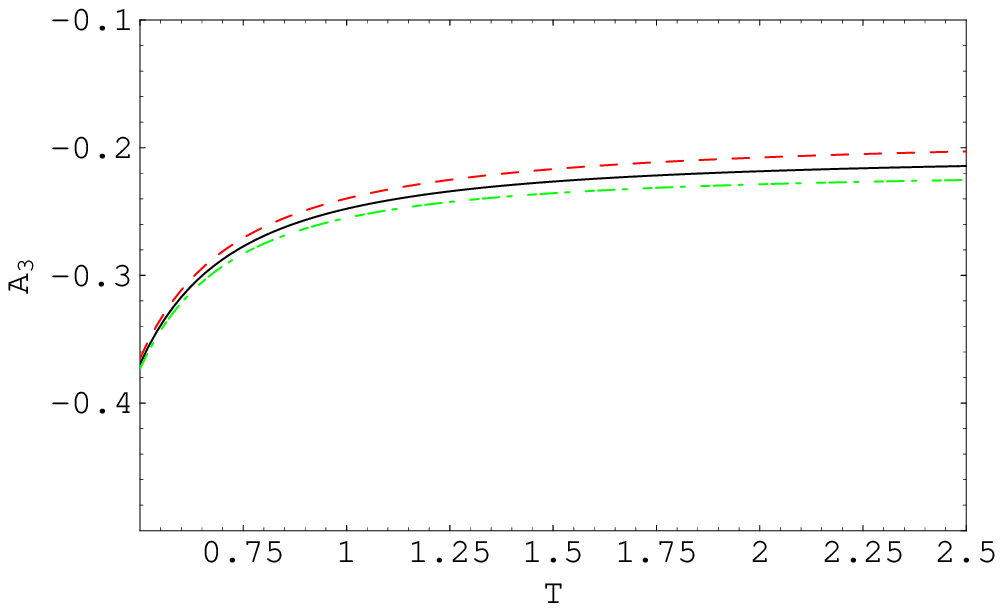}
\hspace{0.2in}
\includegraphics[width=2.2in]{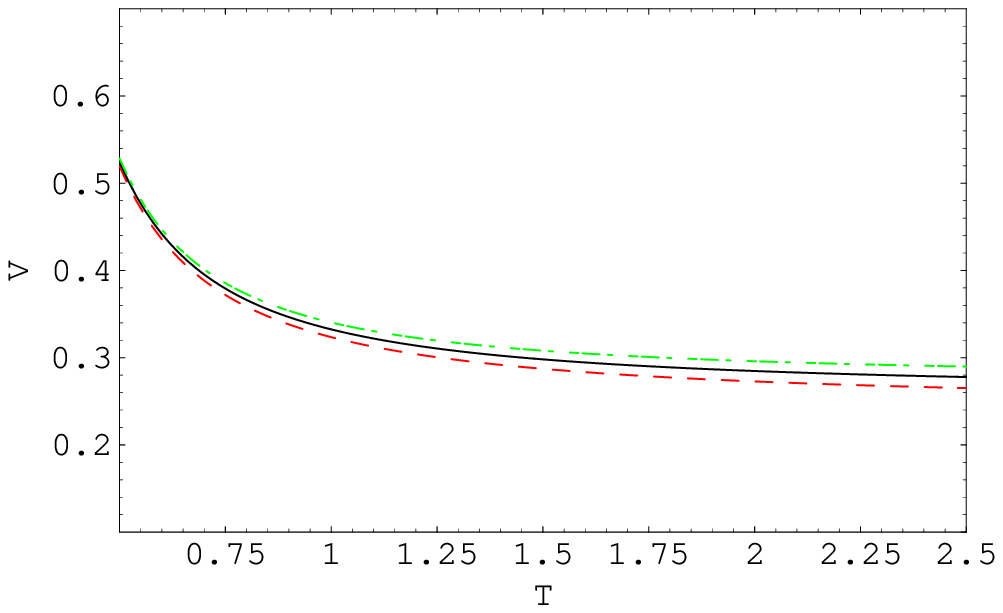}
\\
{\bf Fig.2}: Same as Fig.1 but for $B \rightarrow \rho$ decays.
\end{figure}

\begin{figure}[h]
\centering
\includegraphics[width=2.2in]{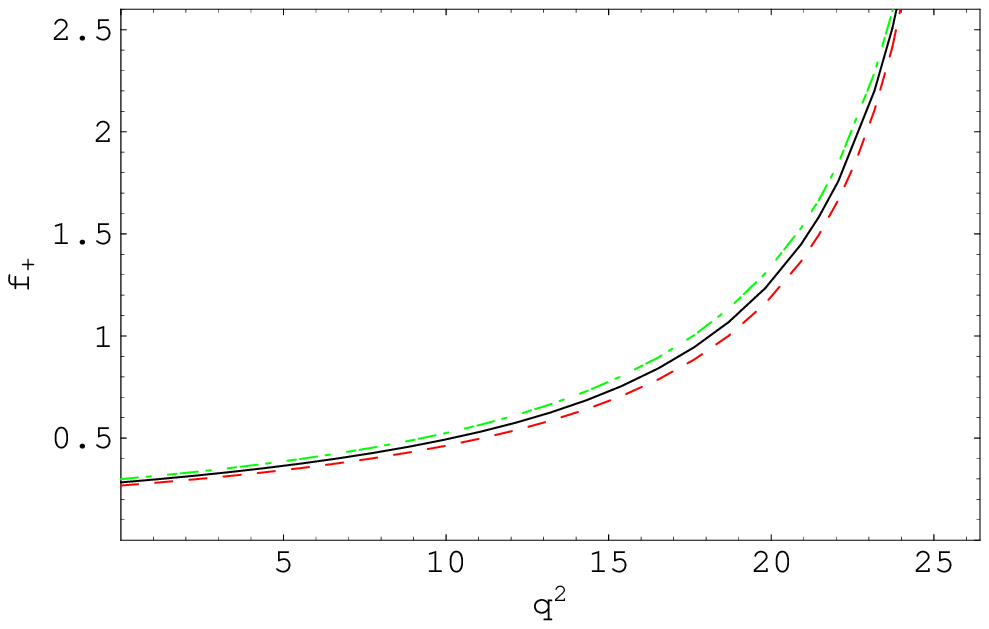}%
\hspace{0.2in}
\includegraphics[width=2.2in]{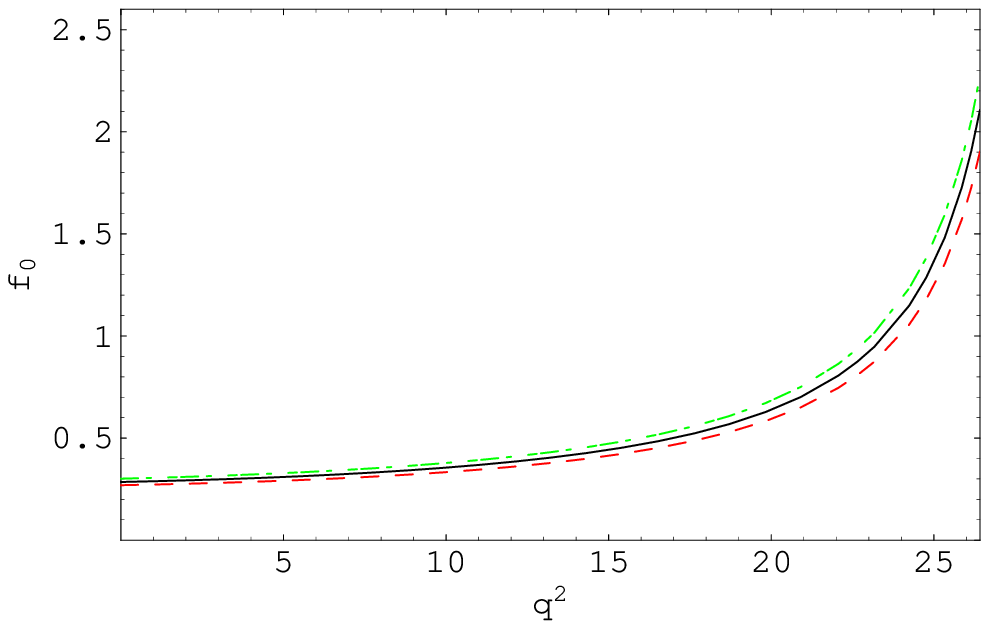}
\\
\begin{flushleft}
{\bf Fig.3}: Form factors of $B \rightarrow \pi$ decays as functions
of $q^2$. The dashed, solid and dot dashed lines correspond to $s_0
= 1.5$Gev, $T = 2.2$Gev; $s_0 = 1.6$Gev, $T=2.0$Gev and
$s_0=1.7$GeV, $T= 1.8$Gev  respectively, which reflect the possible
large uncertainties.
\end{flushleft}
\end{figure}

\begin{figure}[h]
\centering
\includegraphics[width=2.2in]{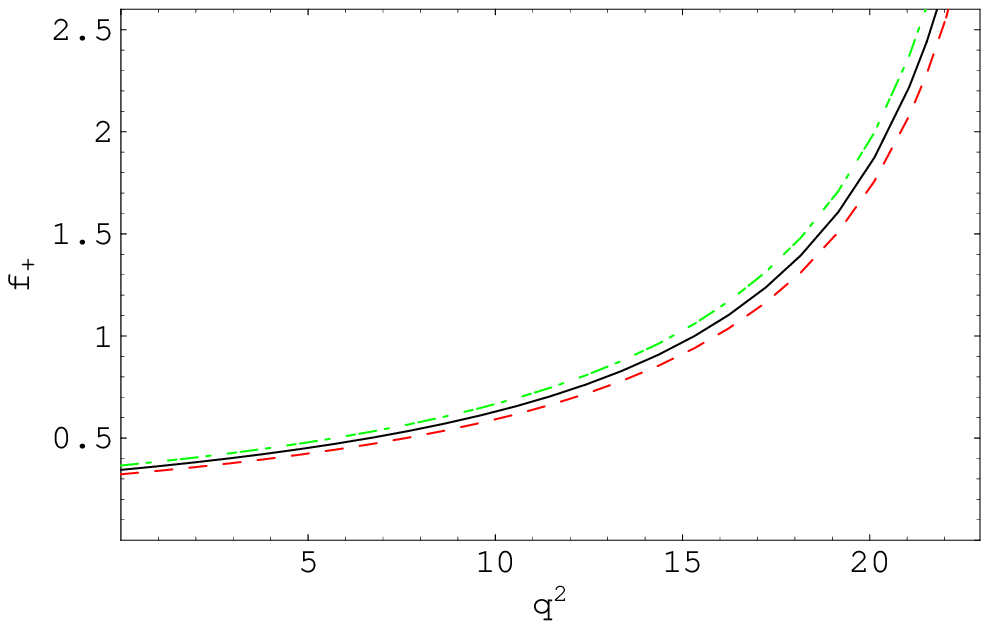}%
\hspace{0.2in}
\includegraphics[width=2.2in]{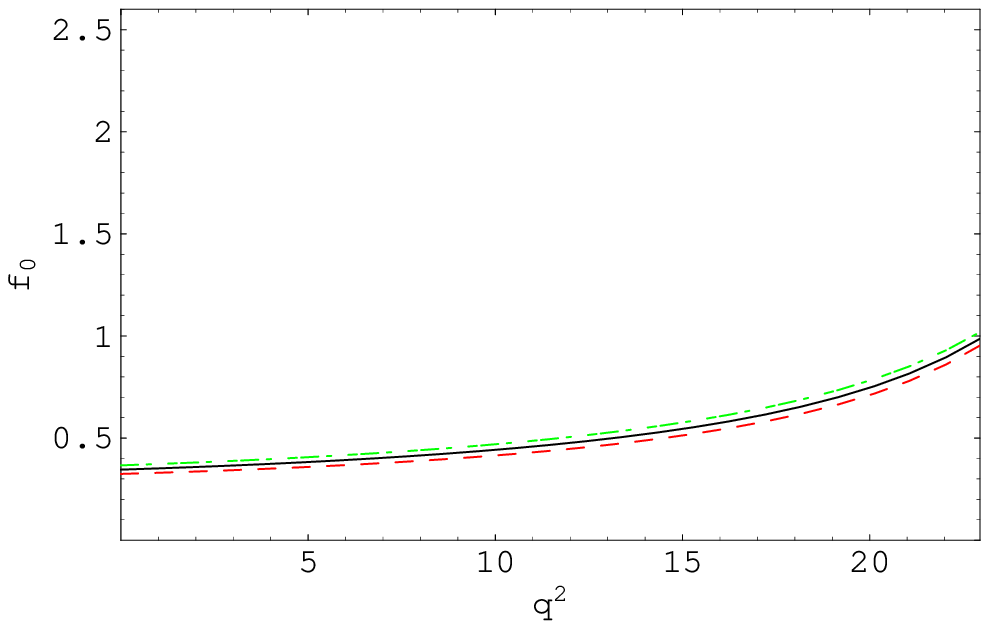}
\\
{\bf Fig.4}: Same as Fig.3, but for $B \rightarrow K$ decays.
\end{figure}

\begin{figure}[h]
\centering
\includegraphics[width=2.2in]{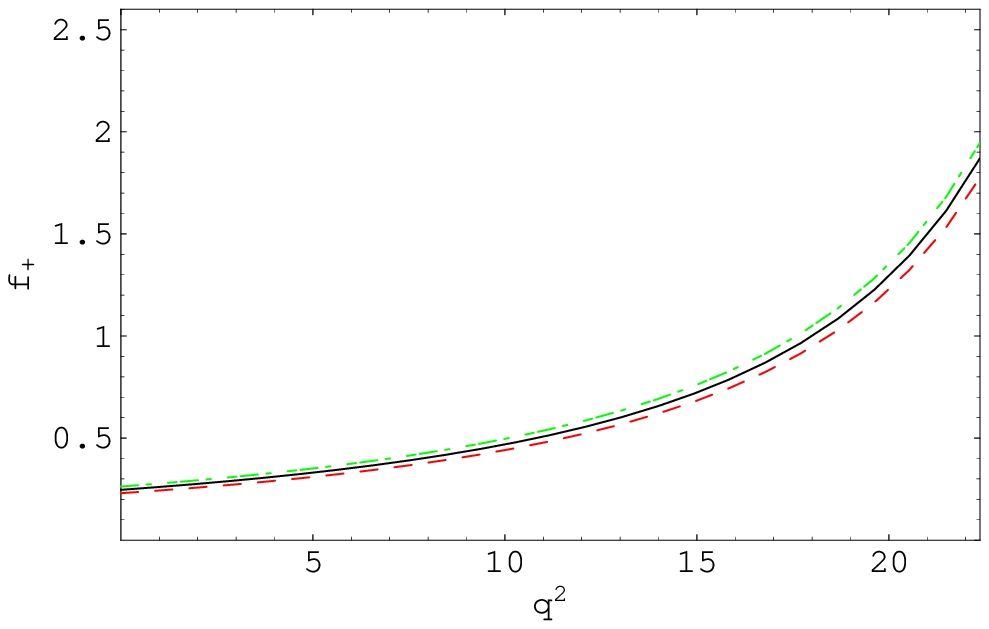}%
\hspace{0.2in}
\includegraphics[width=2.2in]{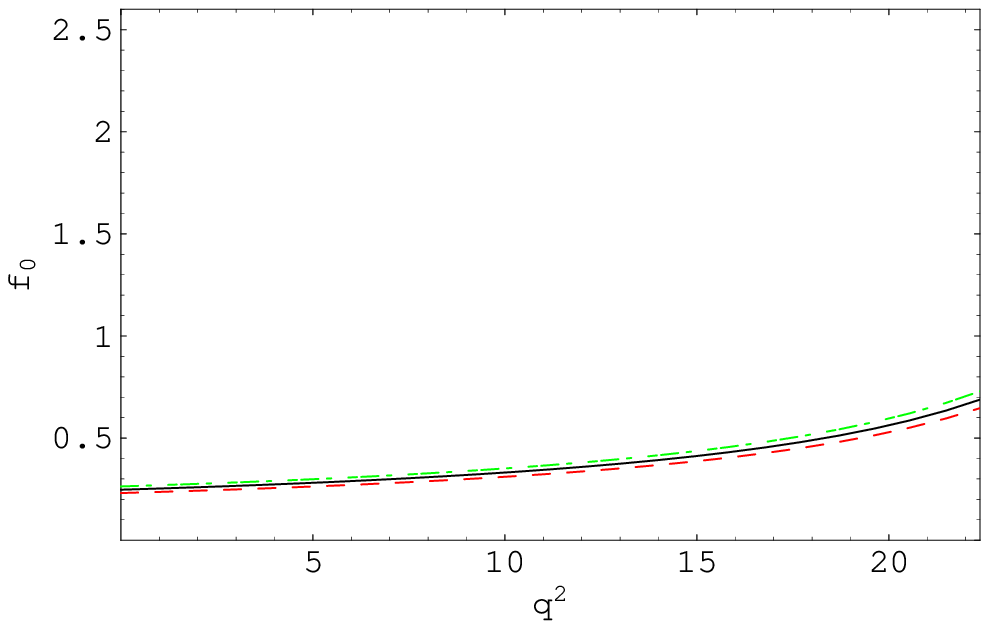}
\\
{\bf Fig.5}: Same as Fig.3, but for $B \rightarrow \eta$ decays.
\end{figure}

\clearpage

\begin{figure}[h]
\centering
\includegraphics[width=2.2in]{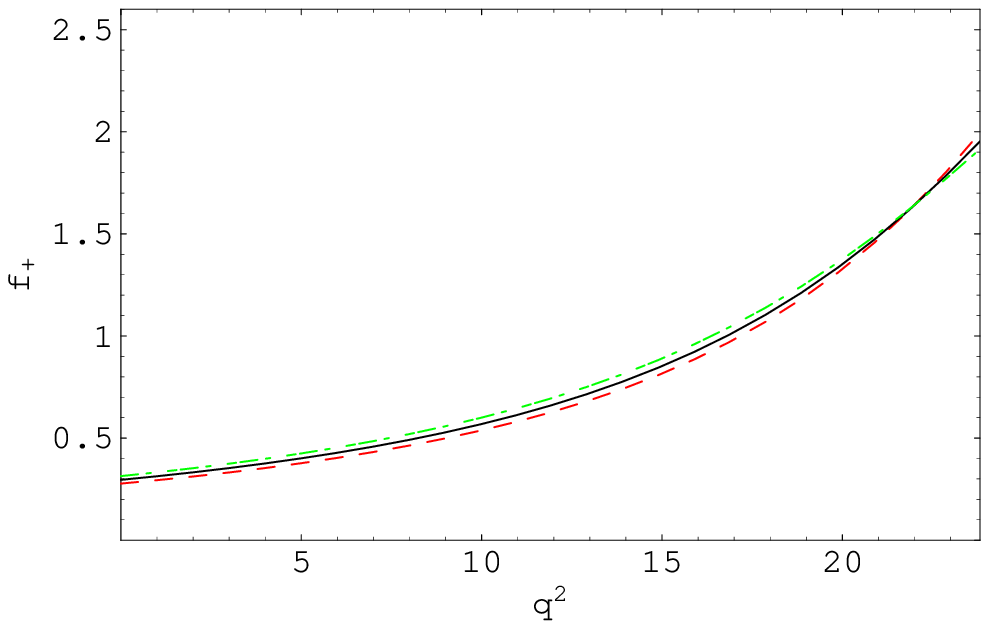}%
\hspace{0.2in}
\includegraphics[width=2.2in]{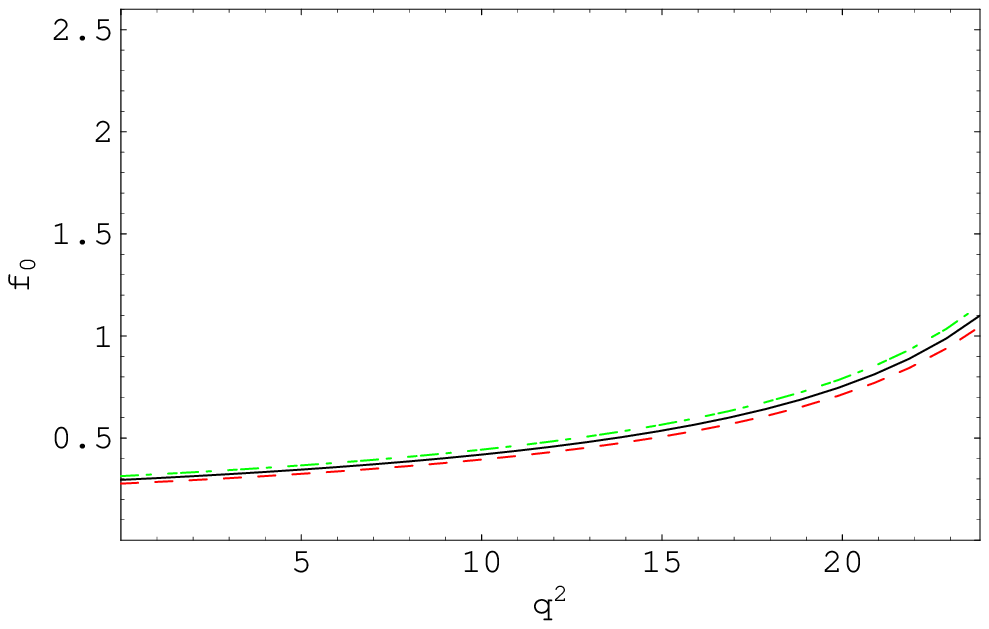}
\\
\flushleft{{\bf Fig.6}: Form factors of $B_s \rightarrow K$ decays
as functions of $q^2$. The dashed, solid and dot dashed lines
correspond to $s_0 = 1.8$Gev, $T=2.2$Gev; $s_0= 1.9$Gev, $T=2.0$Gev
and $s_0=2.0$GeV, $T=1.8$Gev respectively, which reflect the
possible large uncertainties.}
\end{figure}

\begin{figure}[h]
\centering
\includegraphics[width=2.2in]{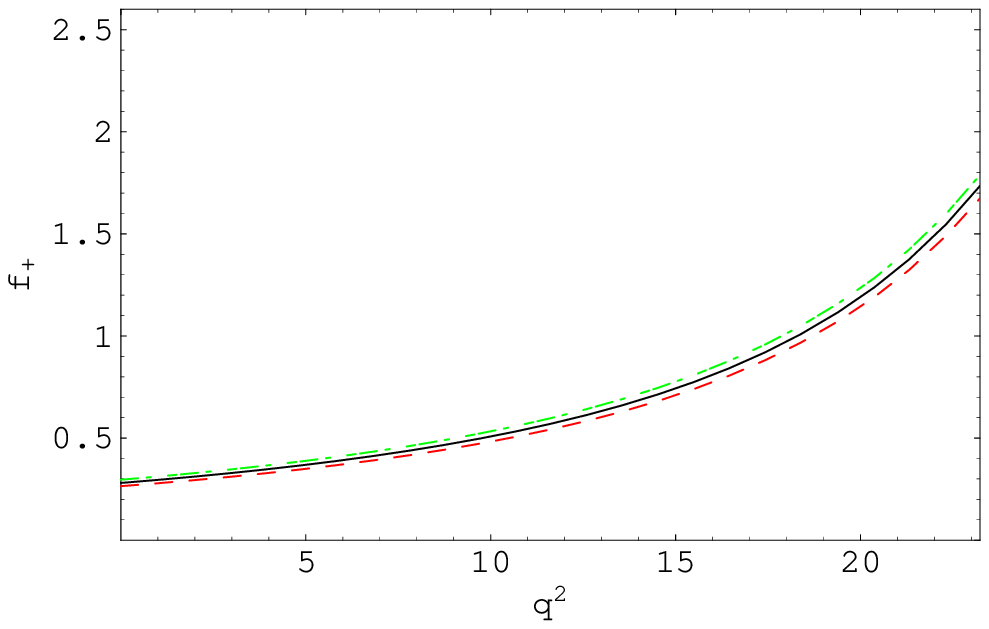}%
\hspace{0.2in}
\includegraphics[width=2.2in]{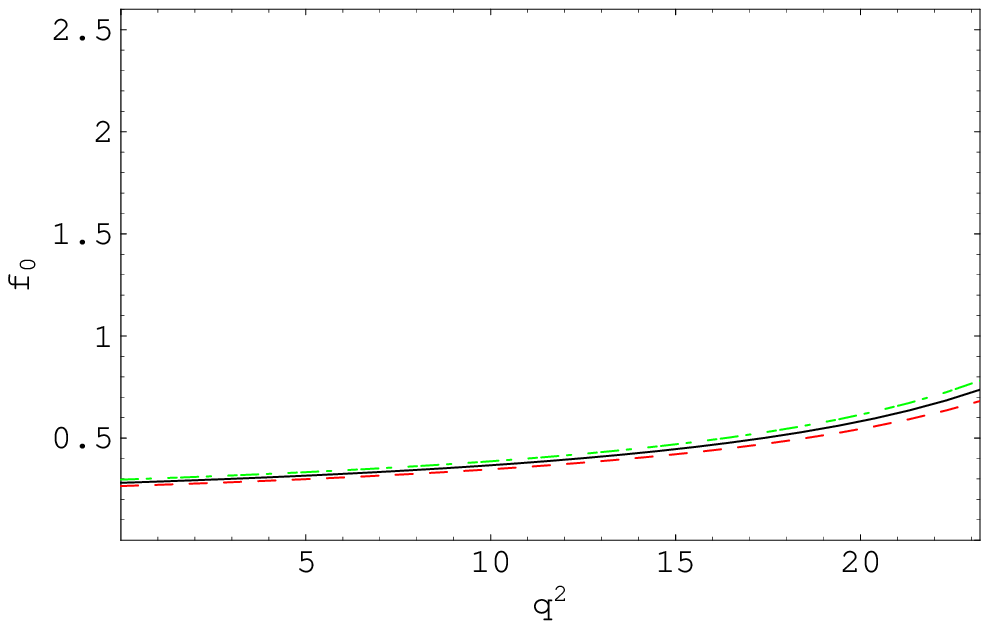}
\\
{\bf Fig.7}: Same as Fig.6, but for $B_s \rightarrow \eta$ decays.
\end{figure}

\clearpage

\begin{figure}[t]
\centering
\includegraphics[width=2.2in]{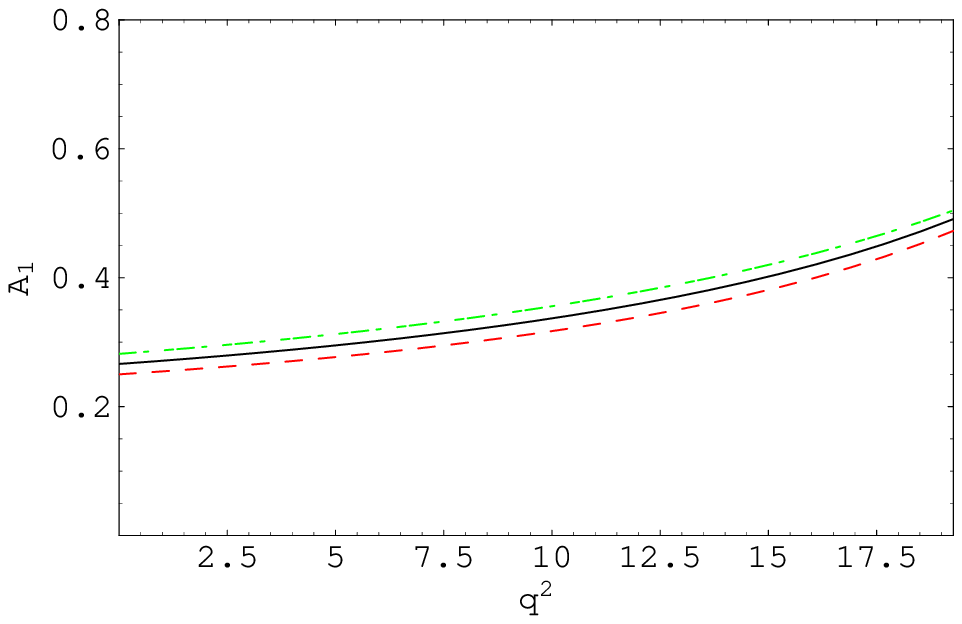}%
\hspace{0.2in}
\includegraphics[width=2.2in]{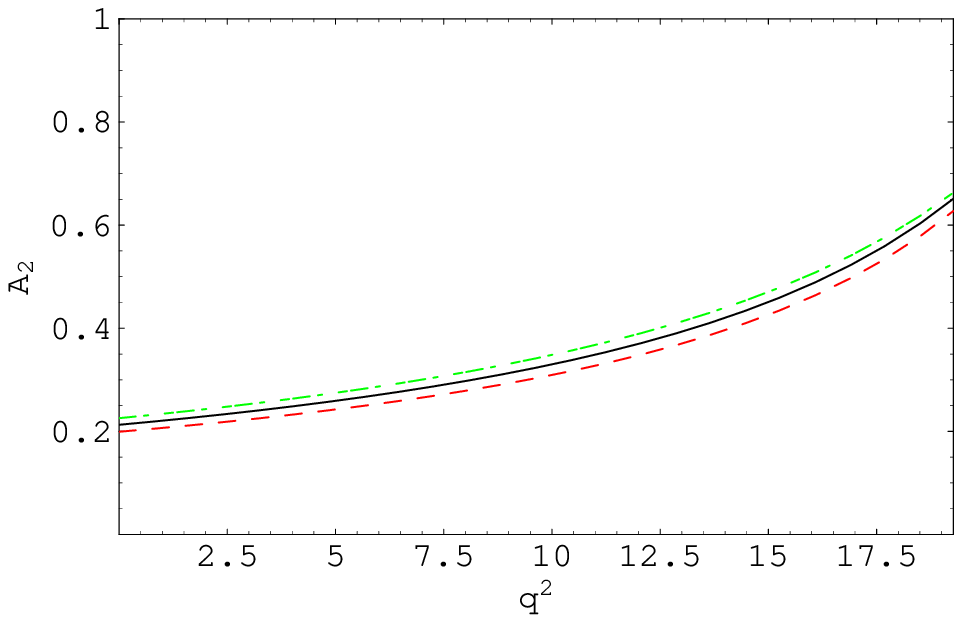}
\hspace{0.2in}
\includegraphics[width=2.2in]{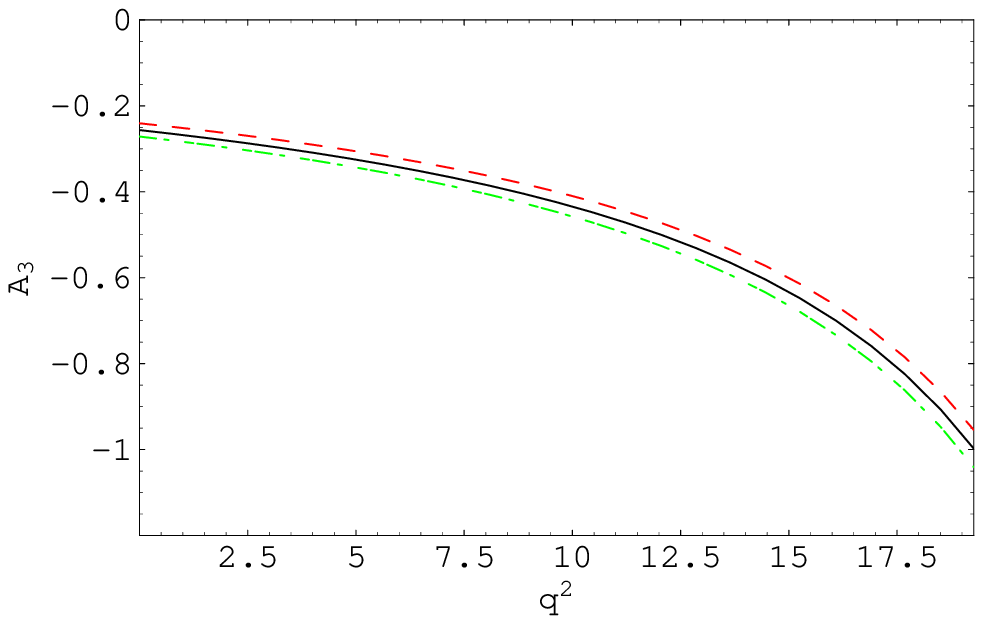}
\hspace{0.2in}
\includegraphics[width=2.2in]{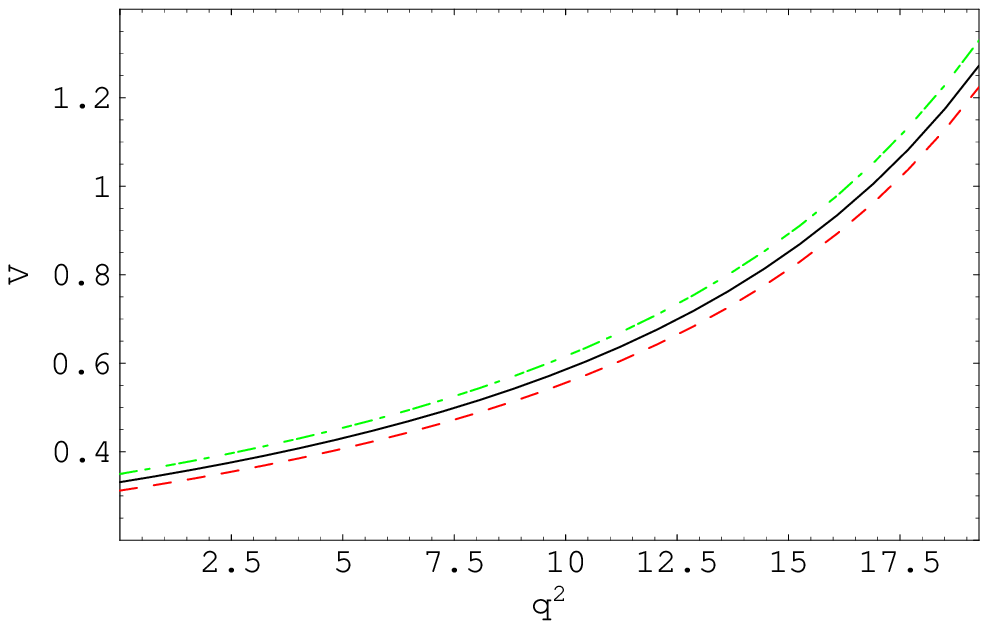}
\\
\flushleft{{\bf Fig.8}: Form factors of $B \rightarrow K^*$ decays
as functions of $q^2$ obtained with considering the meson DAs up to
twist-4. The dashed, solid and dot dashed lines correspond to $s_0 =
1.6$Gev, $T=2.0$Gev; $s_0= 1.7$Gev, $T=1.8$Gev and $s_0=1.8$GeV,
$T=1.6$Gev respectively, which reflect the possible large
uncertainties. }
\end{figure}

\begin{figure}
\centering
\includegraphics[width=2.2in]{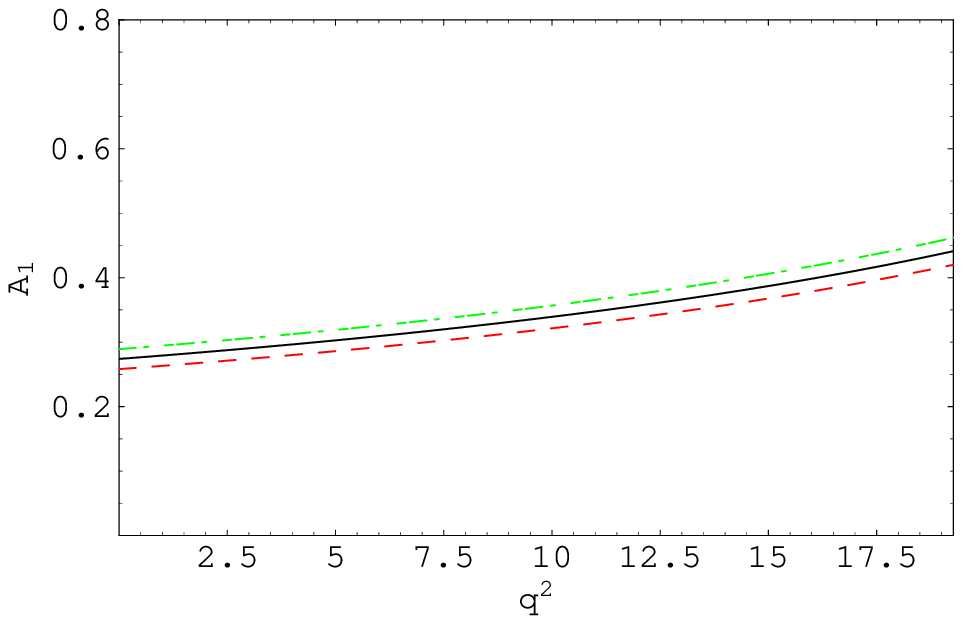}%
\hspace{0.2in}
\includegraphics[width=2.2in]{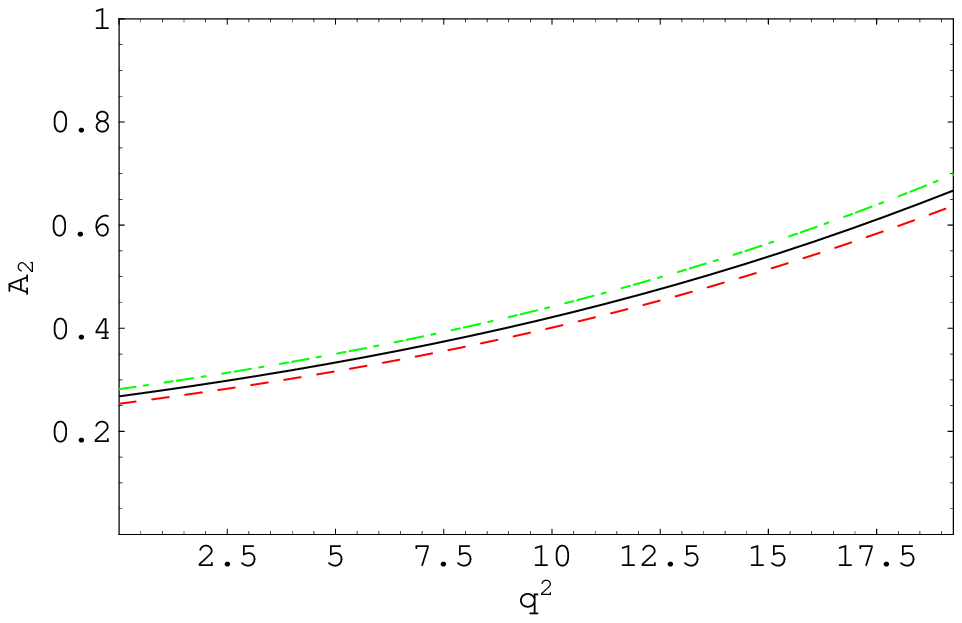}
\hspace{0.2in}
\includegraphics[width=2.2in]{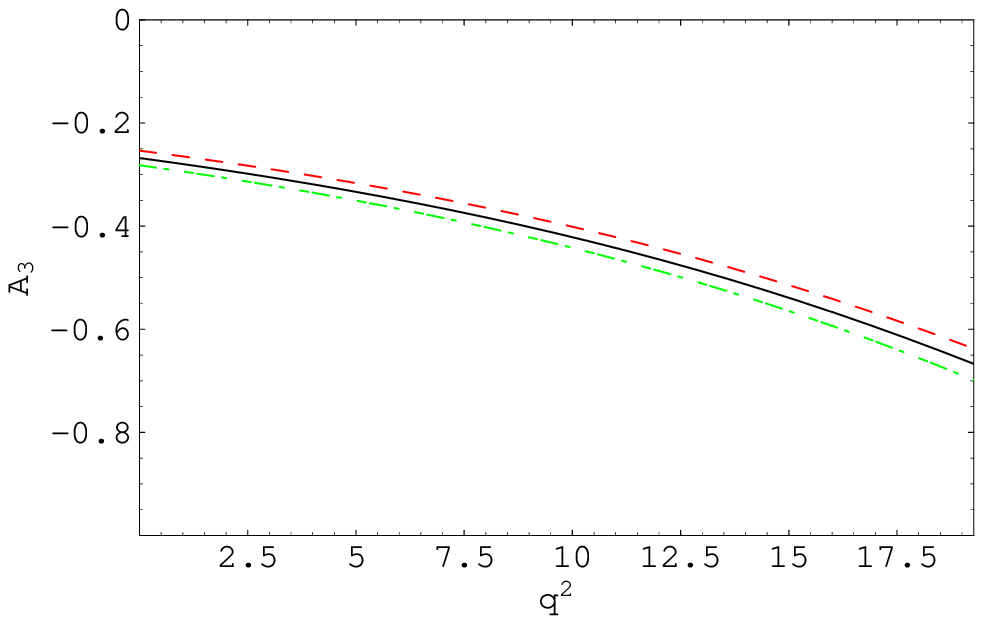}
\hspace{0.2in}
\includegraphics[width=2.2in]{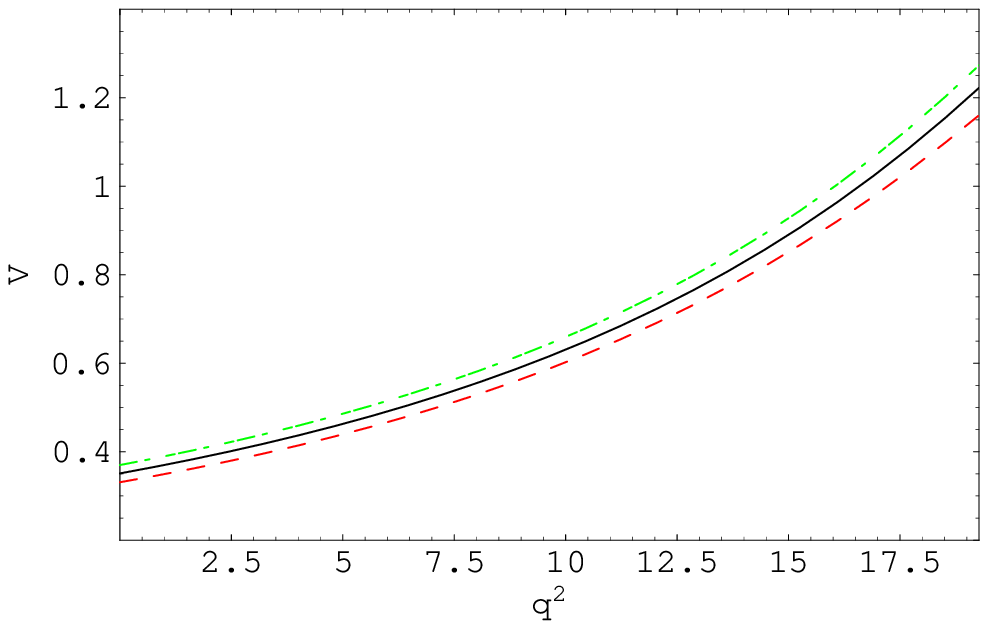}
\flushleft{{\bf Fig.9}: Form factors of $B \rightarrow K^*$ decays
as functions of $q^2$ obtained with only considering the leading
twist meson DAs. The dashed, solid and dot dashed lines correspond
to $s_0 = 1.6$Gev, $T=2.2$Gev; $s_0= 1.7$Gev, $T=2.0$Gev and
$s_0=1.8$GeV, $T=1.8$Gev respectively, which reflect the possible
large uncertainties.}
\end{figure}

\clearpage

\begin{figure}[t]
\centering
\includegraphics[width=2.2in]{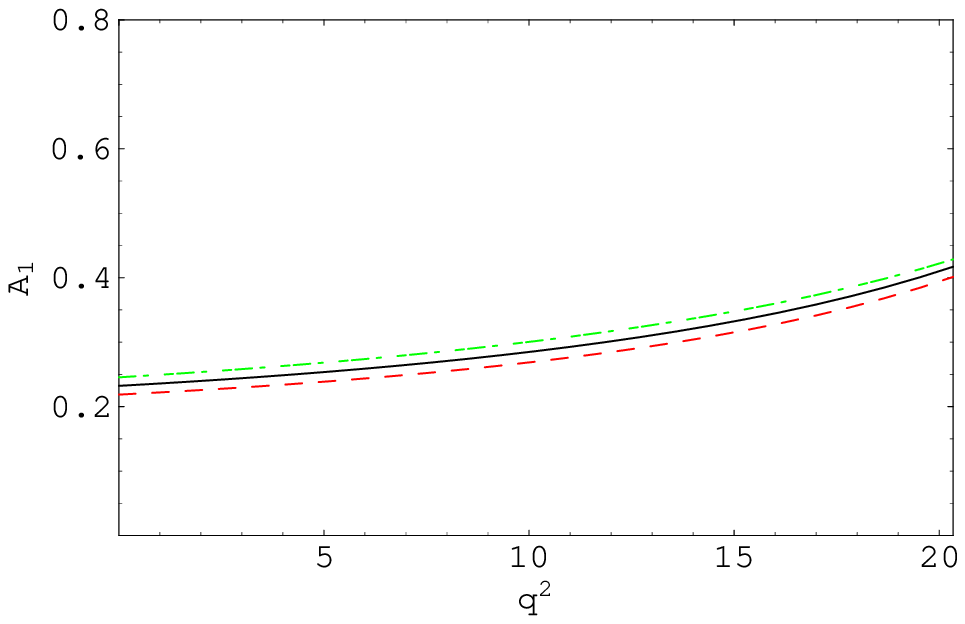}%
\hspace{0.2in}
\includegraphics[width=2.2in]{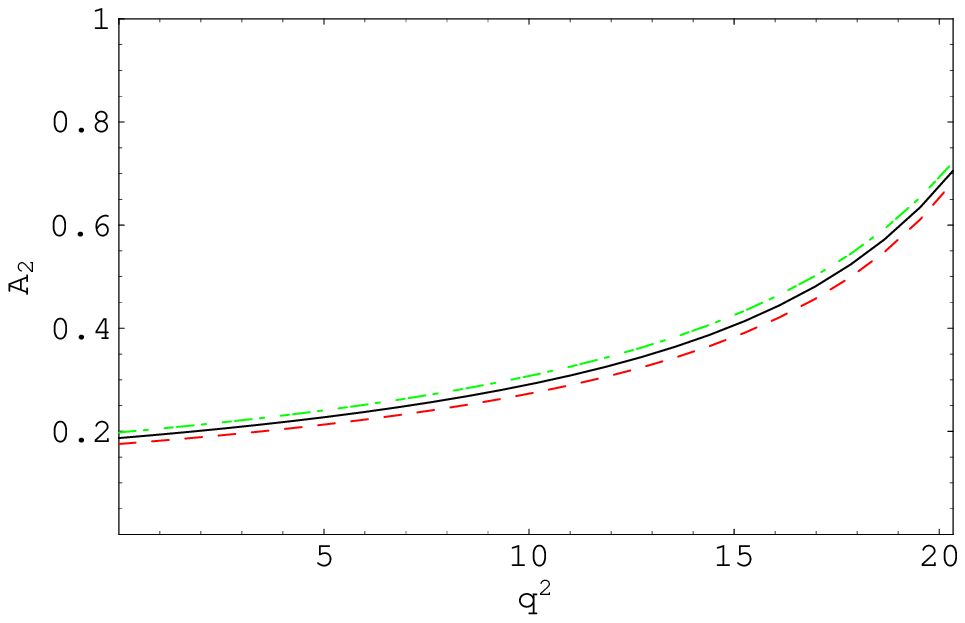}
\hspace{0.2in}
\includegraphics[width=2.2in]{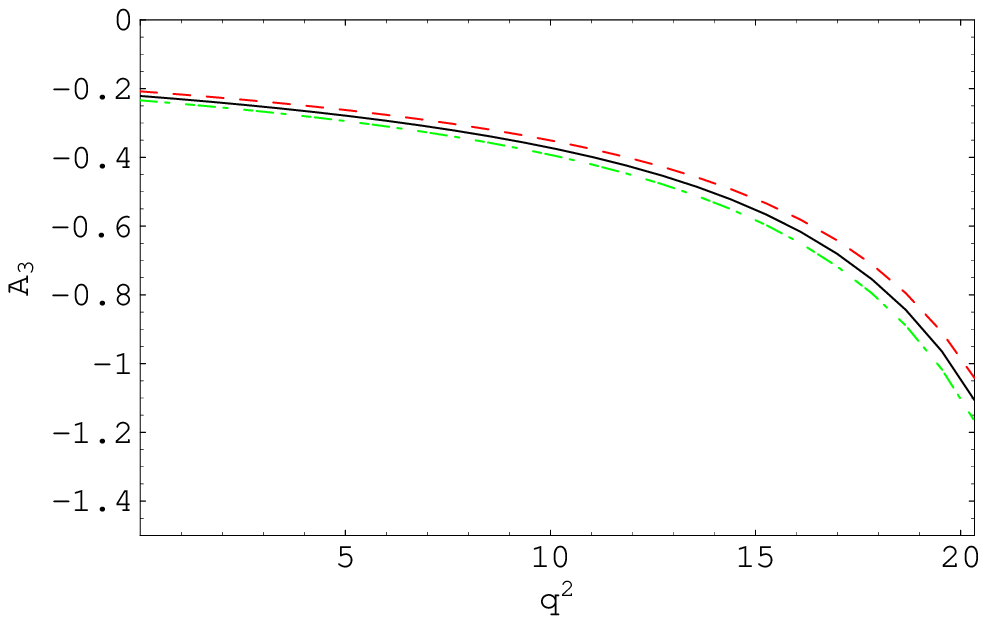}
\hspace{0.2in}
\includegraphics[width=2.2in]{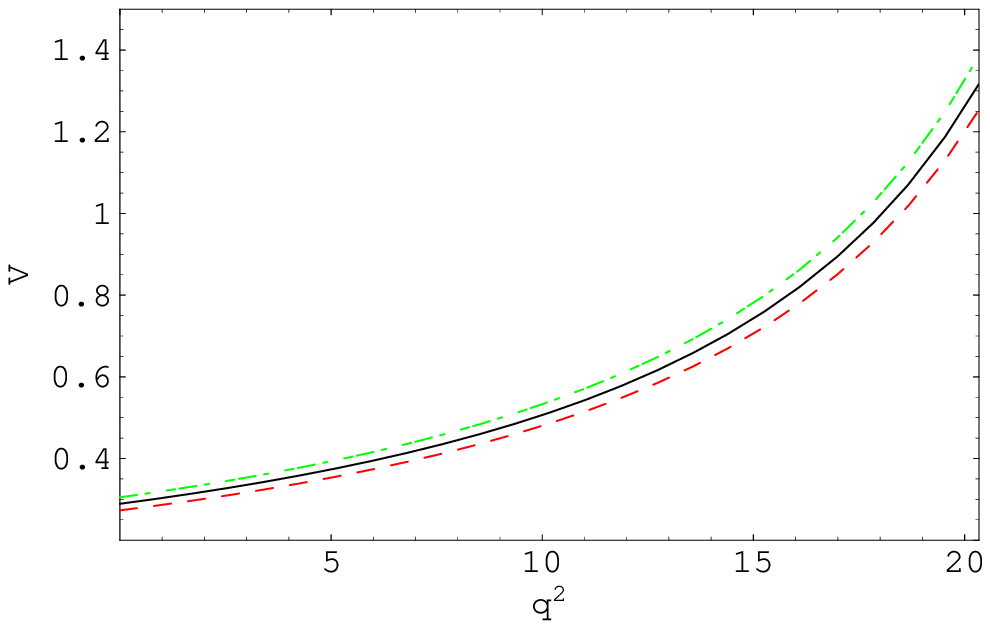}
\\
\flushleft{{\bf Fig.10}: Form factors of $B \rightarrow \rho$ decays
as functions of $q^2$ obtained with considering the meson DAs up to
twist-4. The dashed, solid and dot dashed lines correspond to $s_0 =
1.6$Gev, $T=2.0$Gev; $s_0= 1.7$Gev, $T=1.8$Gev and $s_0=1.8$GeV,
$T=1.6$Gev respectively, which reflect the possible large
uncertainties. }
\end{figure}

\begin{figure}
\centering
\includegraphics[width=2.2in]{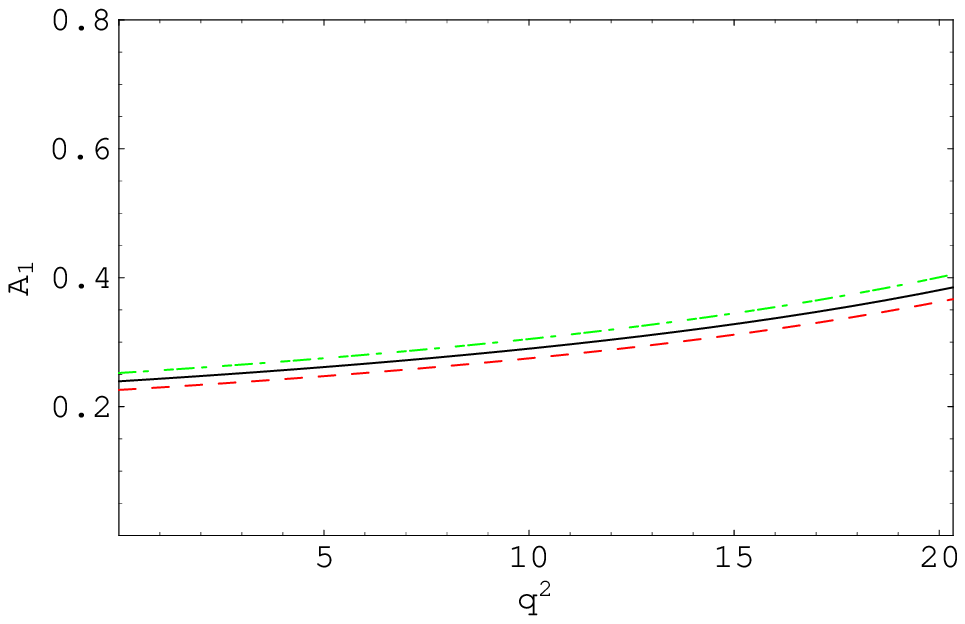}%
\hspace{0.2in}
\includegraphics[width=2.2in]{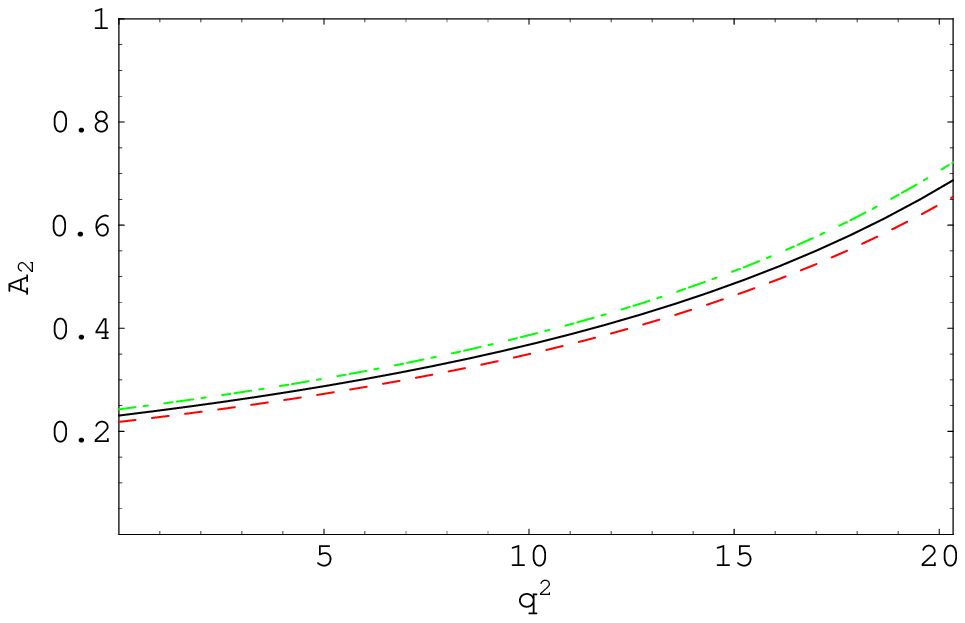}
\hspace{0.2in}
\includegraphics[width=2.2in]{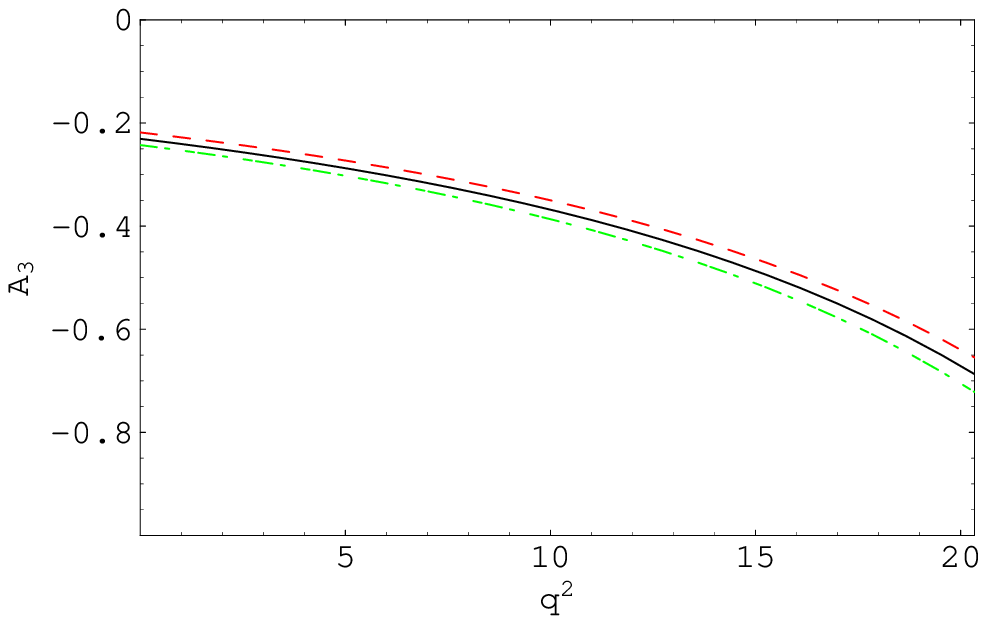}
\hspace{0.2in}
\includegraphics[width=2.2in]{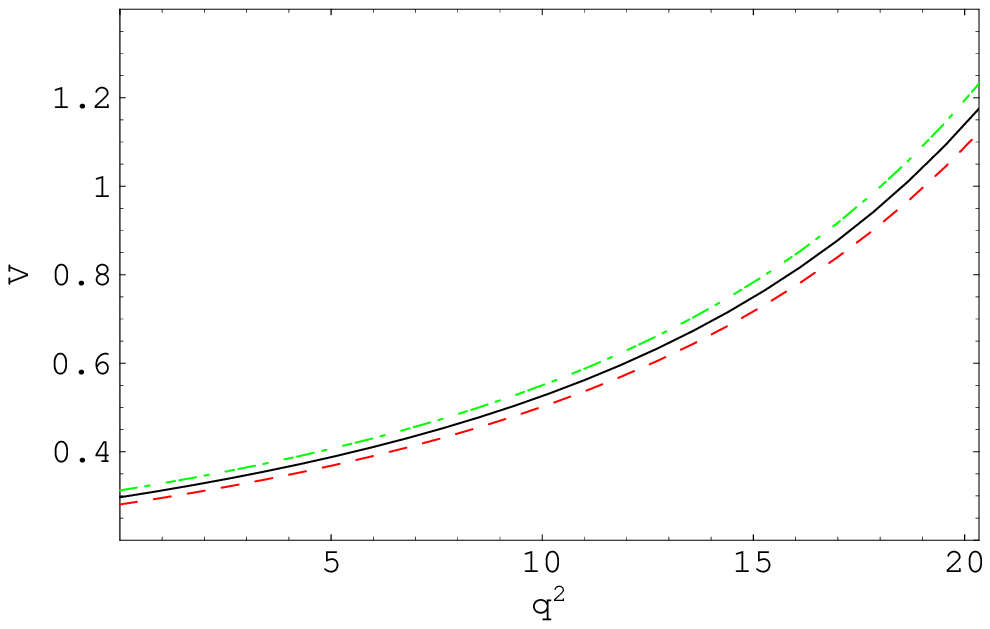}
\\
\flushleft{{\bf Fig.11}: Form factors of $B \rightarrow \rho$ decays
as functions of $q^2$ obtained with only considering the leading
twist meson DAs. The dashed, solid and dot dashed lines correspond
to $s_0 = 1.6$Gev, $T=2.2$Gev; $s_0= 1.7$Gev, $T=2.0$Gev and
$s_0=1.8$GeV, $T=1.8$Gev respectively, which reflect the possible
large uncertainties.}
\end{figure}

\clearpage

\begin{figure}[t]
\centering
\includegraphics[width=2.2in]{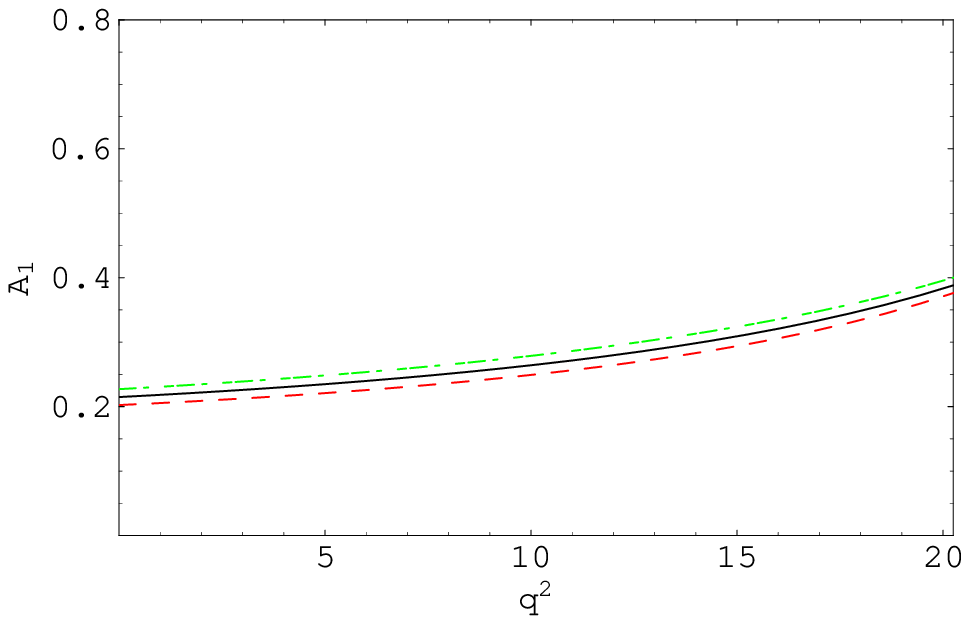}%
\hspace{0.2in}
\includegraphics[width=2.2in]{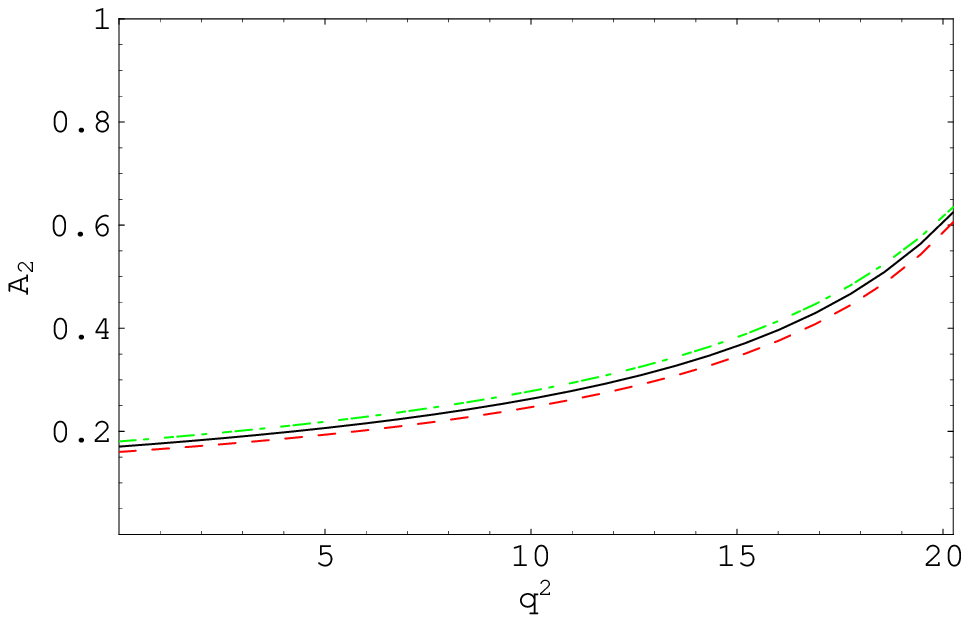}
\hspace{0.2in}
\includegraphics[width=2.2in]{fightl/Fig10c.eps}
\hspace{0.2in}
\includegraphics[width=2.2in]{fightl/Fig10d.eps}
\\
\flushleft{{\bf Fig.12}: Form factors of $B \rightarrow \omega$
decays as functions of $q^2$ obtained with considering the meson DAs
up to twist-4. The dashed, solid and dot dashed lines correspond to
$s_0 = 1.6$Gev, $T=2.0$Gev; $s_0= 1.7$Gev, $T=1.8$Gev and
$s_0=1.8$GeV, $T=1.6$Gev respectively, which reflect the possible
large uncertainties. }
\end{figure}

\begin{figure}
\centering
\includegraphics[width=2.2in]{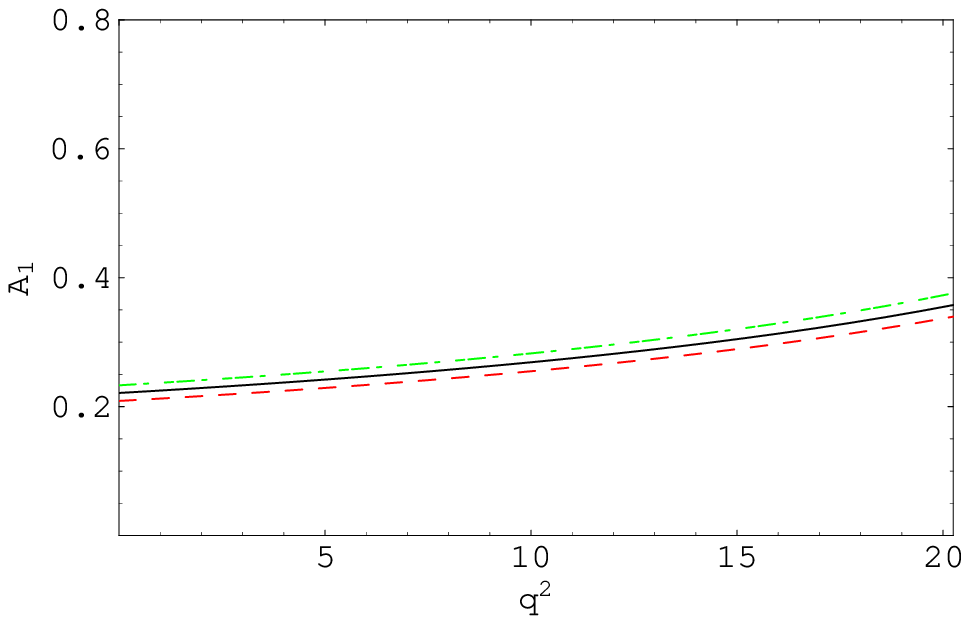}%
\hspace{0.2in}
\includegraphics[width=2.2in]{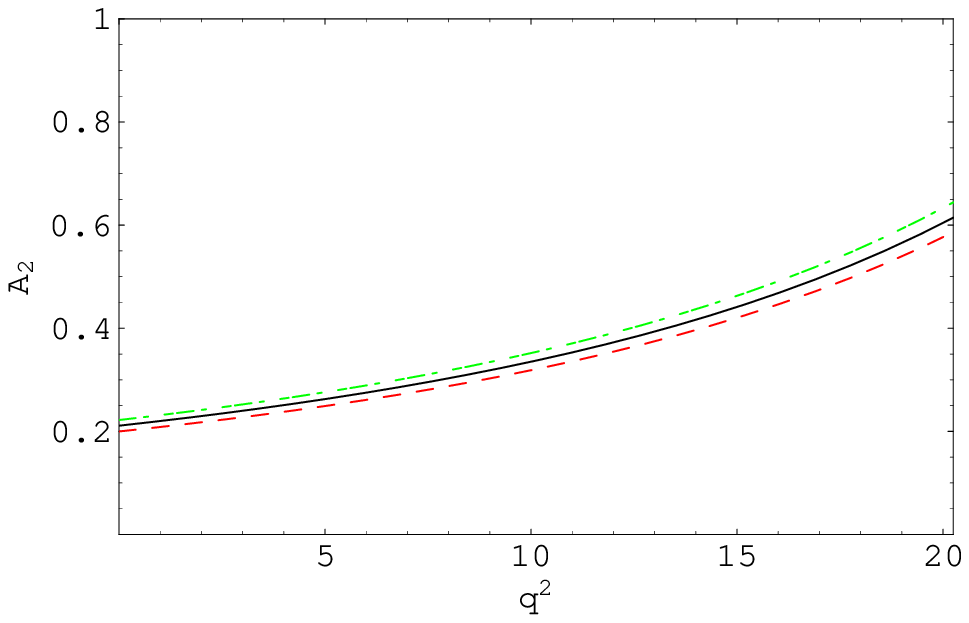}
\hspace{0.2in}
\includegraphics[width=2.2in]{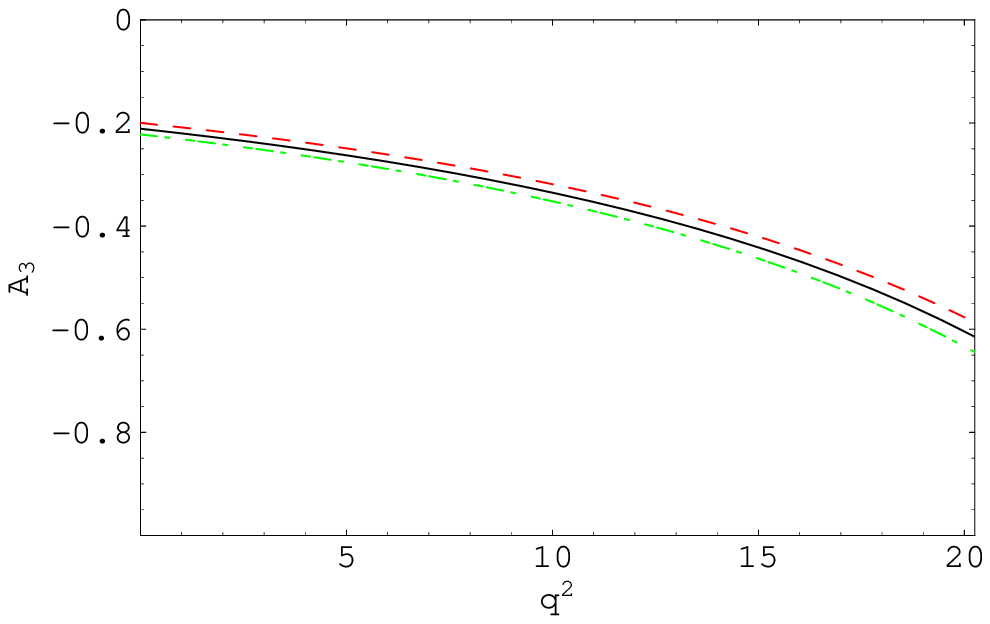}
\hspace{0.2in}
\includegraphics[width=2.2in]{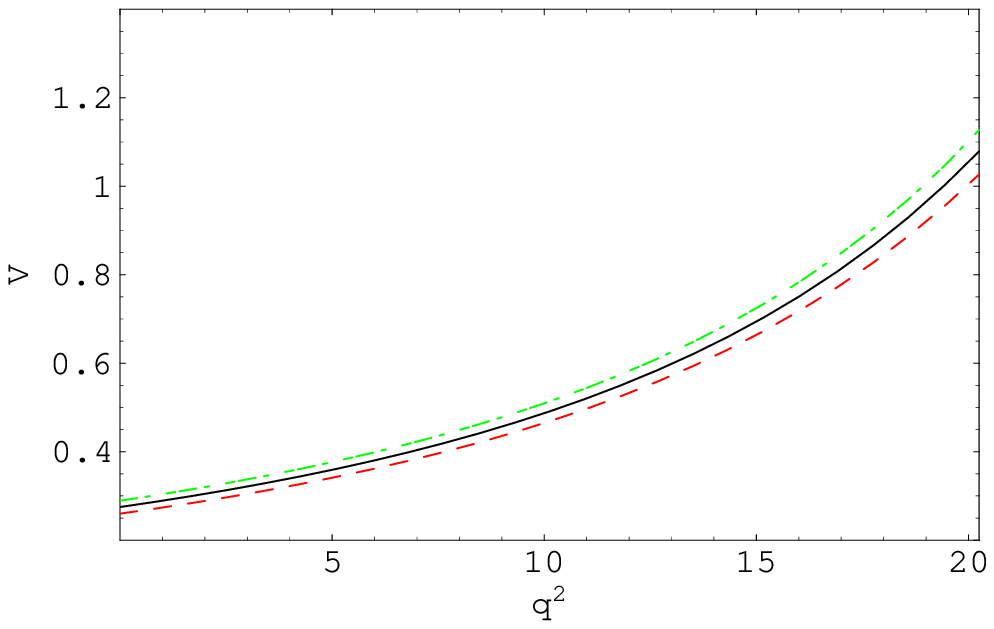}
\\
\flushleft{{\bf Fig.13}: Form factors of $B \rightarrow \omega$
decays as functions of $q^2$ obtained with only considering the
leading twist meson DAs. The dashed, solid and dot dashed lines
correspond to $s_0 = 1.6$Gev, $T=2.2$Gev; $s_0= 1.7$Gev, $T=2.0$Gev
and $s_0=1.8$GeV, $T=1.8$Gev respectively, which reflect the
possible large uncertainties.}
\end{figure}

\clearpage

\begin{figure}[t]
\centering
\includegraphics[width=2.2in]{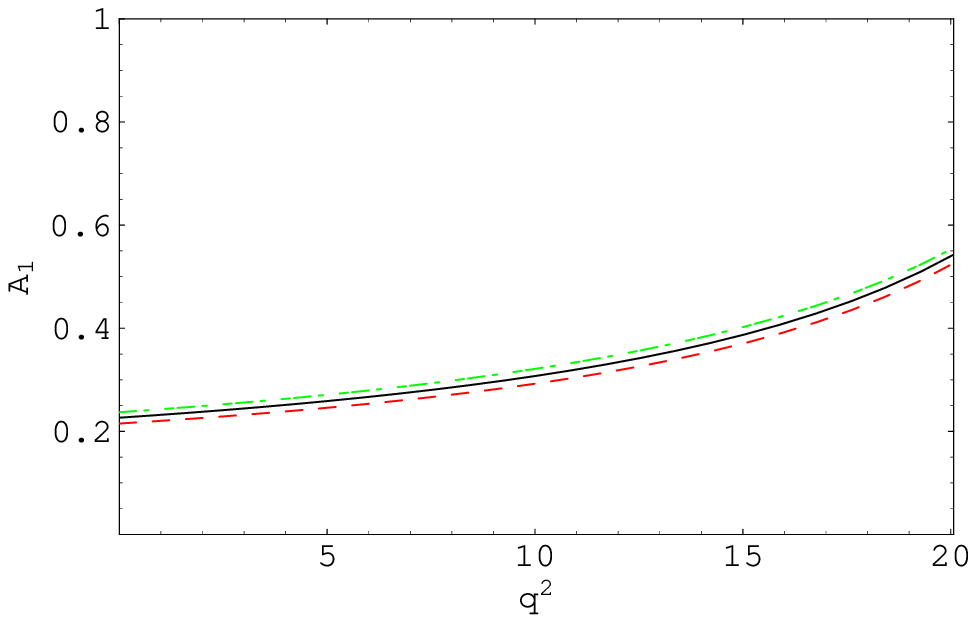}%
\hspace{0.2in}
\includegraphics[width=2.2in]{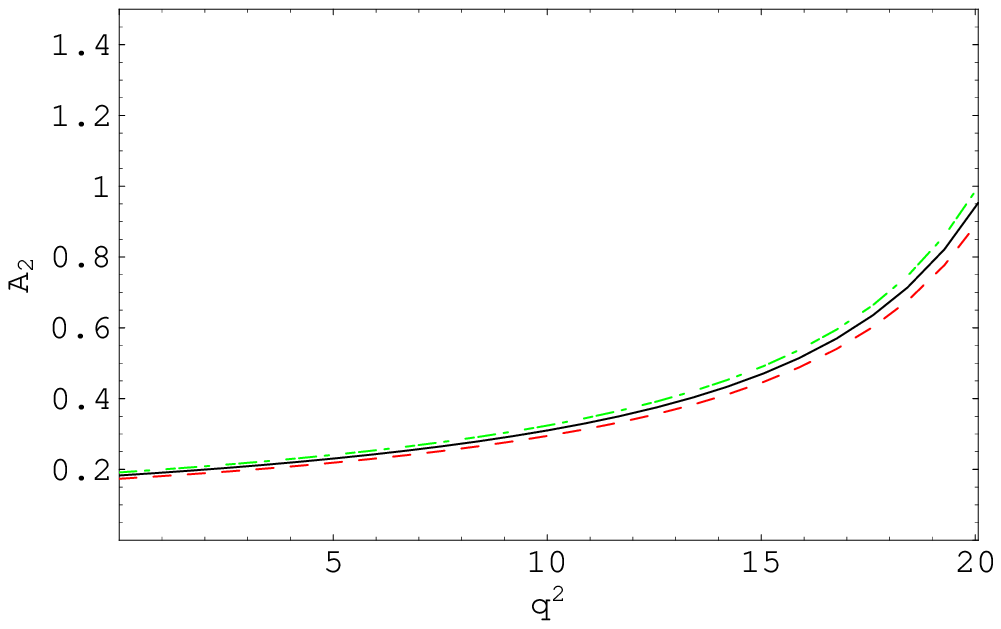}
\hspace{0.2in}
\includegraphics[width=2.2in]{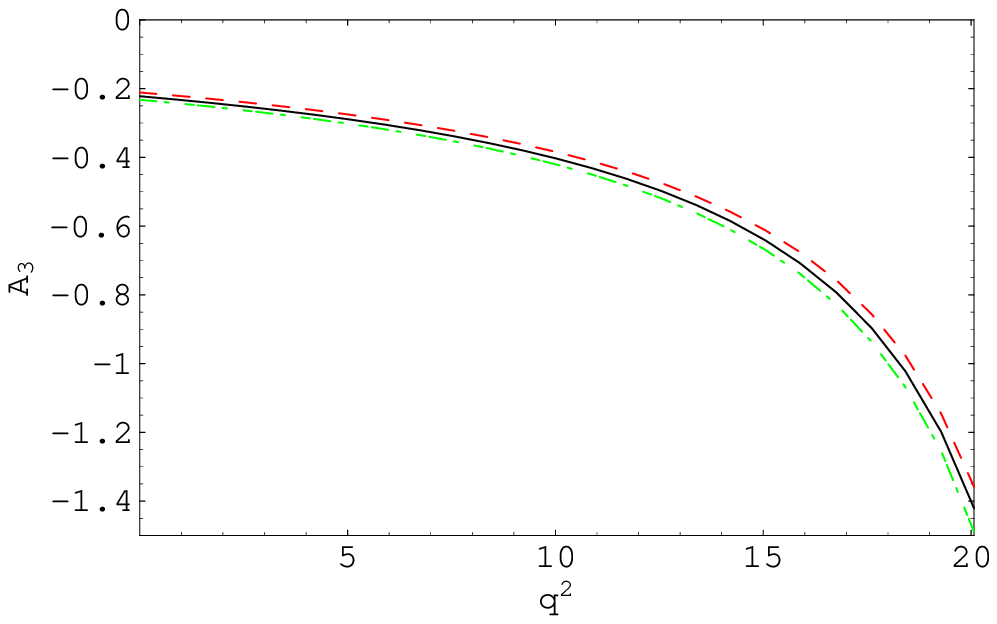}
\hspace{0.2in}
\includegraphics[width=2.2in]{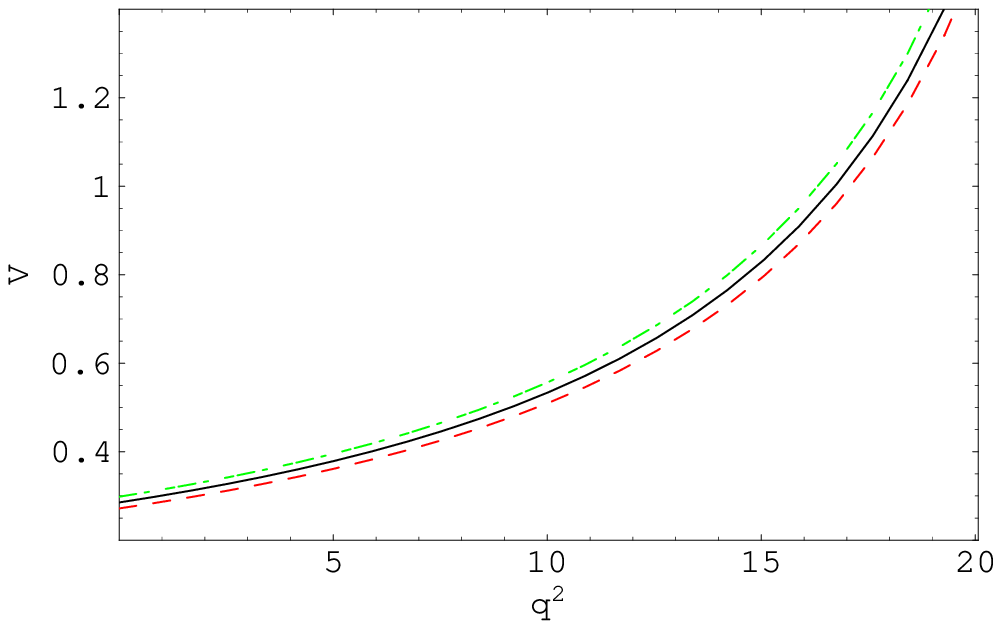}
\flushleft{{\bf Fig.14}: Form factors of $B_s \rightarrow K^*$
decays as functions of $q^2$ obtained with considering the meson DAs
up to twist-4. The dashed, solid and dot dashed lines correspond to
$s_0 = 2.0$Gev, $T=2.2$Gev; $s_0= 2.1$Gev, $T=2.0$Gev and
$s_0=2.2$GeV, $T=1.8$Gev respectively, which reflect the possible
large uncertainties.}
\end{figure}

\begin{figure}
\centering
\includegraphics[width=2.2in]{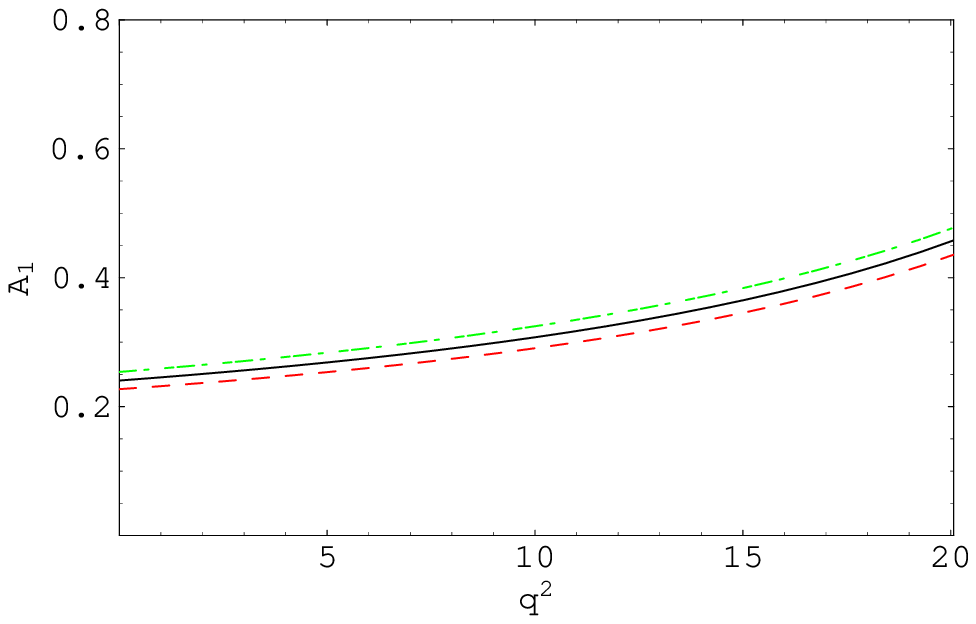}%
\hspace{0.2in}
\includegraphics[width=2.2in]{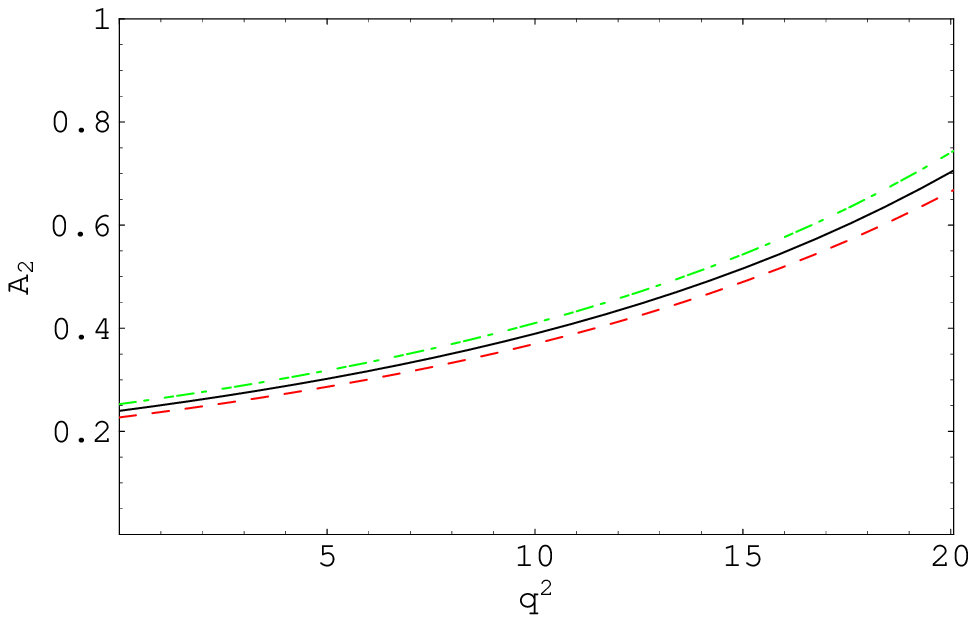}
\hspace{0.2in}
\includegraphics[width=2.2in]{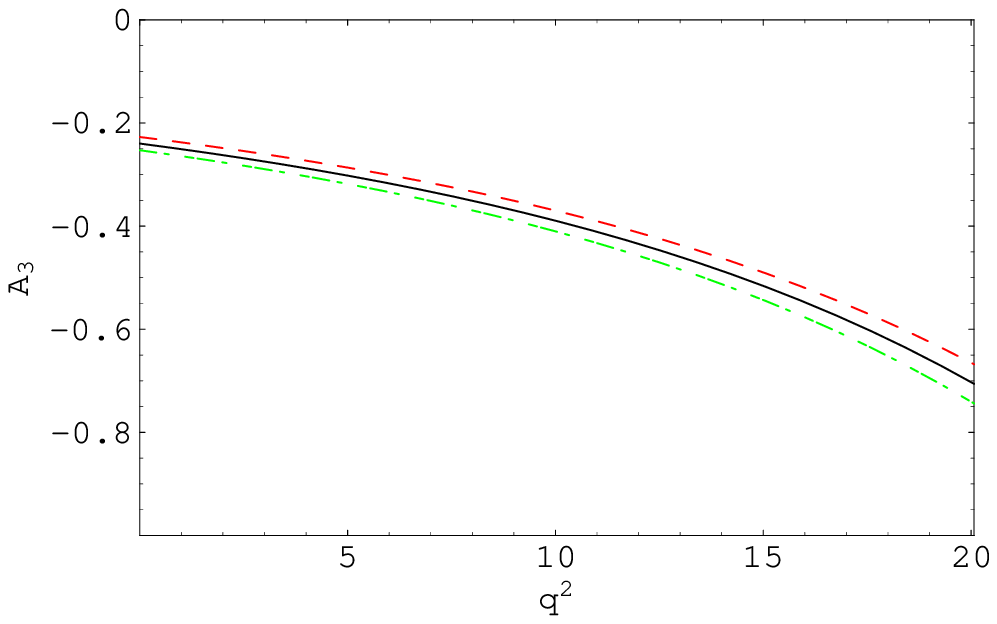}
\hspace{0.2in}
\includegraphics[width=2.2in]{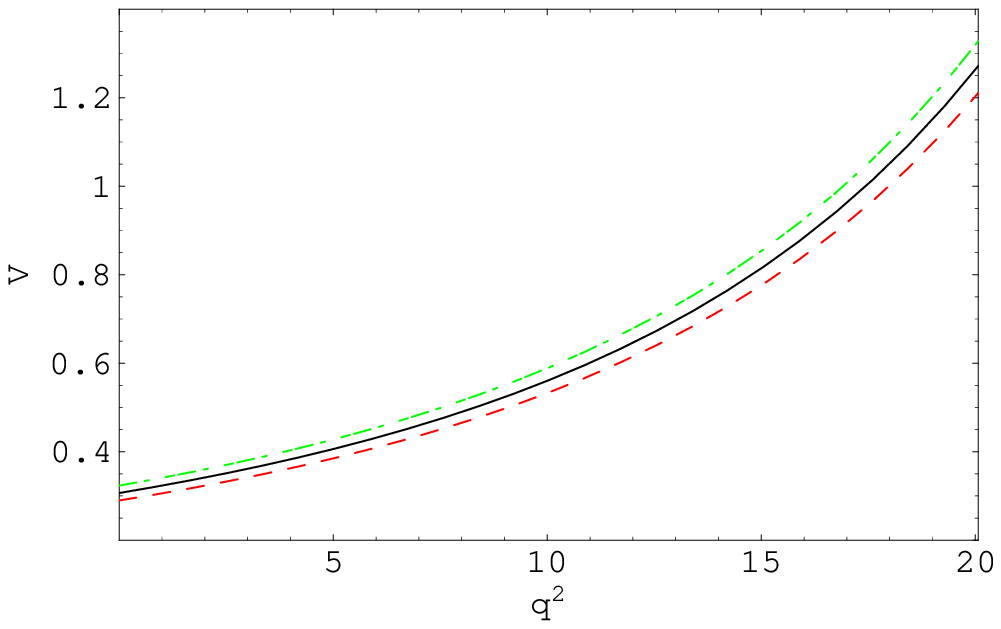}
\flushleft{{\bf Fig.15}: Form factors of $B_s \rightarrow K^*$
decays as functions of $q^2$ obtained with only considering the
leading twist meson DAs. The dashed, solid and dot dashed lines
correspond to $s_0 = 1.8$Gev, $T=2.2$Gev; $s_0= 1.9$Gev, $T=2.0$Gev
and $s_0=2.0$GeV, $T=1.8$Gev respectively, which reflect the
possible large uncertainties.}
\end{figure}

\clearpage

\begin{figure}[t]
\centering
\includegraphics[width=2.2in]{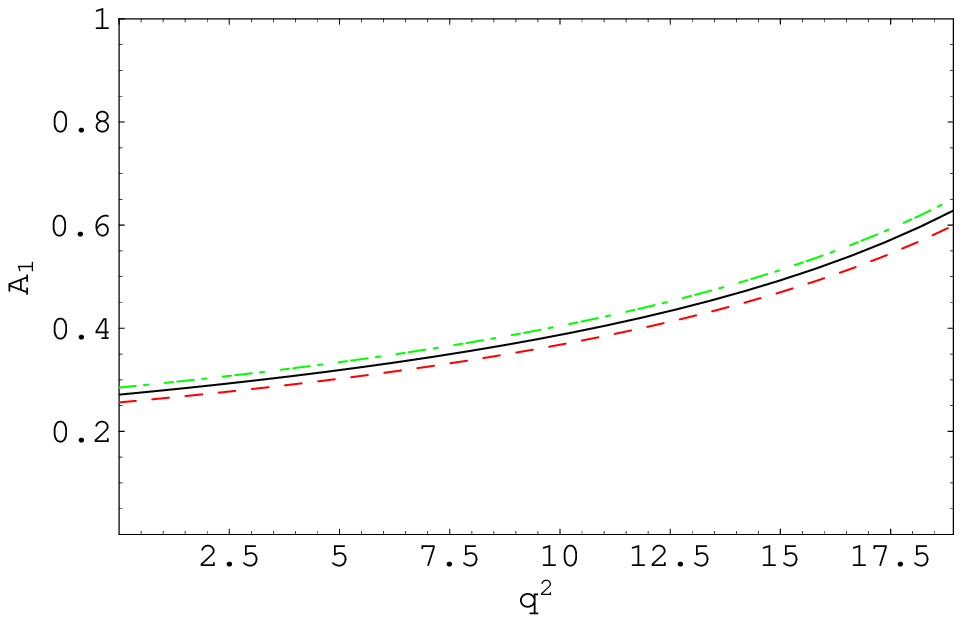}%
\hspace{0.2in}
\includegraphics[width=2.2in]{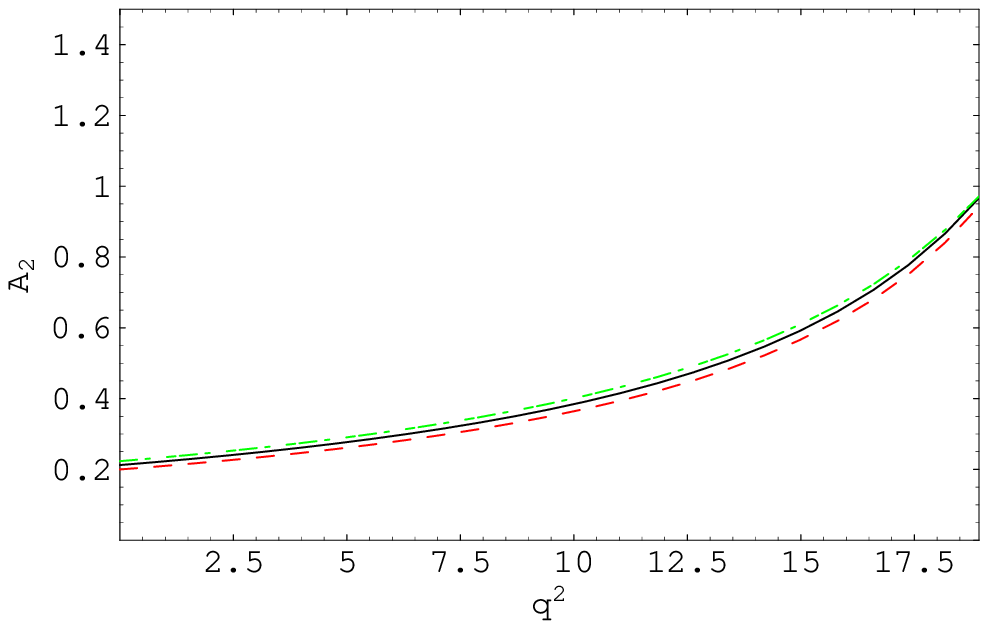}
\hspace{0.2in}
\includegraphics[width=2.2in]{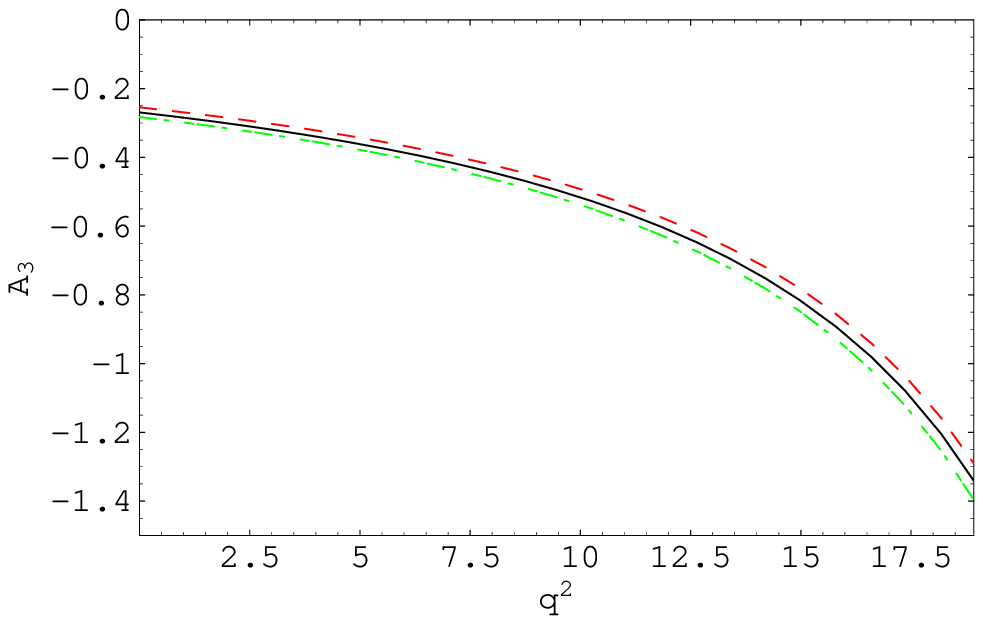}
\hspace{0.2in}
\includegraphics[width=2.2in]{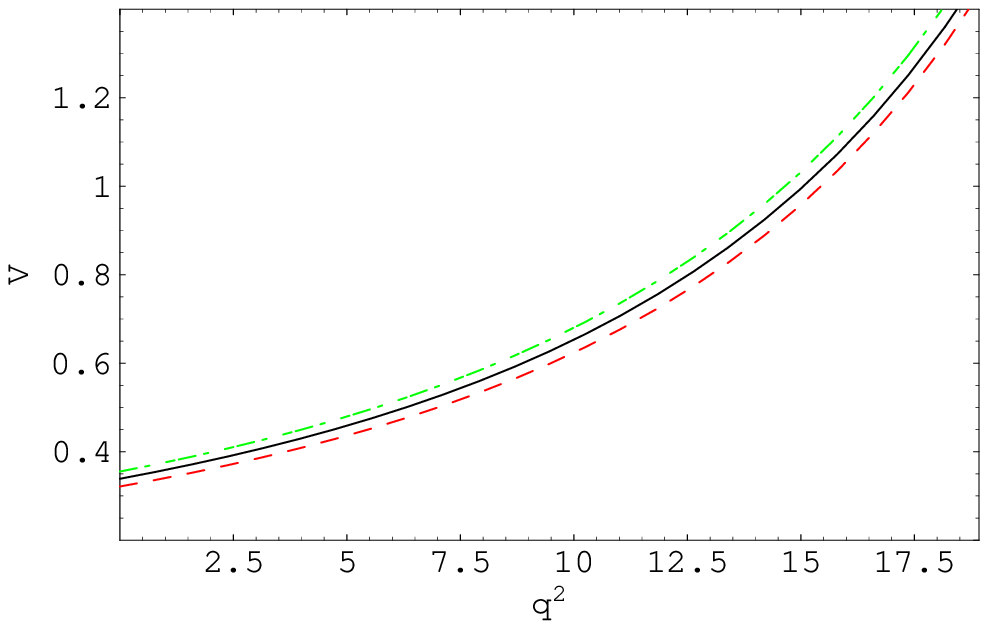}
\\
\flushleft{{\bf Fig.16}: Form factors of $B_s \rightarrow \phi$
decays as functions of $q^2$ obtained with considering the meson DAs
up to twist-4. The dashed, solid and dot dashed lines correspond to
$s_0 = 2.0$Gev, $T=2.2$Gev; $s_0= 2.1$Gev, $T=2.0$Gev and
$s_0=2.2$GeV, $T=1.8$Gev respectively, which reflect the possible
large uncertainties.}
\end{figure}

\begin{figure}
\centering
\includegraphics[width=2.2in]{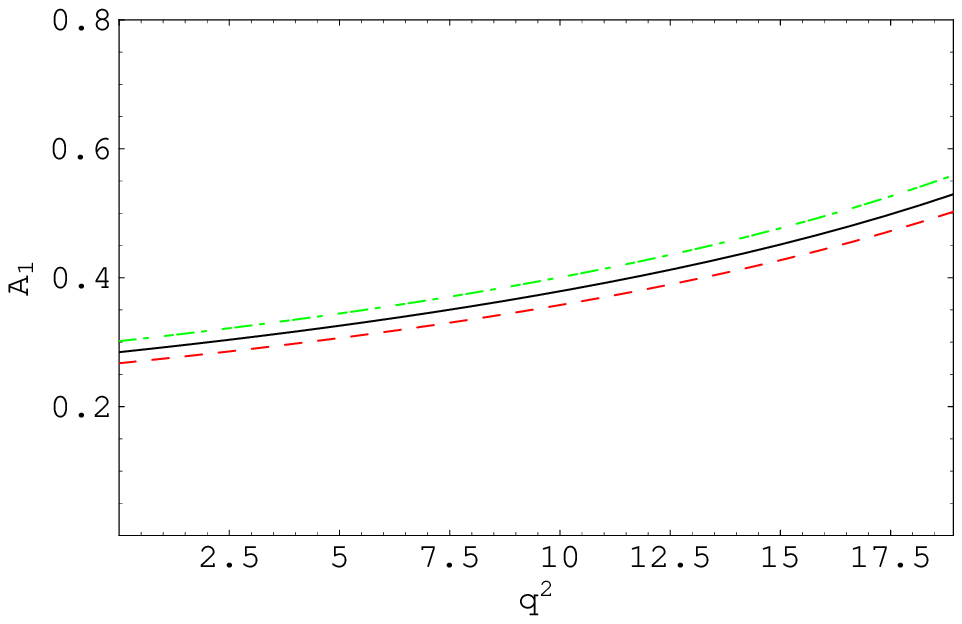}%
\hspace{0.2in}
\includegraphics[width=2.2in]{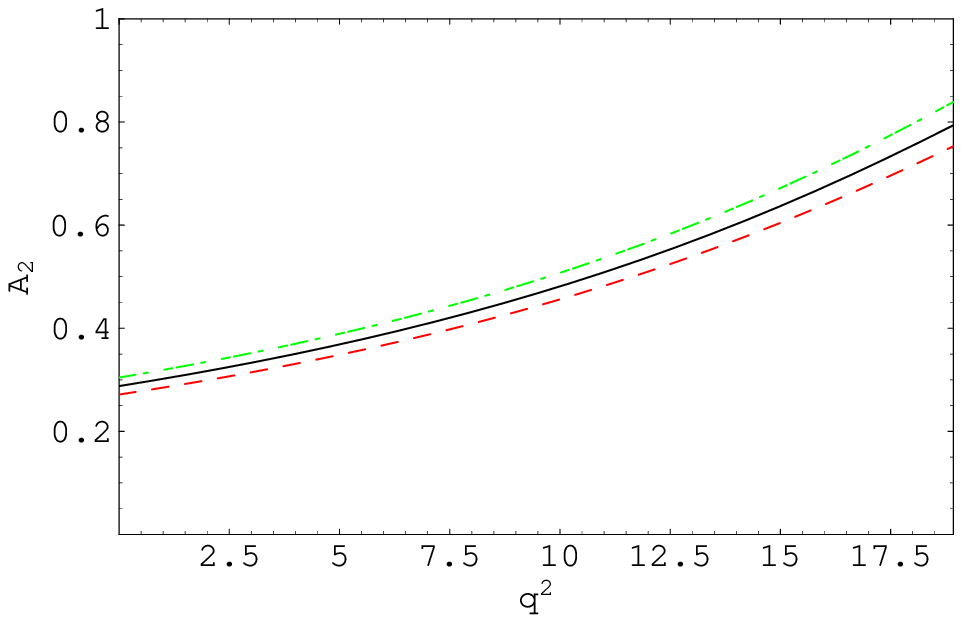}
\hspace{0.2in}
\includegraphics[width=2.2in]{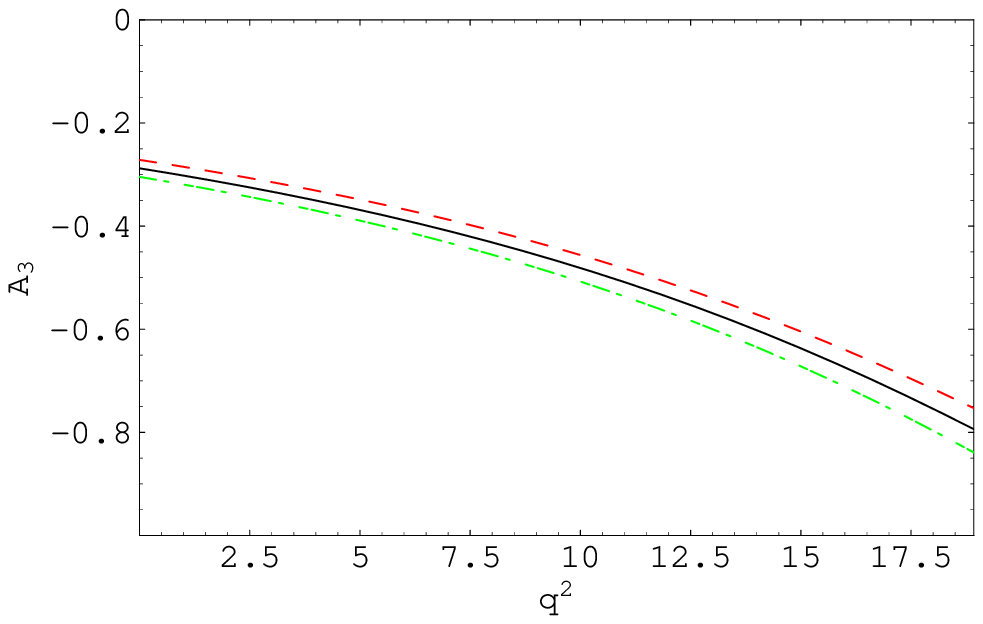}
\hspace{0.2in}
\includegraphics[width=2.2in]{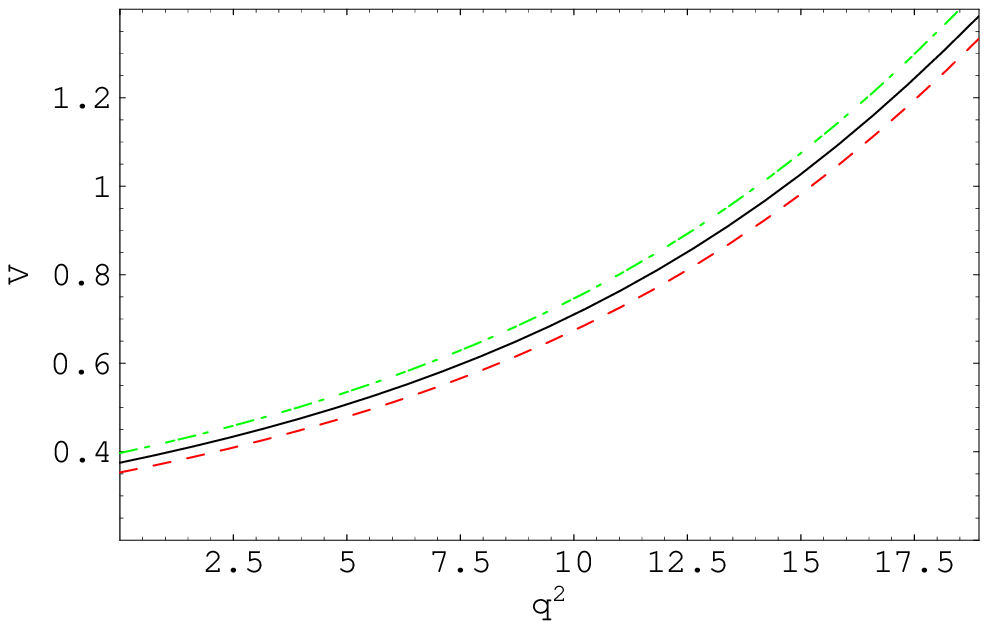}
\\
\flushleft{{\bf Fig.17}: Form factors of $B_s \rightarrow \phi$
decays as functions of $q^2$ obtained with only considering the
leading twist meson DAs. The dashed, solid and dot dashed lines
correspond to $s_0 = 1.8$Gev, $T=2.2$Gev; $s_0= 1.9$Gev, $T=2.0$Gev
and $s_0=2.0$GeV, $T=1.8$Gev respectively, which reflect the
possible large uncertainties.}
\end{figure}

\clearpage

\begin{figure}[h]
\centering
\includegraphics[width=2.2in]{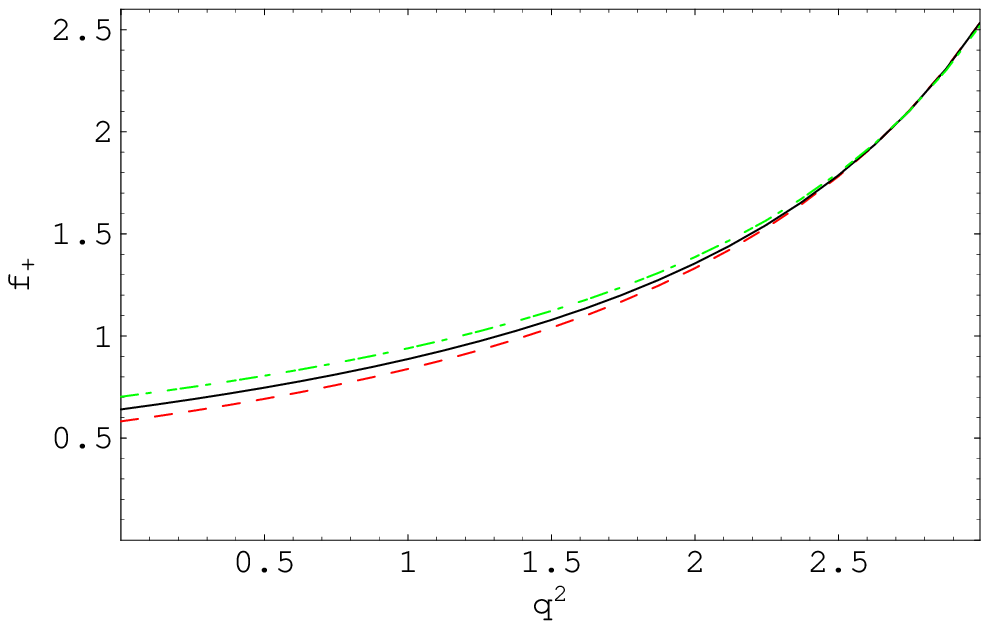}%
\hspace{0.2in}
\includegraphics[width=2.2in]{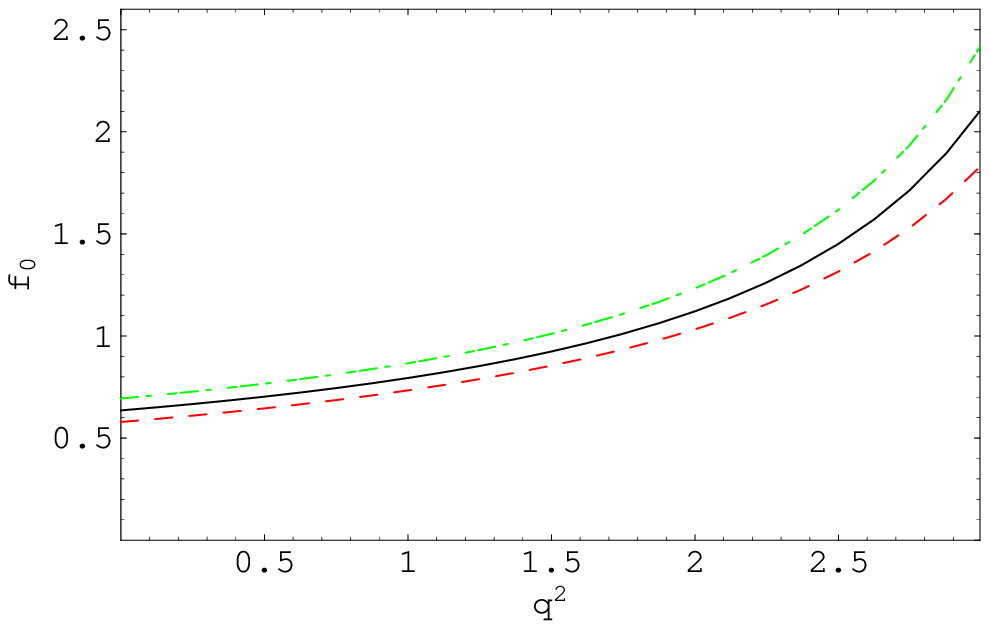}
\flushleft{{\bf Fig.18}: Form factors of $D \rightarrow \pi$ decays
as functions of $q^2$. The dashed, solid and dot dashed lines
correspond to $s_0 = 1.1$Gev, $T= 1.7$Gev; $s_0=1.2$Gev, $T=1.5$Gev
and $s_0=1.3$GeV, $T=1.3$Gev respectively, which reflect the
possible large uncertainties.}
\end{figure}

\begin{figure}[h]
\centering
\includegraphics[width=2.2in]{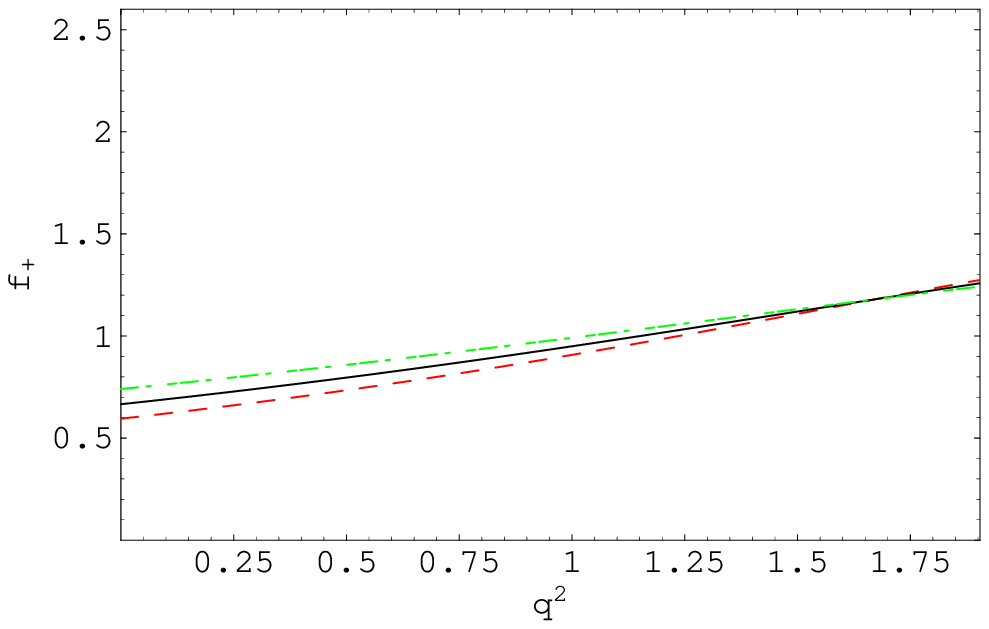}%
\hspace{0.2in}
\includegraphics[width=2.2in]{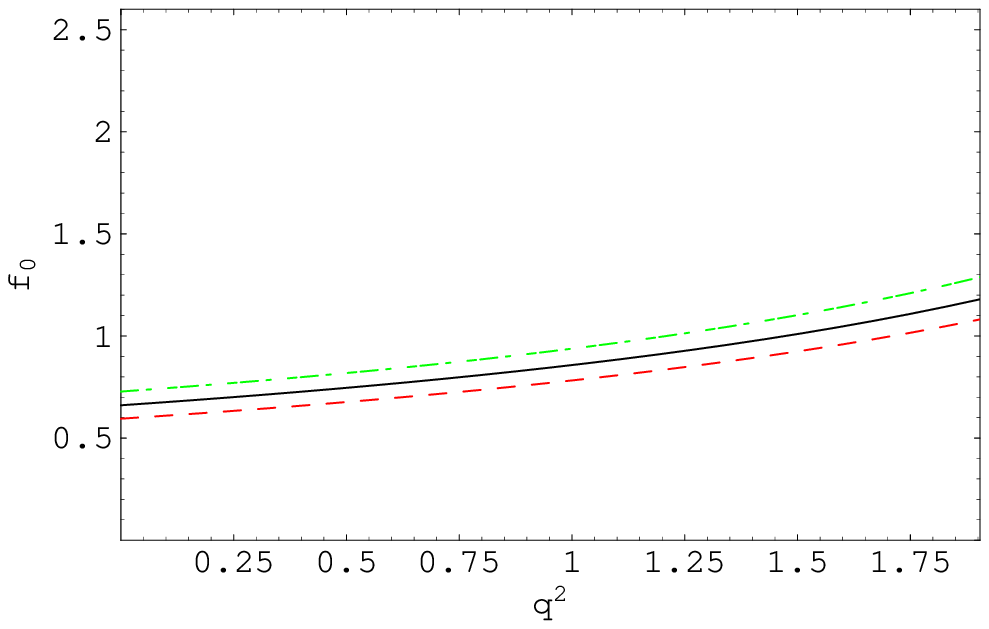}
\\
{\bf Fig.19}: Same as Fig.18, but for $D \rightarrow K$ decays.
\end{figure}

\begin{figure}[h]
\centering
\includegraphics[width=2.2in]{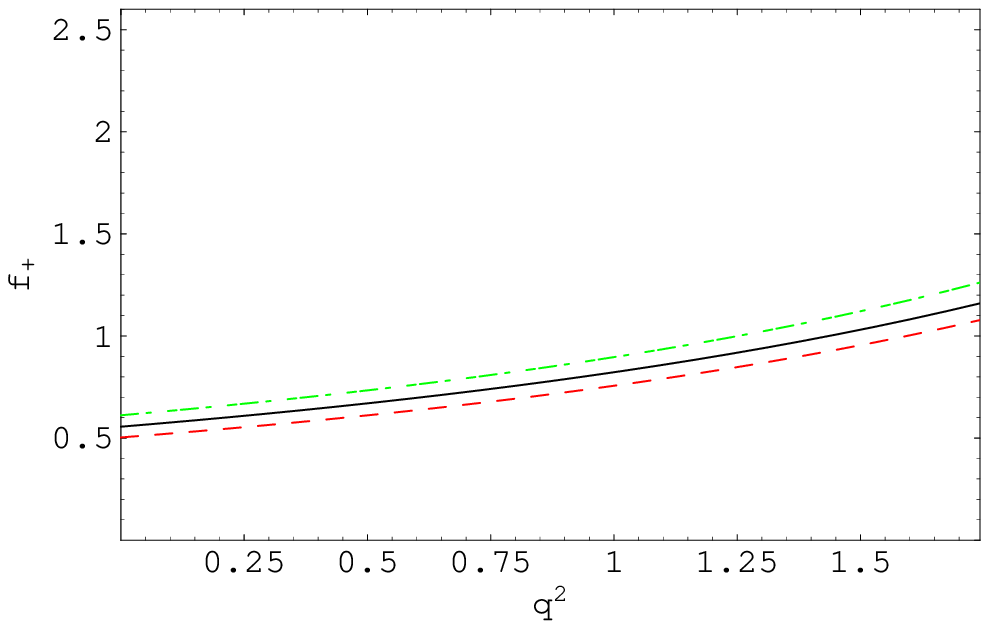}%
\hspace{0.2in}
\includegraphics[width=2.2in]{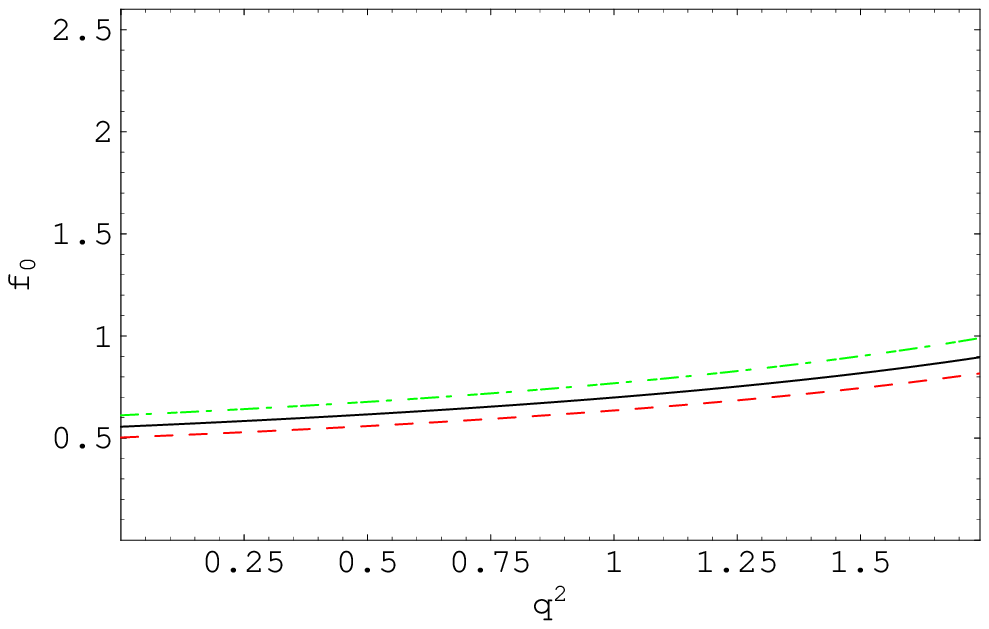}
\\
{\bf Fig.20}: Same as Fig.18, but for $D \rightarrow \eta$ decays.
\end{figure}

\newpage

\begin{figure}[h]
\centering
\includegraphics[width=2.2in]{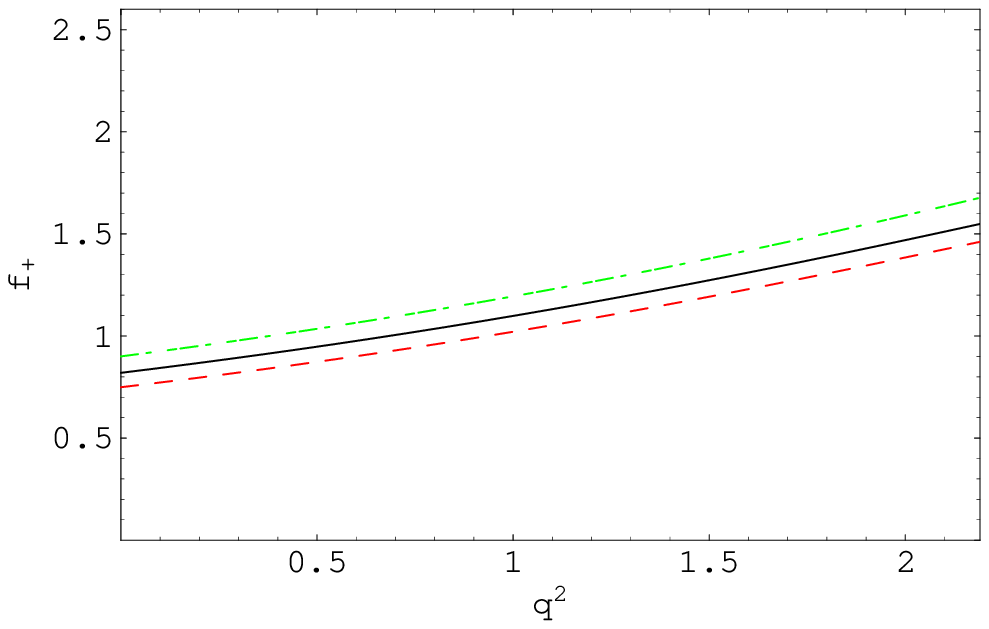}%
\hspace{0.2in}
\includegraphics[width=2.2in]{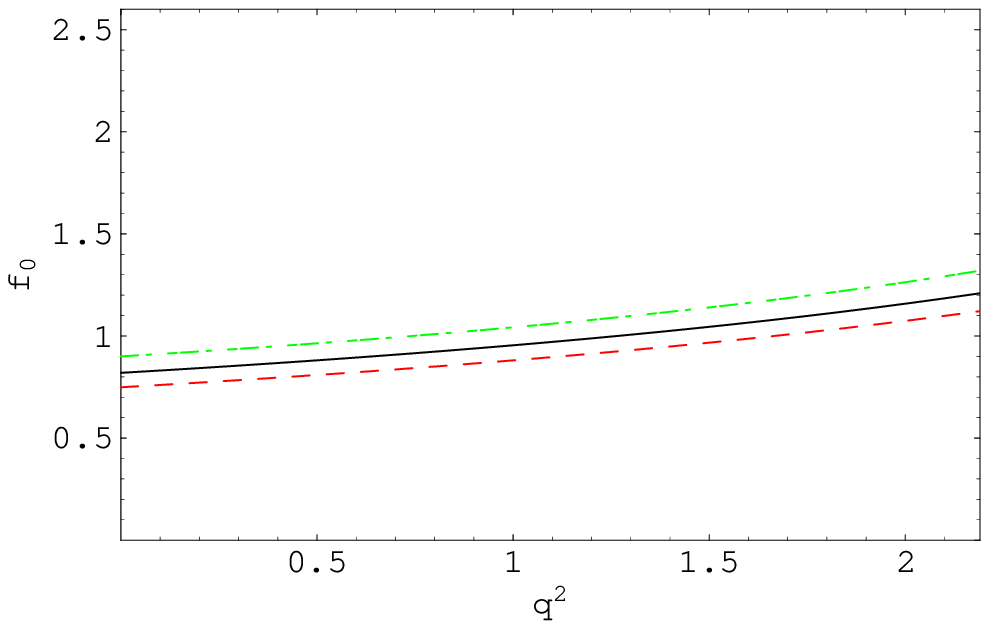}
\flushleft{{\bf Fig.21}: Form factors of $D_s \rightarrow K$ decays
as functions of $q^2$. The dashed, solid and dot dashed lines
correspond to $s_0 = 1.3$Gev, $T= 1.7$Gev; $s_0=1.4$Gev, $T=1.5$Gev
and $s_0=1.5$GeV, $T=1.3$Gev respectively, which reflect the
possible large uncertainties.}
\end{figure}

\begin{figure}[h]
\centering
\includegraphics[width=2.2in]{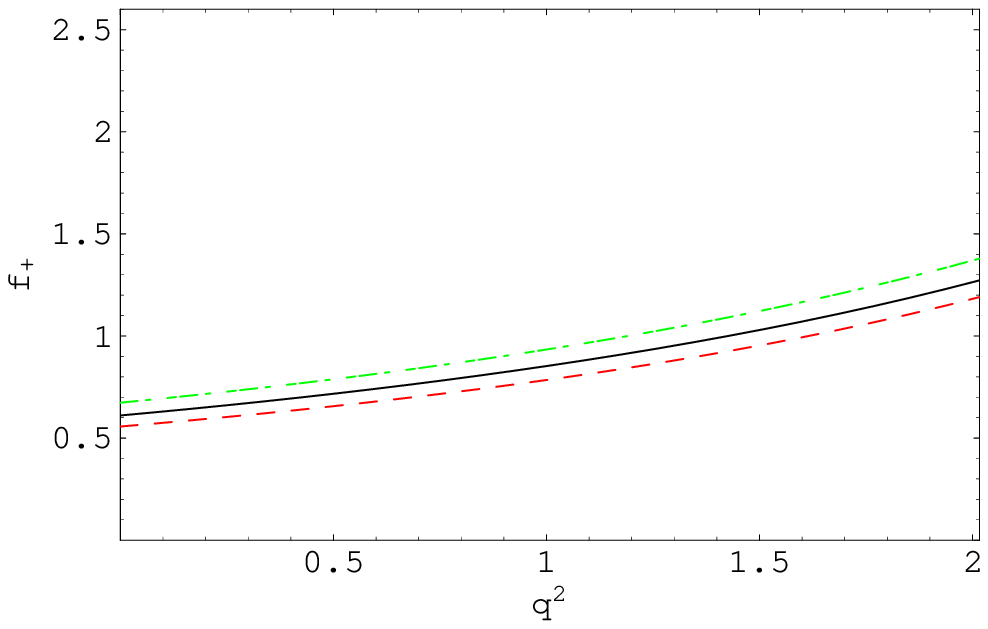}%
\hspace{0.2in}
\includegraphics[width=2.2in]{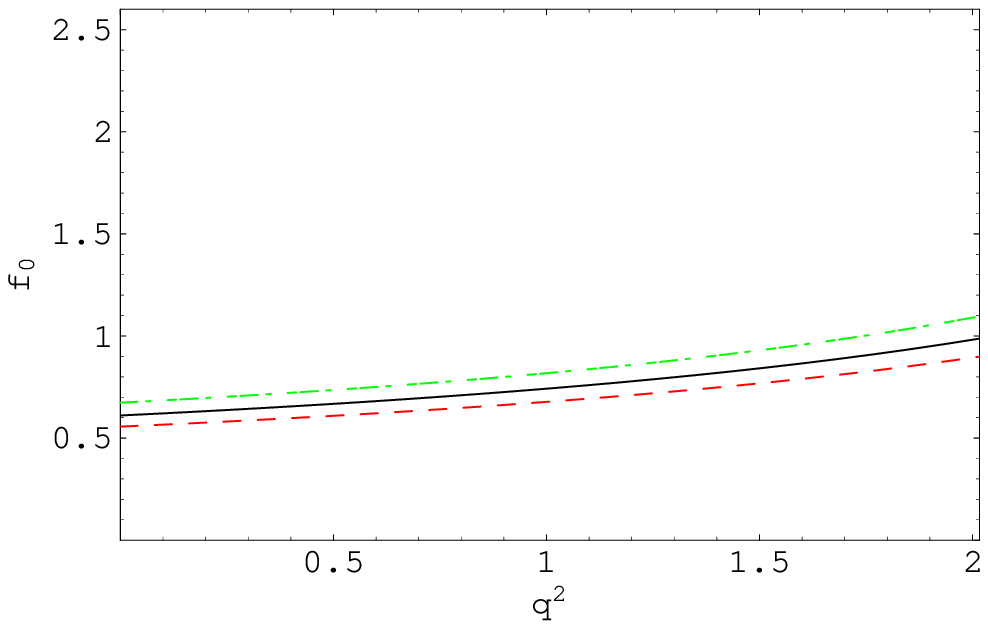}
\\
{\bf Fig.22}: Same as Fig.21, but for $D_s \rightarrow \eta$ decays.
\end{figure}

\clearpage

\begin{figure}[t]
\centering
\includegraphics[width=2.2in]{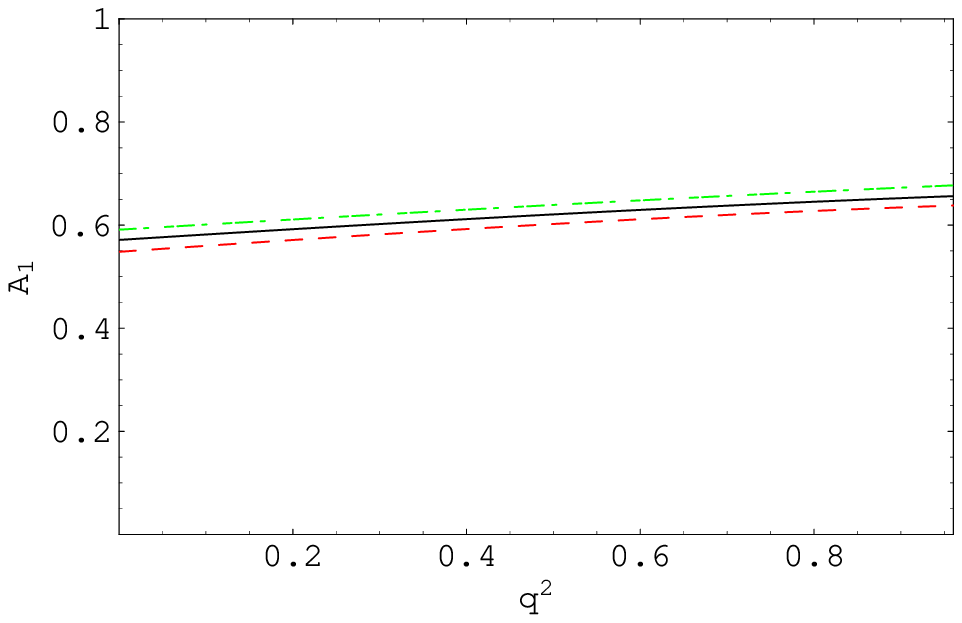}%
\hspace{0.2in}
\includegraphics[width=2.2in]{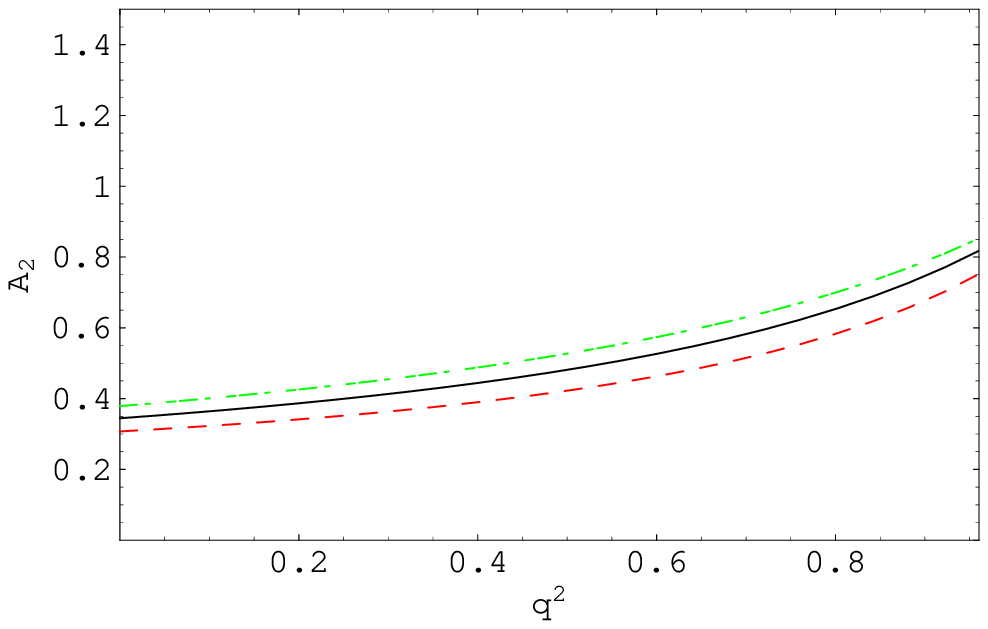}
\hspace{0.2in}
\includegraphics[width=2.2in]{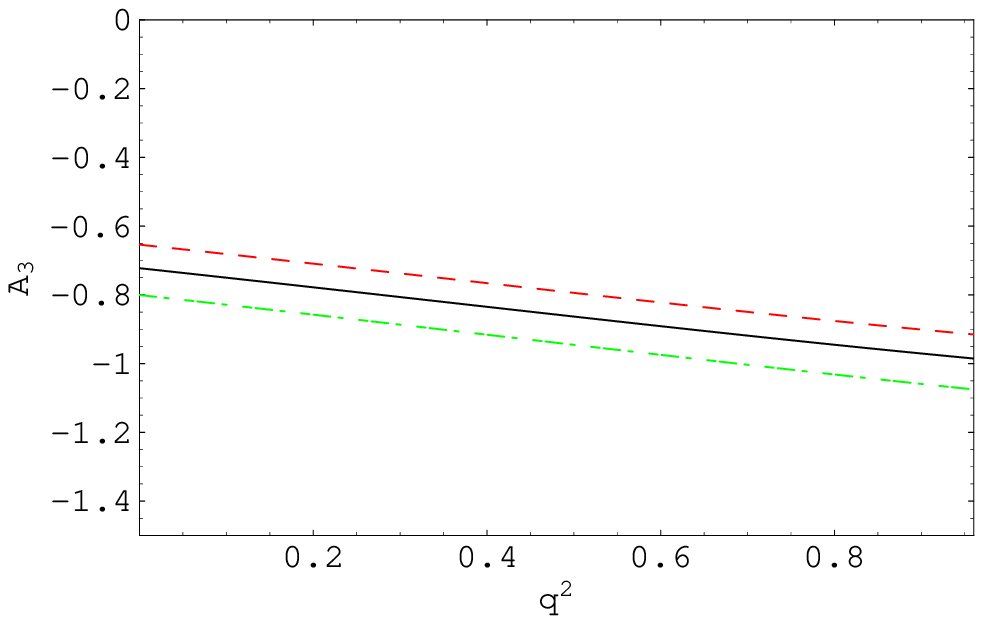}
\hspace{0.2in}
\includegraphics[width=2.2in]{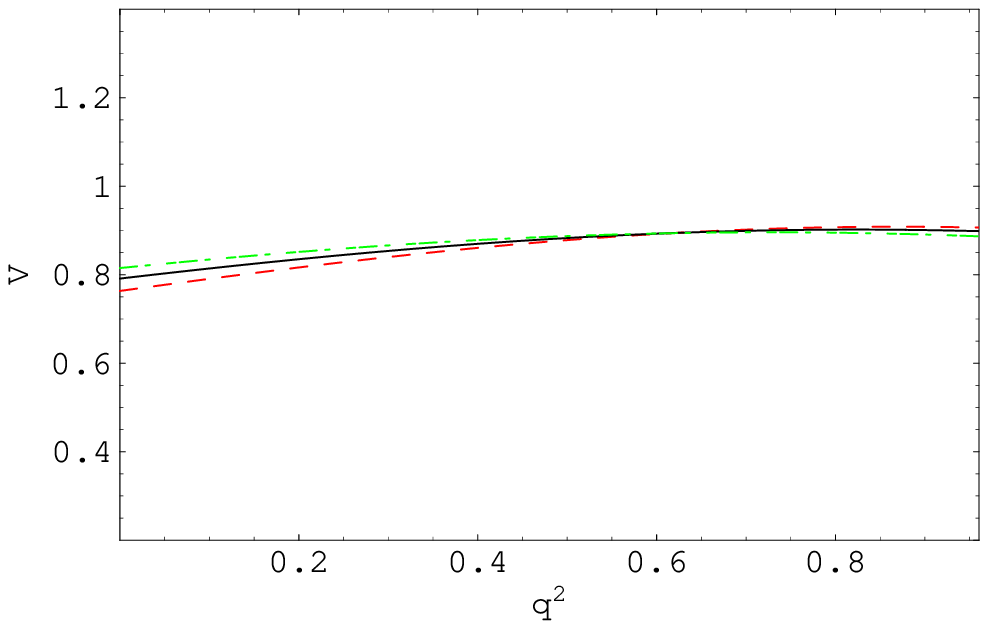}
\flushleft{{\bf Fig.23}: Form factors of $D \rightarrow K^*$ decays
as functions of $q^2$ obtained with considering the meson DAs up to
twist-4. The dashed, solid and dot dashed lines correspond to $s_0 =
1.8$Gev, $T=1.4$Gev; $s_0=1.9$Gev, $T=1.2$Gev and $s_0=2.0$GeV,
$T=1.0$Gev respectively, which reflect the possible large
uncertainties.}
\end{figure}

\begin{figure}
\centering
\includegraphics[width=2.2in]{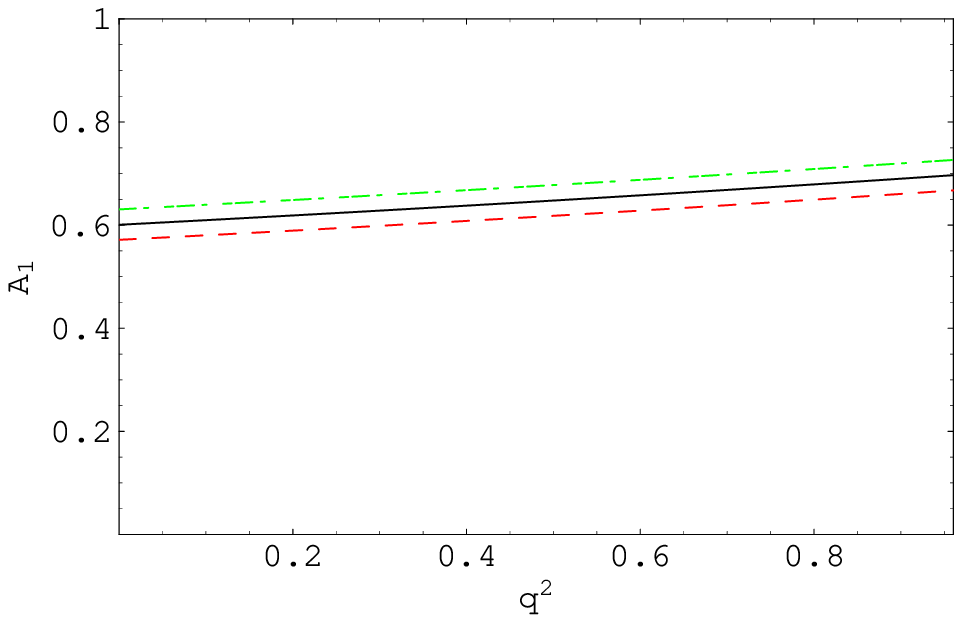}%
\hspace{0.2in}
\includegraphics[width=2.2in]{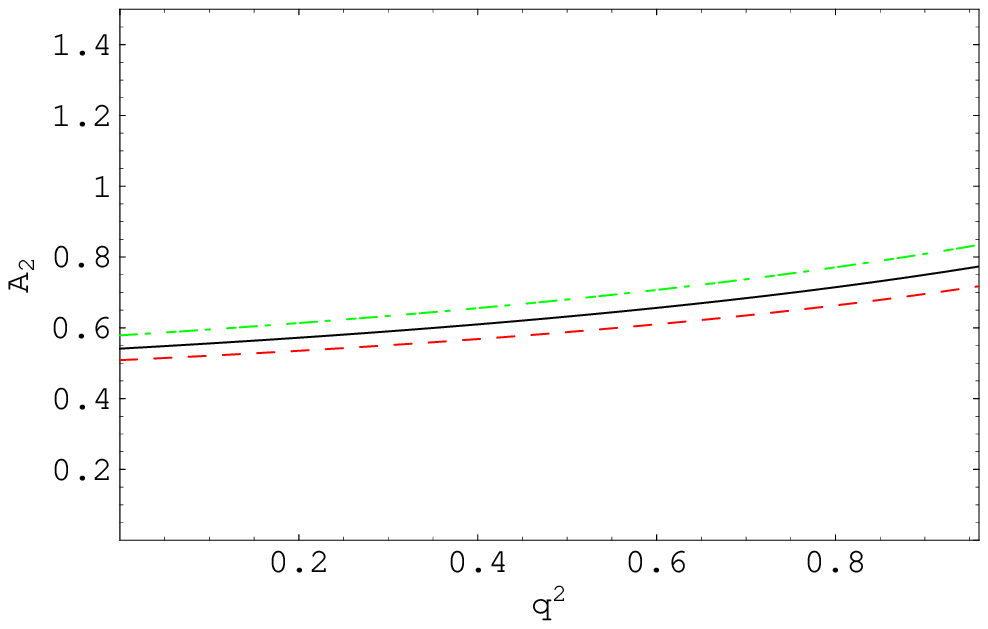}
\hspace{0.2in}
\includegraphics[width=2.2in]{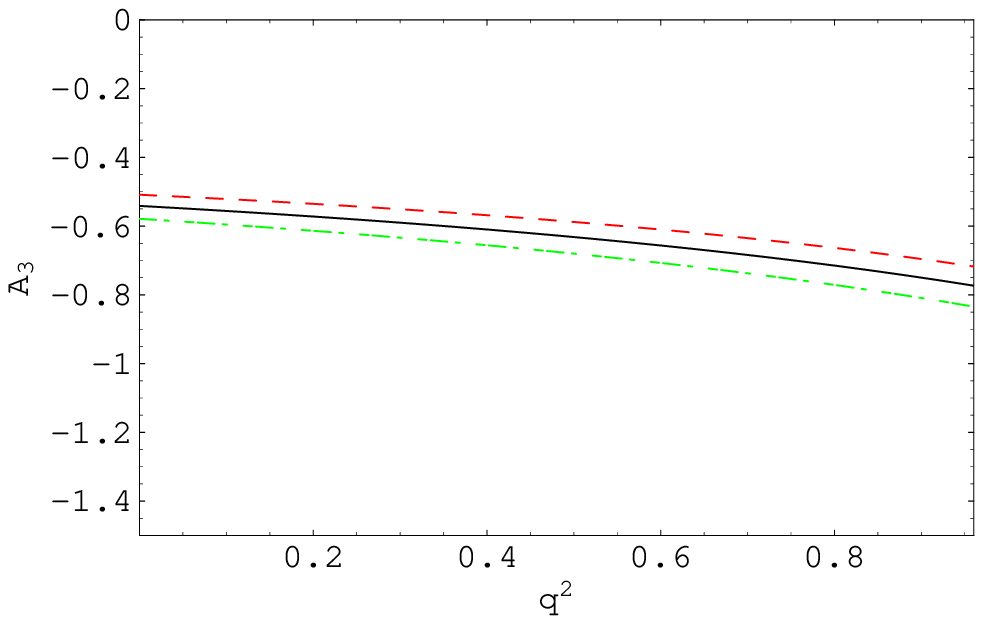}
\hspace{0.2in}
\includegraphics[width=2.2in]{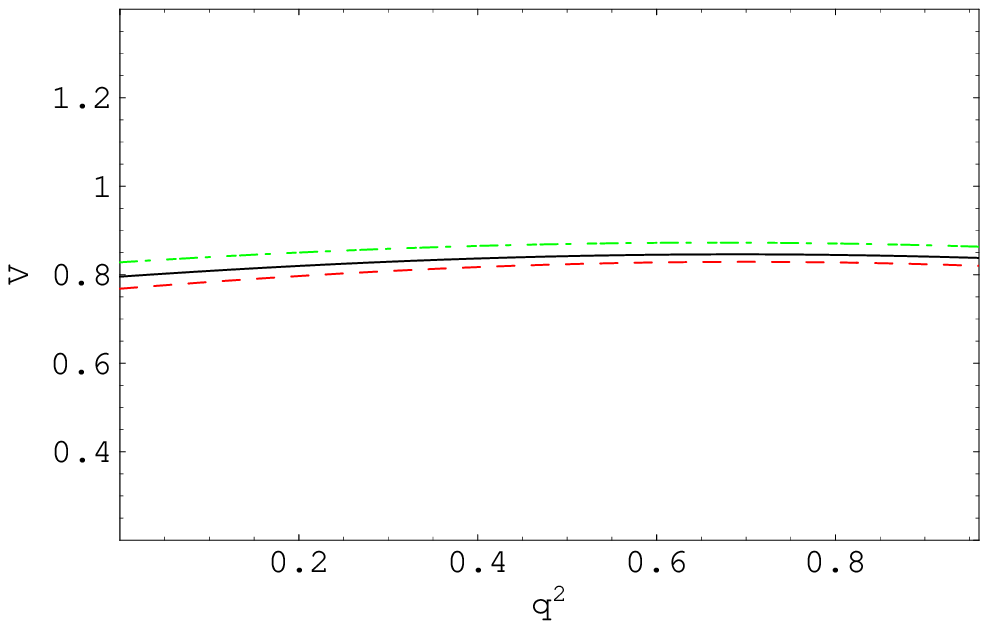}
\flushleft{{\bf Fig.24}: Form factors of $D \rightarrow K^*$ decays
as functions of $q^2$ obtained with only considering the leading
twist meson DAs. The dashed, solid and dot dashed lines correspond
to $s_0 = 1.9$Gev, $T=1.7$Gev; $s_0=2.0$Gev, $T=1.5$Gev and
$s_0=2.1$GeV, $T=1.3$Gev respectively, which reflect the possible
large uncertainties.}
\end{figure}

\clearpage

\begin{figure}[t]
\centering
\includegraphics[width=2.2in]{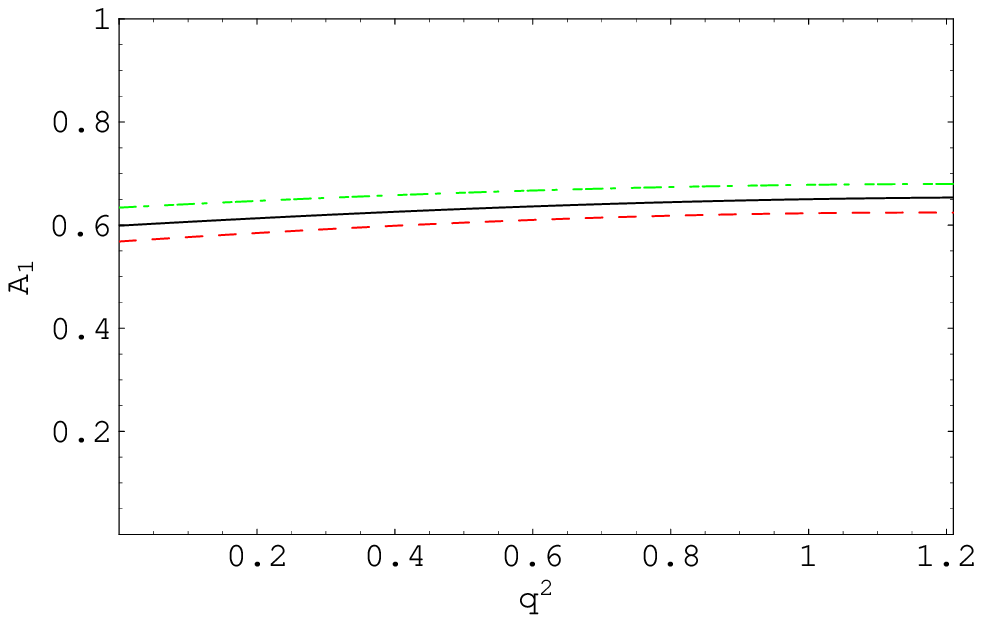}%
\hspace{0.2in}
\includegraphics[width=2.2in]{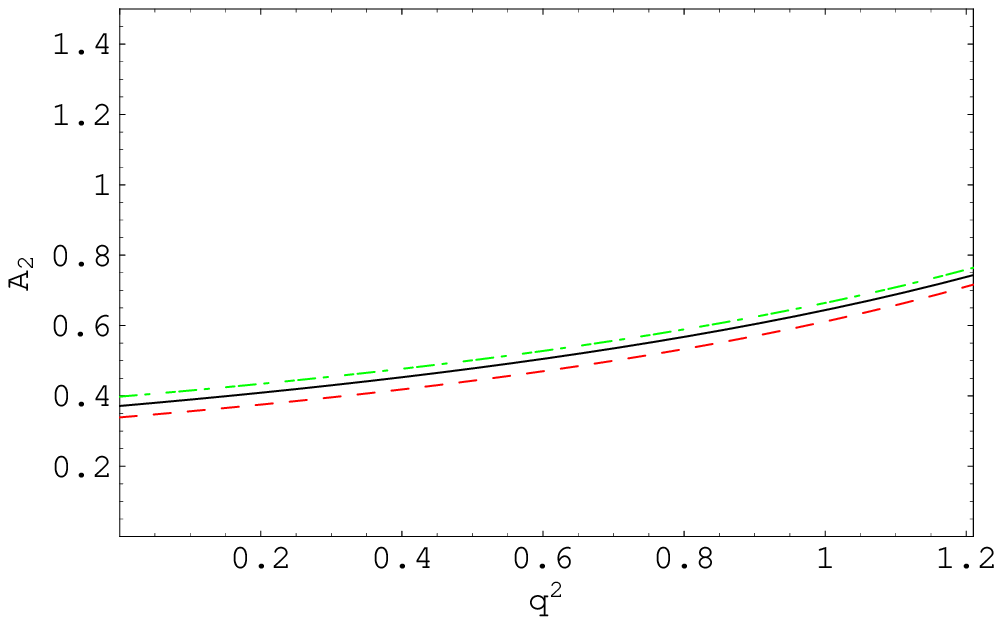}
\hspace{0.2in}
\includegraphics[width=2.2in]{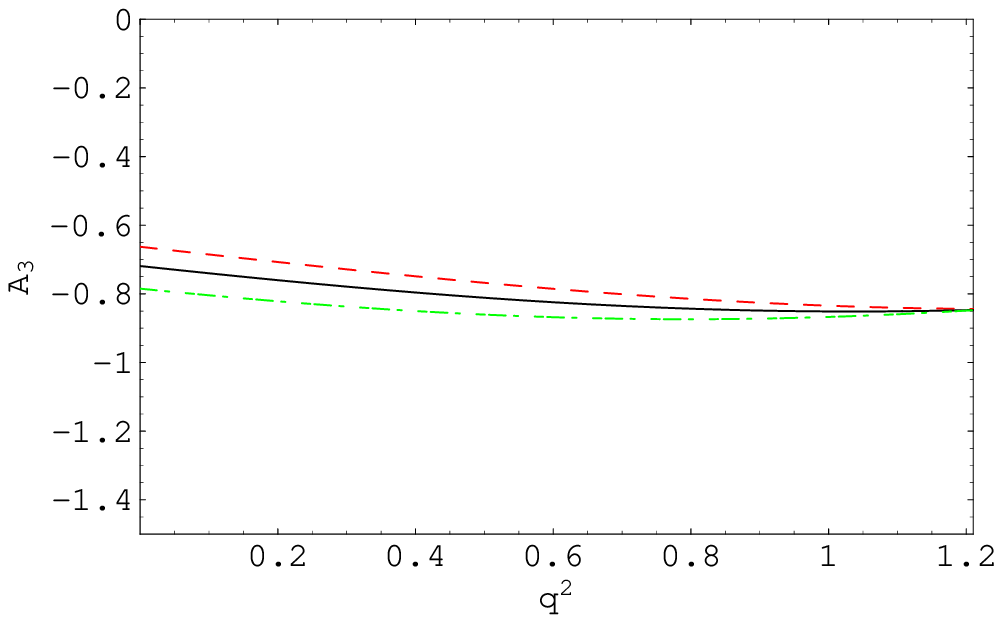}
\hspace{0.2in}
\includegraphics[width=2.2in]{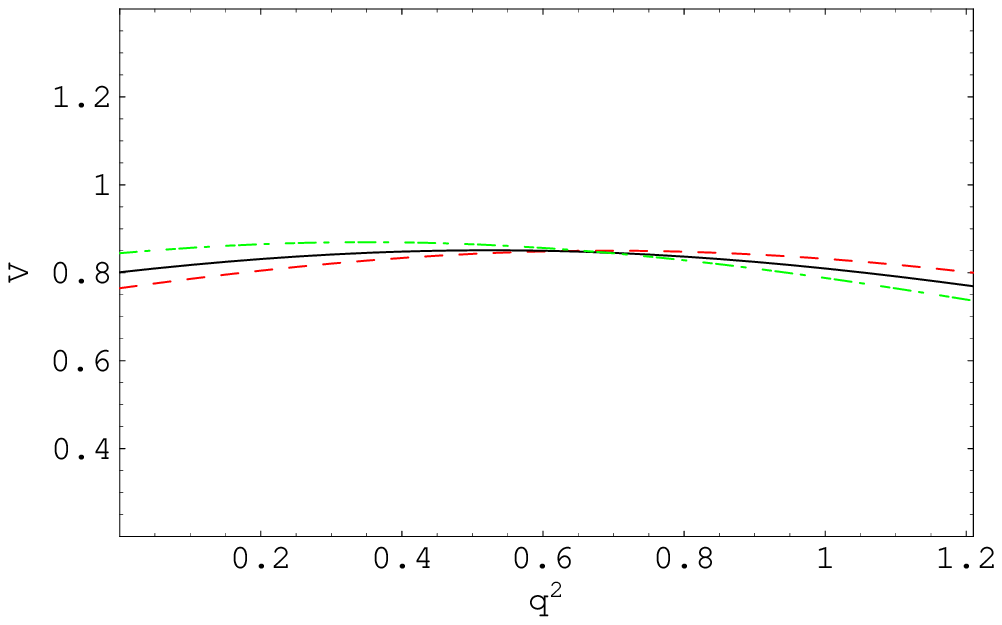}
\\
\flushleft{{\bf Fig.25}: Form factors of $D \rightarrow \rho$ decays
as functions of $q^2$ obtained with considering the meson DAs up to
twist-4. The dashed, solid and dot dashed lines correspond to $s_0 =
1.8$Gev, $T=1.4$Gev; $s_0=1.9$Gev, $T=1.2$Gev and $s_0=2.0$GeV,
$T=1.0$Gev respectively, which reflect the possible large
uncertainties.}
\end{figure}

\begin{figure}
\centering
\includegraphics[width=2.2in]{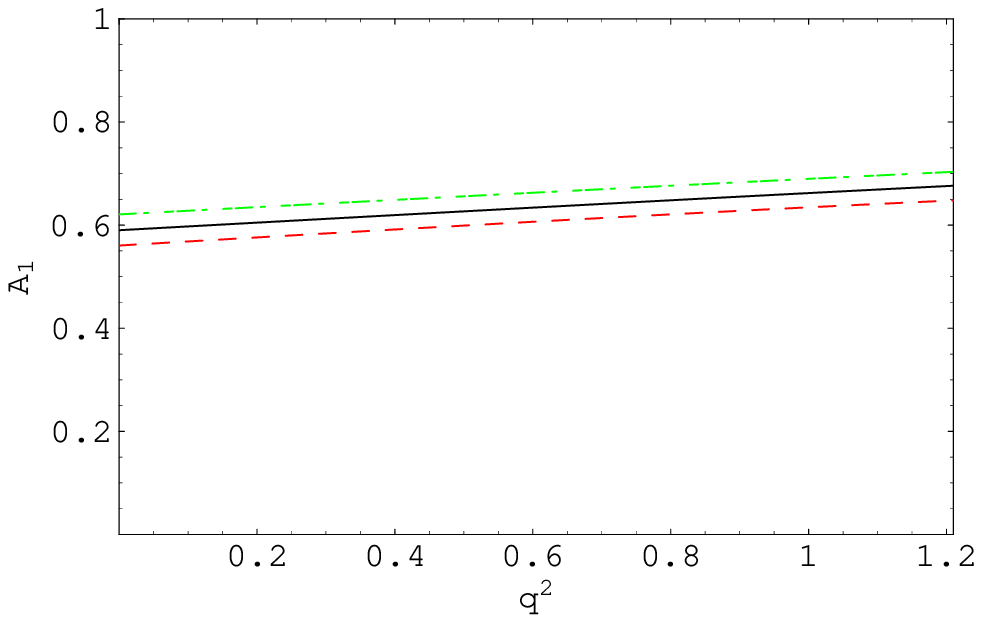}%
\hspace{0.2in}
\includegraphics[width=2.2in]{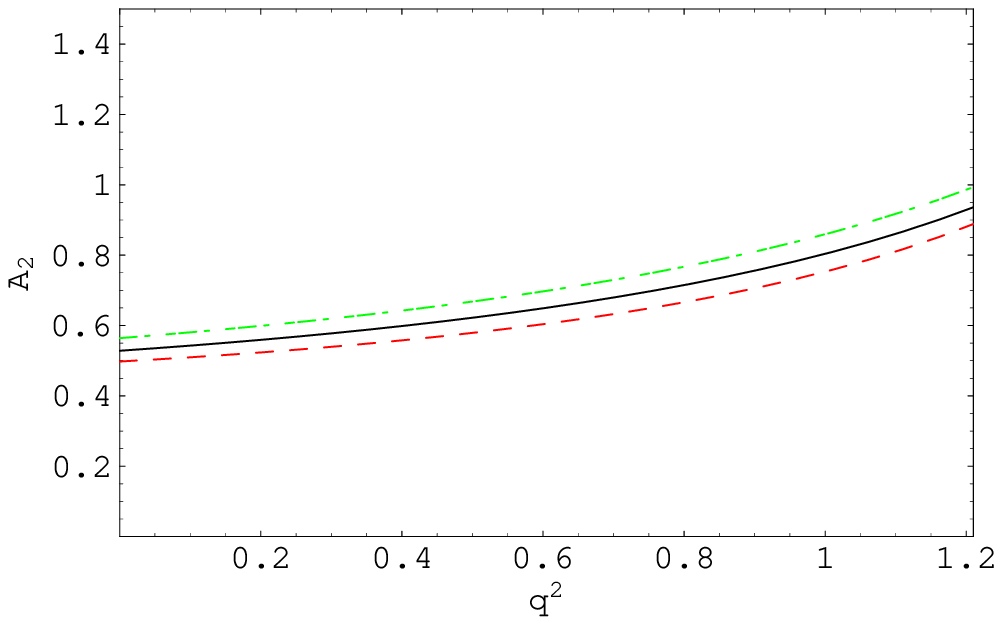}
\hspace{0.2in}
\includegraphics[width=2.2in]{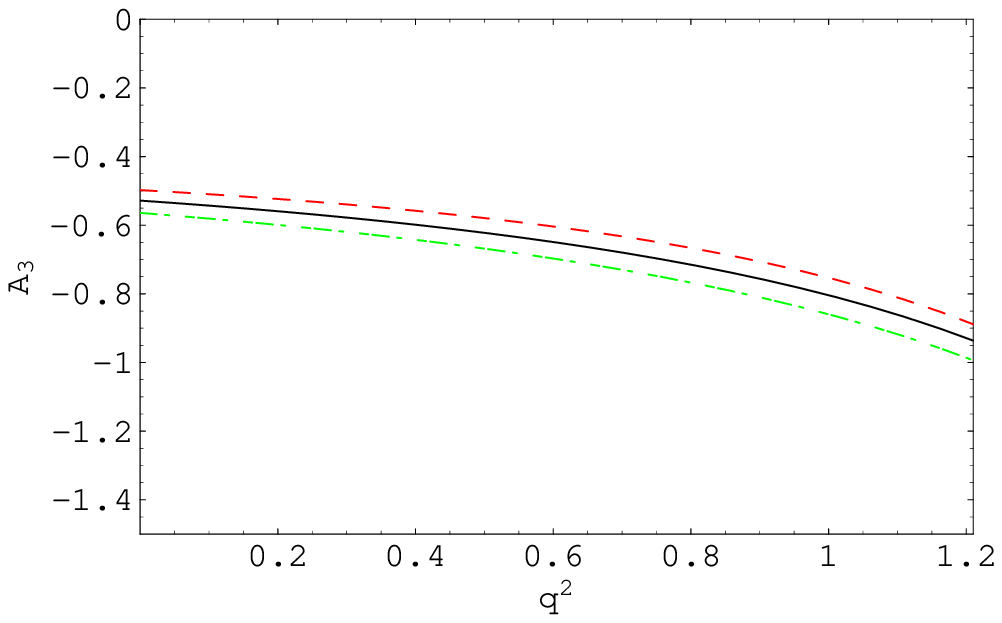}
\hspace{0.2in}
\includegraphics[width=2.2in]{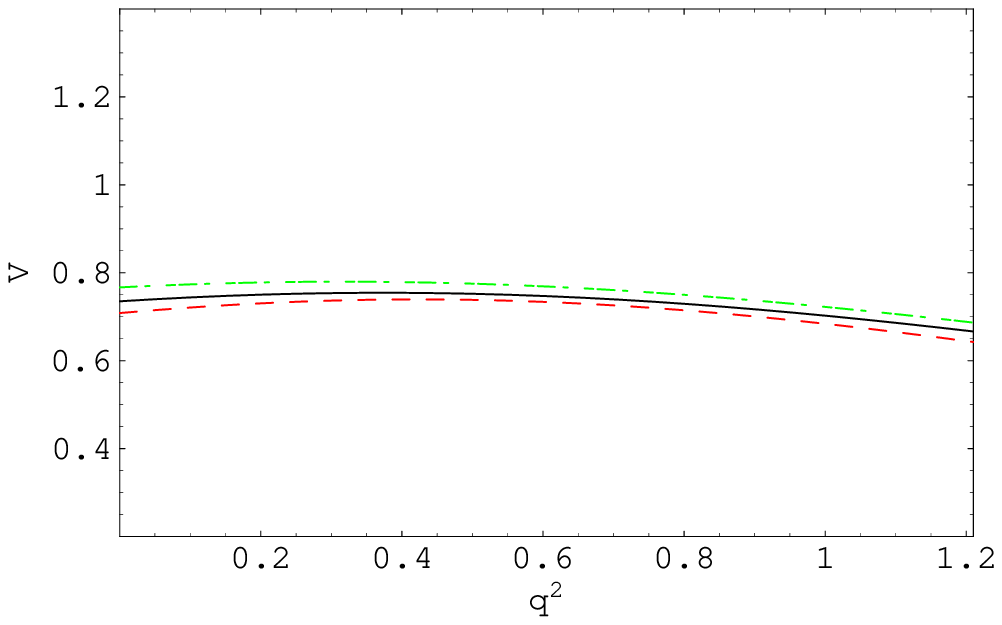}
\\
\flushleft{{\bf Fig.26}: Form factors of $D \rightarrow \rho$ decays
as functions of $q^2$ obtained with only considering the leading
twist meson DAs. The dashed, solid and dot dashed lines correspond
to $s_0 = 1.9$Gev, $T=1.7$Gev; $s_0=2.0$Gev, $T=1.5$Gev and
$s_0=2.1$GeV, $T=1.3$Gev respectively, which reflect the possible
large uncertainties.}
\end{figure}

\clearpage

\begin{figure}[t]
\centering
\includegraphics[width=2.2in]{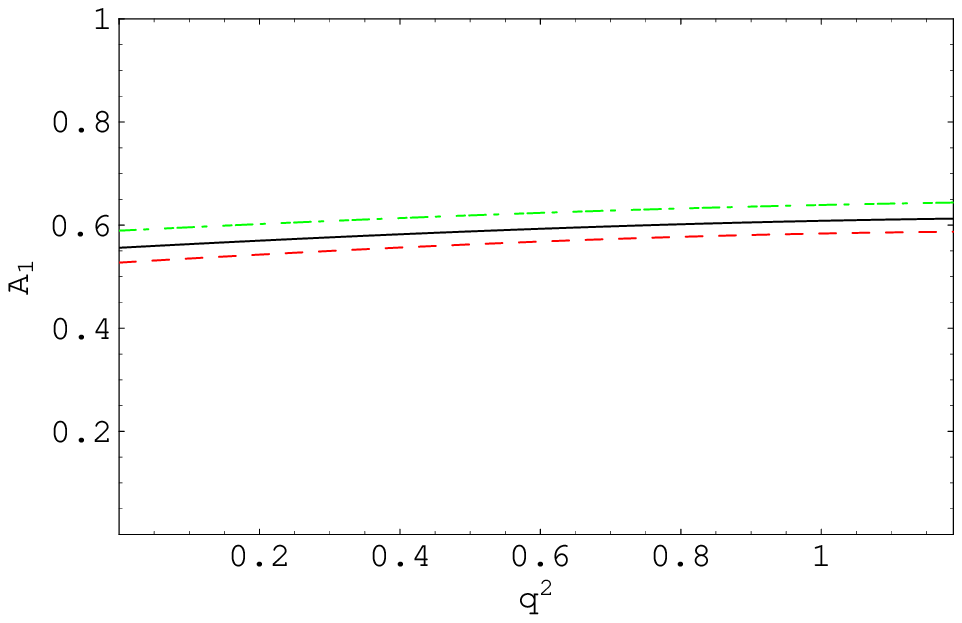}%
\hspace{0.2in}
\includegraphics[width=2.2in]{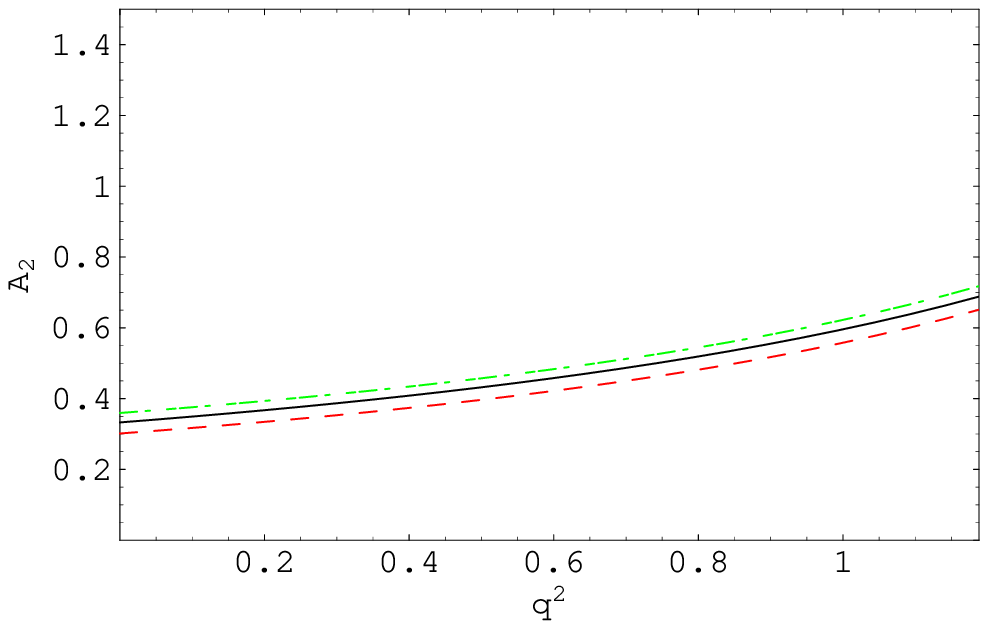}
\hspace{0.2in}
\includegraphics[width=2.2in]{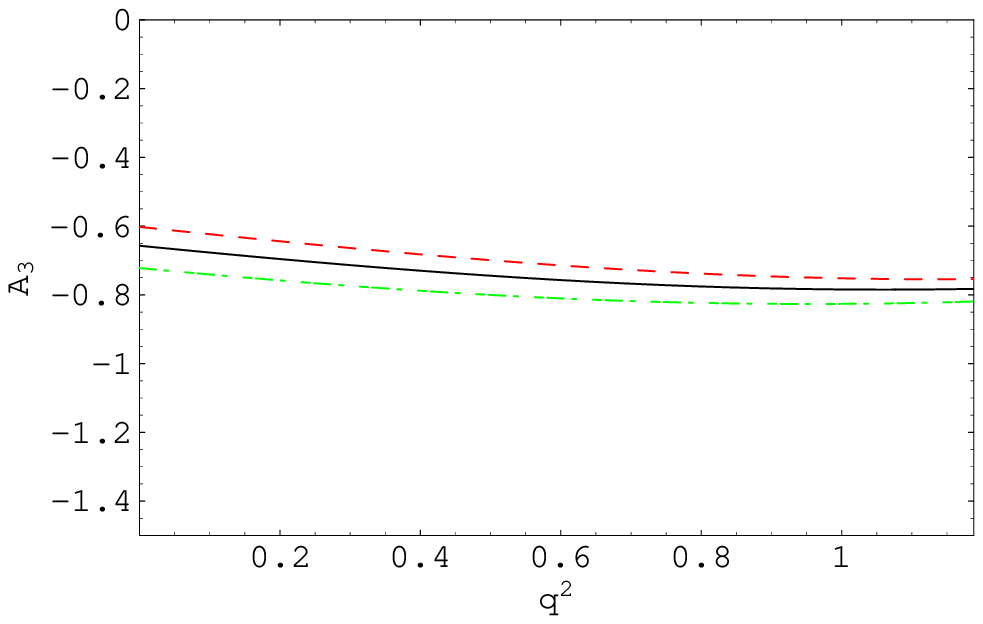}
\hspace{0.2in}
\includegraphics[width=2.2in]{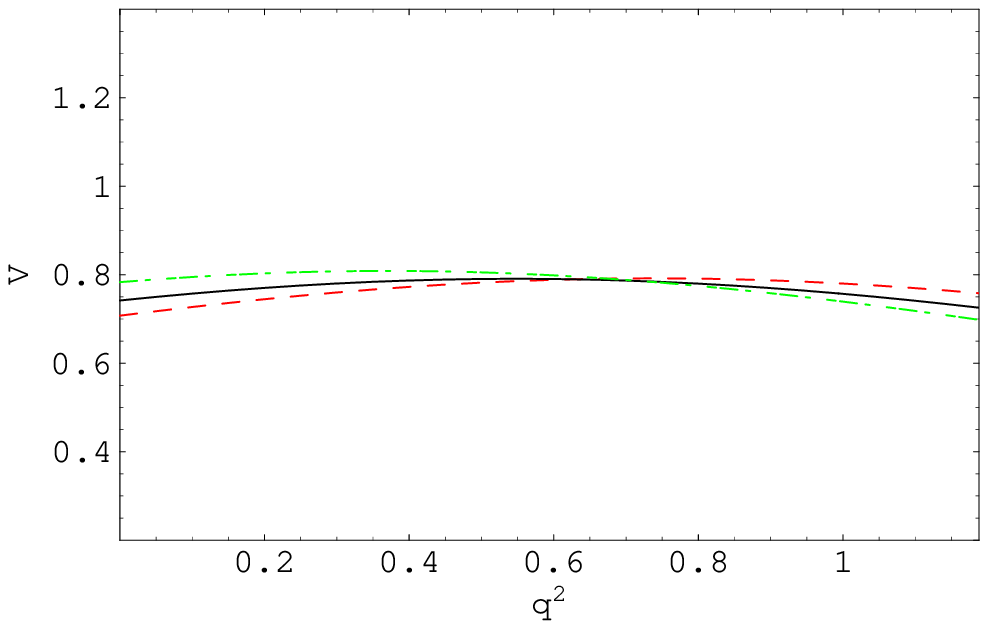}
\\
\flushleft{{\bf Fig.27}: Form factors of $D \rightarrow \omega$
decays as functions of $q^2$ obtained with considering the meson DAs
up to twist-4. The dashed, solid and dot dashed lines correspond to
$s_0 = 1.8$Gev, $T=1.4$Gev; $s_0=1.9$Gev, $T=1.2$Gev and
$s_0=2.0$GeV, $T=1.0$Gev respectively, which reflect the possible
large uncertainties.}
\end{figure}

\begin{figure}
\centering
\includegraphics[width=2.2in]{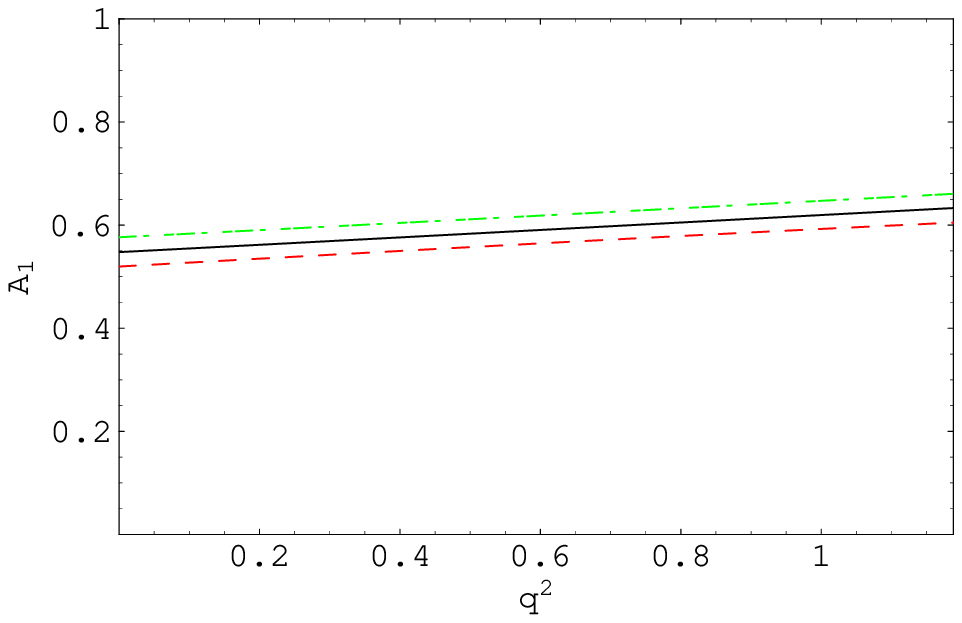}%
\hspace{0.2in}
\includegraphics[width=2.2in]{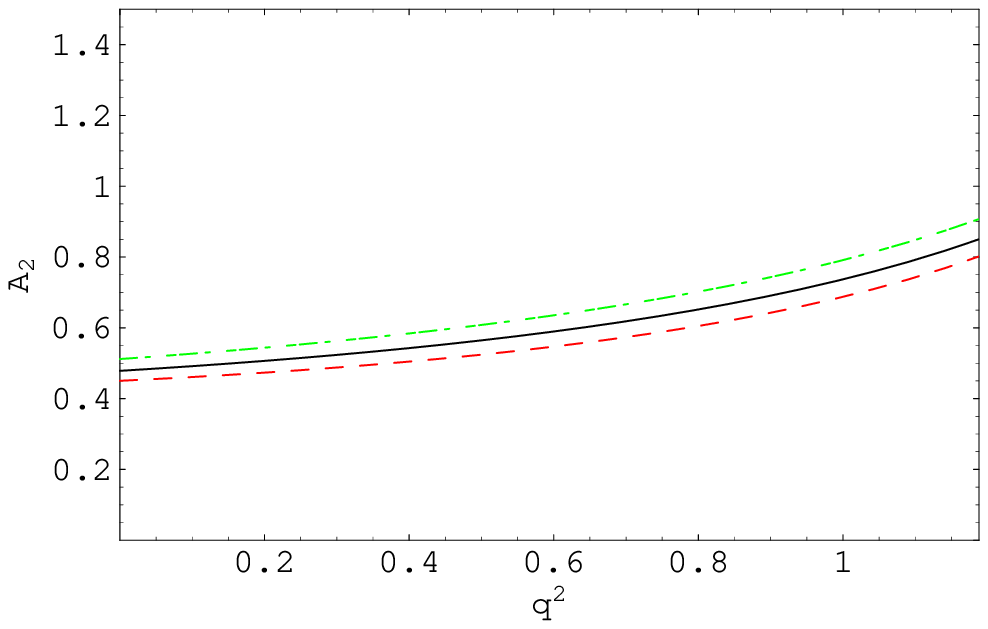}
\hspace{0.2in}
\includegraphics[width=2.2in]{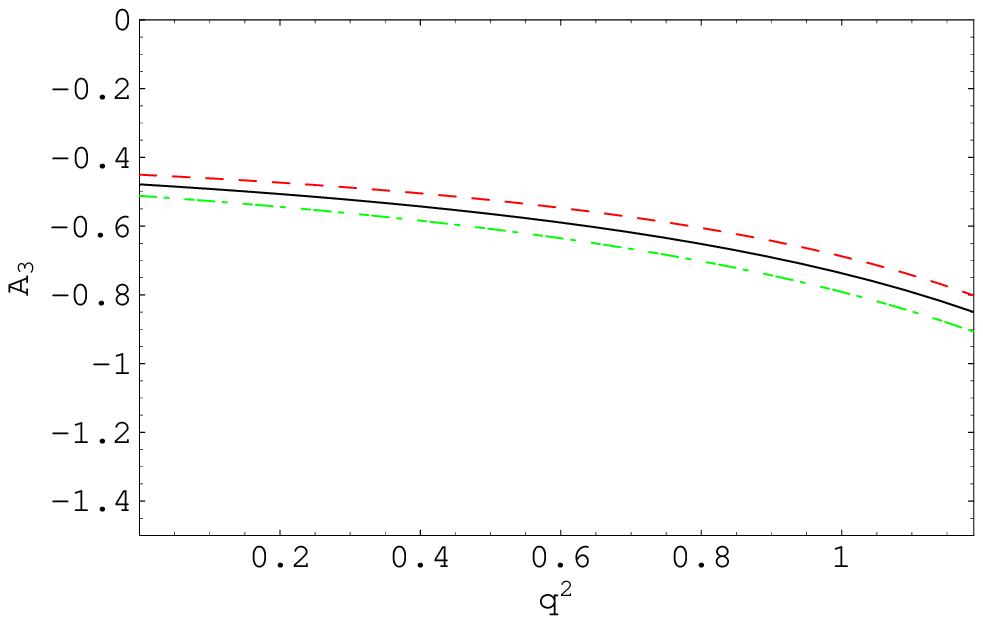}
\hspace{0.2in}
\includegraphics[width=2.2in]{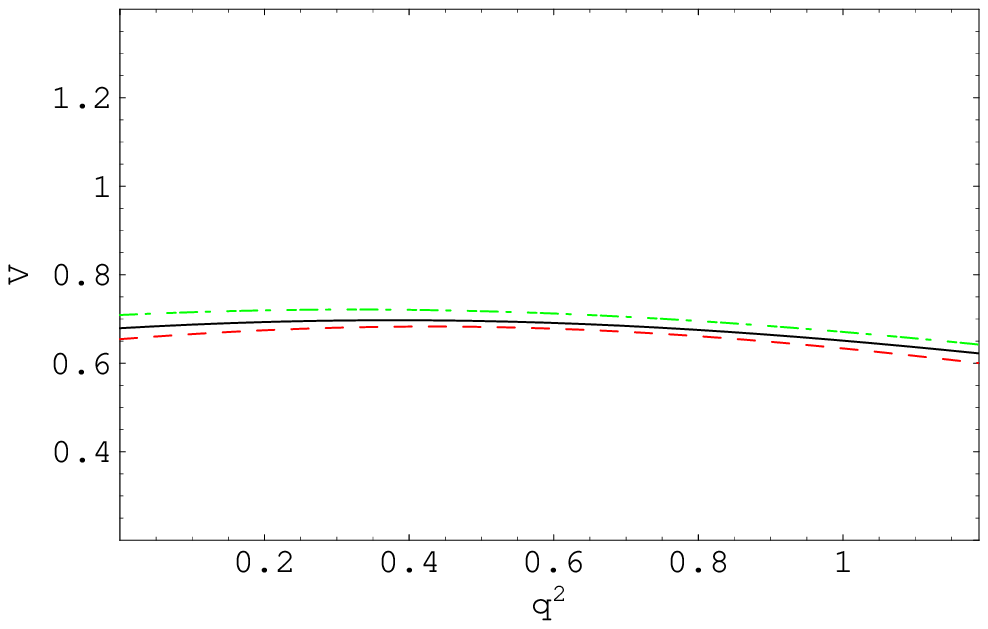}
\\
\flushleft{{\bf Fig.28}: Form factors of $D \rightarrow \omega$
decays as functions of $q^2$ obtained with only considering the
leading twist meson DAs. The dashed, solid and dot dashed lines
correspond to $s_0 = 1.9$Gev, $T=1.7$Gev; $s_0=2.0$Gev, $T=1.5$Gev
and $s_0=2.1$GeV, $T=1.3$Gev respectively, which reflect the
possible large uncertainties.}
\end{figure}

\clearpage

\begin{figure}[t]
\centering
\includegraphics[width=2.2in]{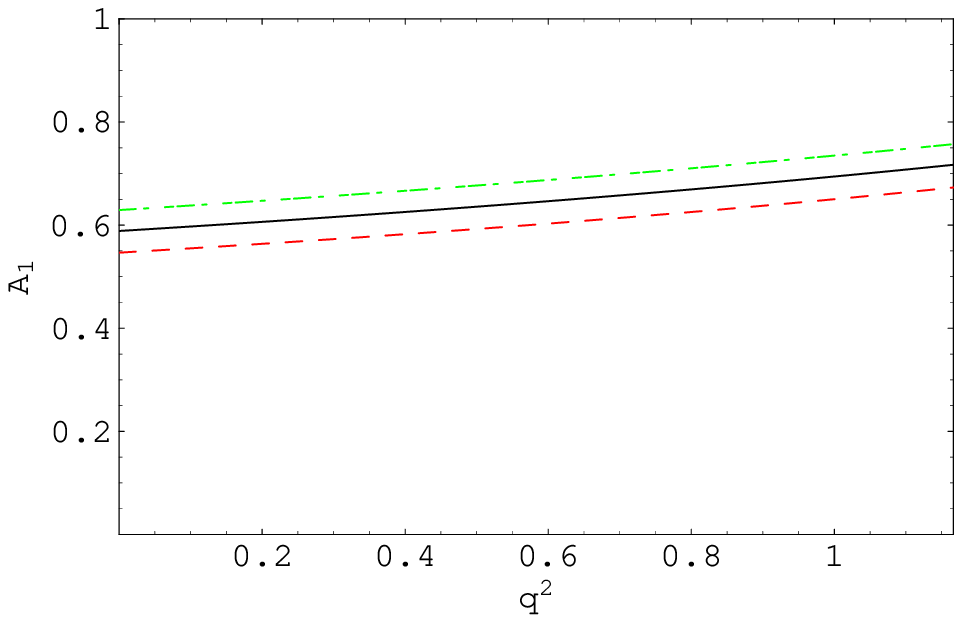}%
\hspace{0.2in}
\includegraphics[width=2.2in]{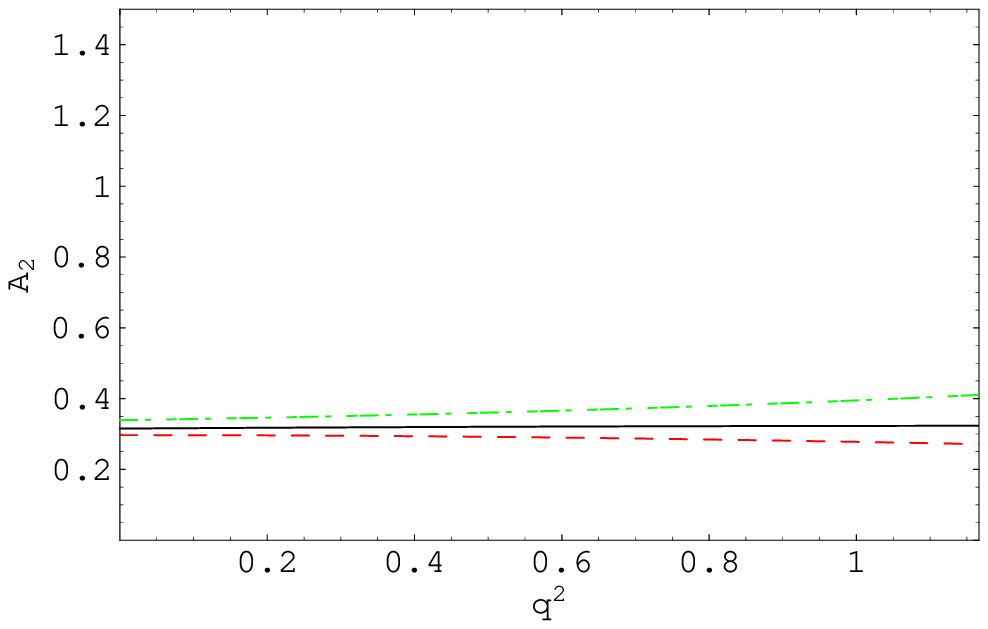}
\hspace{0.2in}
\includegraphics[width=2.2in]{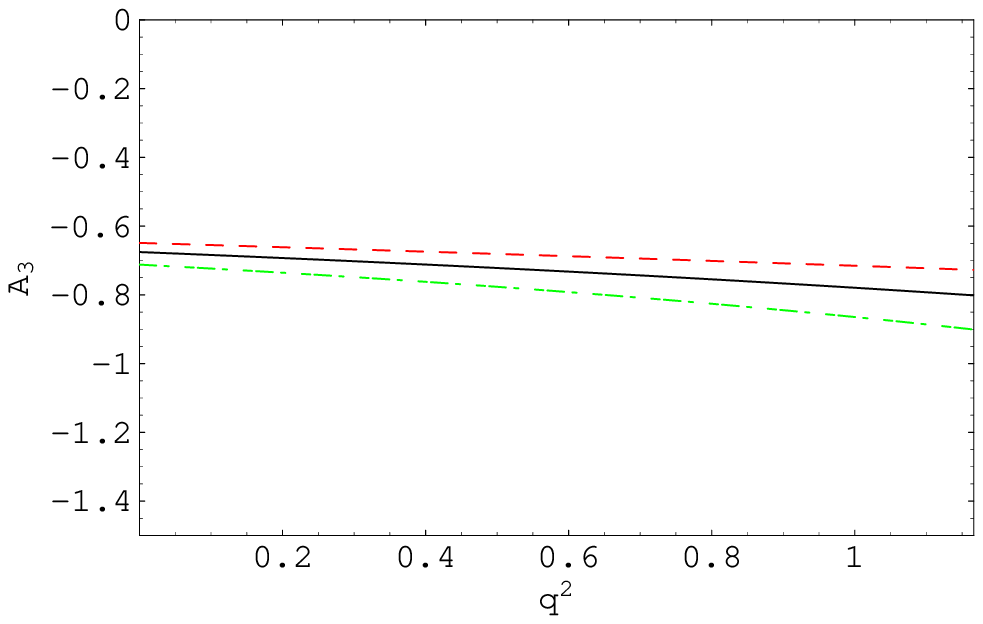}
\hspace{0.2in}
\includegraphics[width=2.2in]{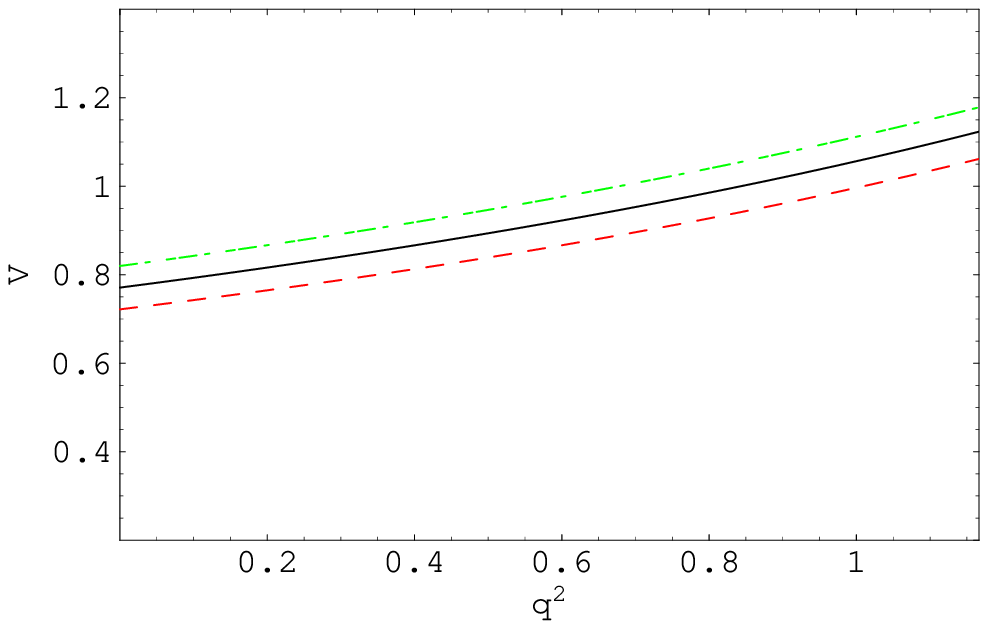}
\flushleft{{\bf Fig.29}: Form factors of $D_s \rightarrow K^*$
decays as functions of $q^2$ obtained with considering the meson DAs
up to twist-4. The dashed, solid and dot dashed lines correspond to
$s_0 = 1.4$Gev, $T=1.6$Gev; $s_0=1.5$Gev, $T= 1.5$Gev and
$s_0=1.6$Gev, $T=1.4$GeV respectively, which reflect the possible
large uncertainties. }
\end{figure}

\begin{figure}
\centering
\includegraphics[width=2.2in]{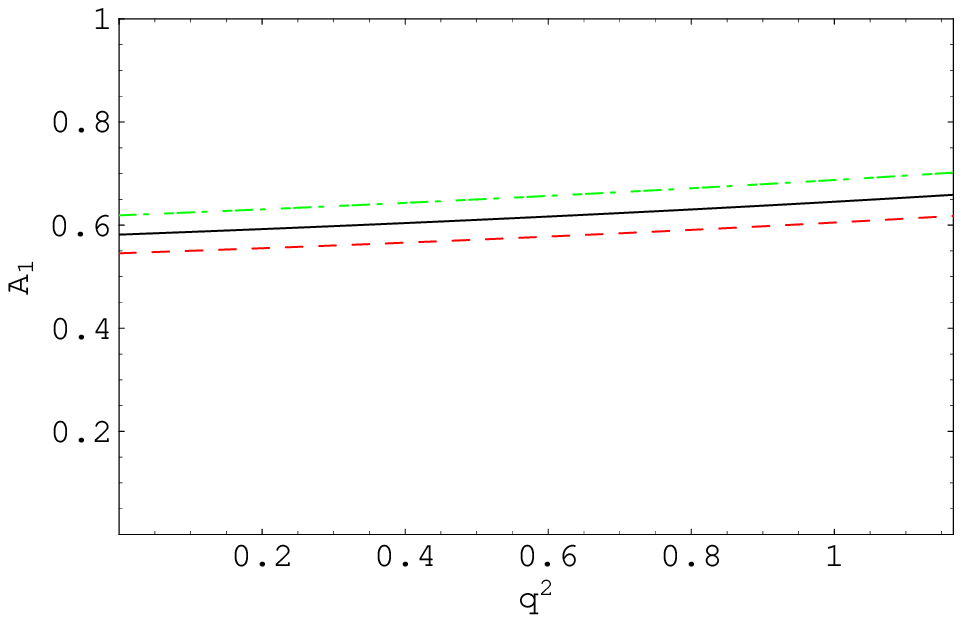}%
\hspace{0.2in}
\includegraphics[width=2.2in]{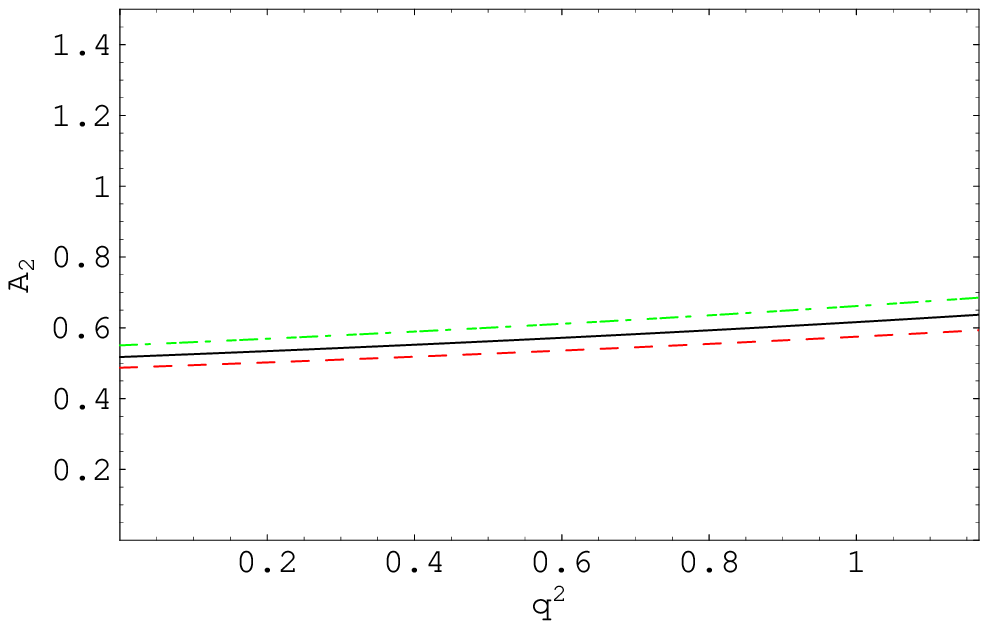}
\hspace{0.2in}
\includegraphics[width=2.2in]{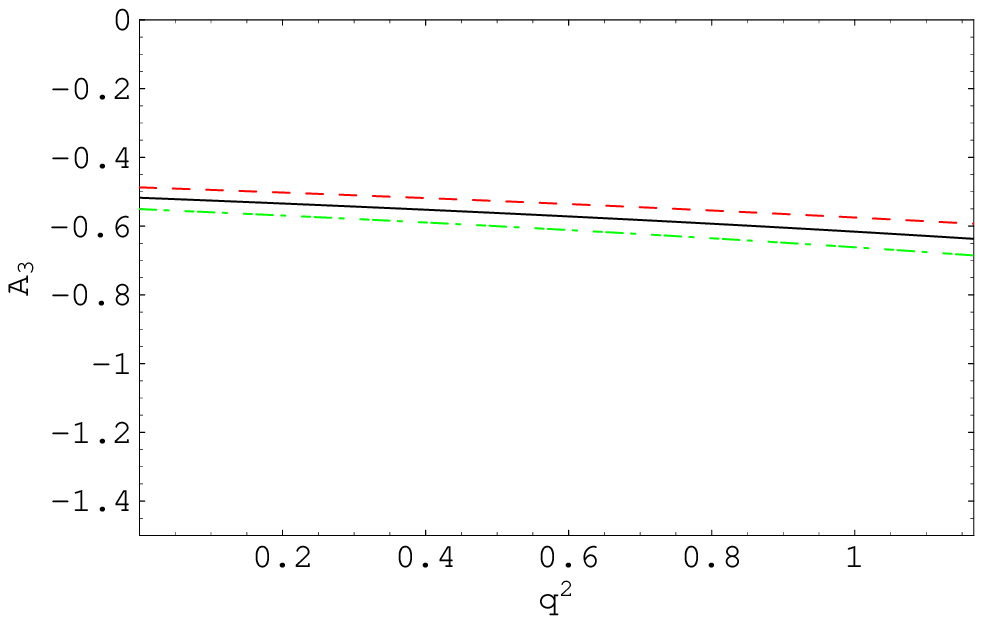}
\hspace{0.2in}
\includegraphics[width=2.2in]{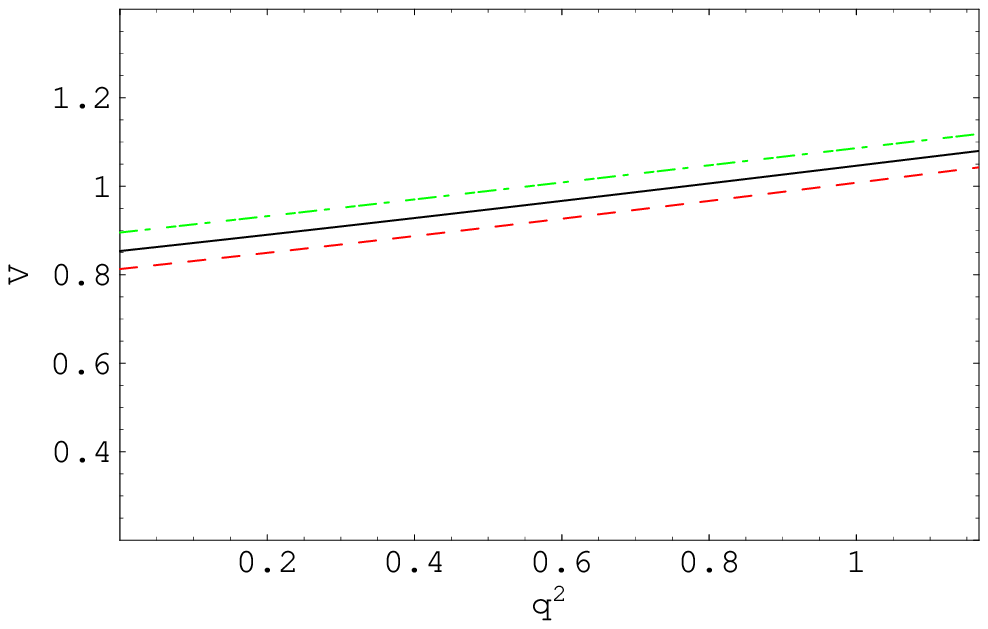}
\flushleft{{\bf Fig.30}: Form factors of $D_s \rightarrow K^*$
decays as functions of $q^2$ obtained with only considering the
leading twist meson DAs. The dashed, solid and dot dashed lines
correspond to $s_0 = 1.4$Gev, $T=1.6$Gev; $s_0=1.5$Gev, $T= 1.5$Gev
and $s_0=1.6$Gev, $T=1.4$GeV respectively, which reflect the
possible large uncertainties. }
\end{figure}

\clearpage

\begin{figure}[t]
\centering
\includegraphics[width=2.2in]{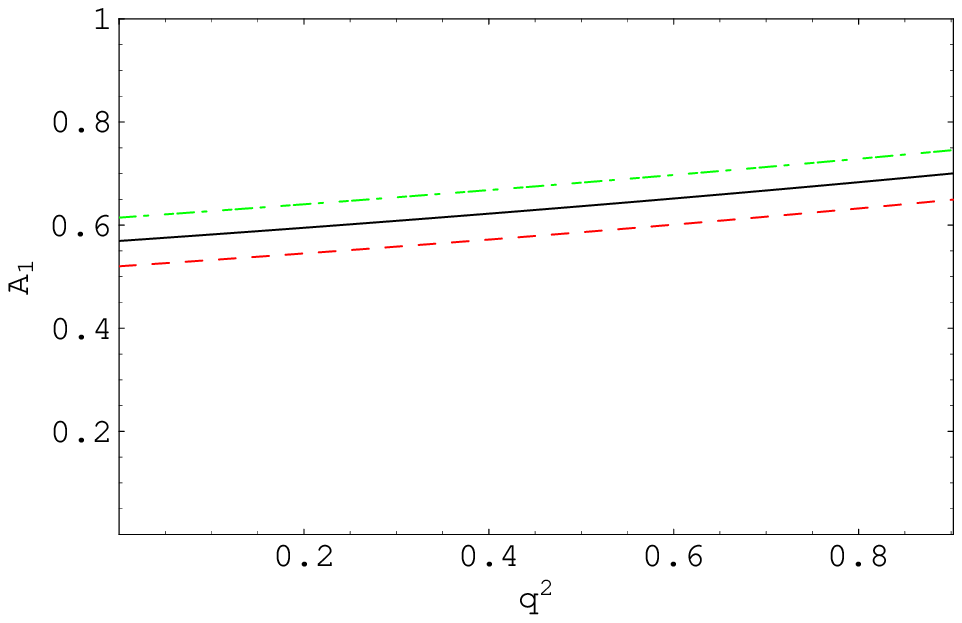}%
\hspace{0.2in}
\includegraphics[width=2.2in]{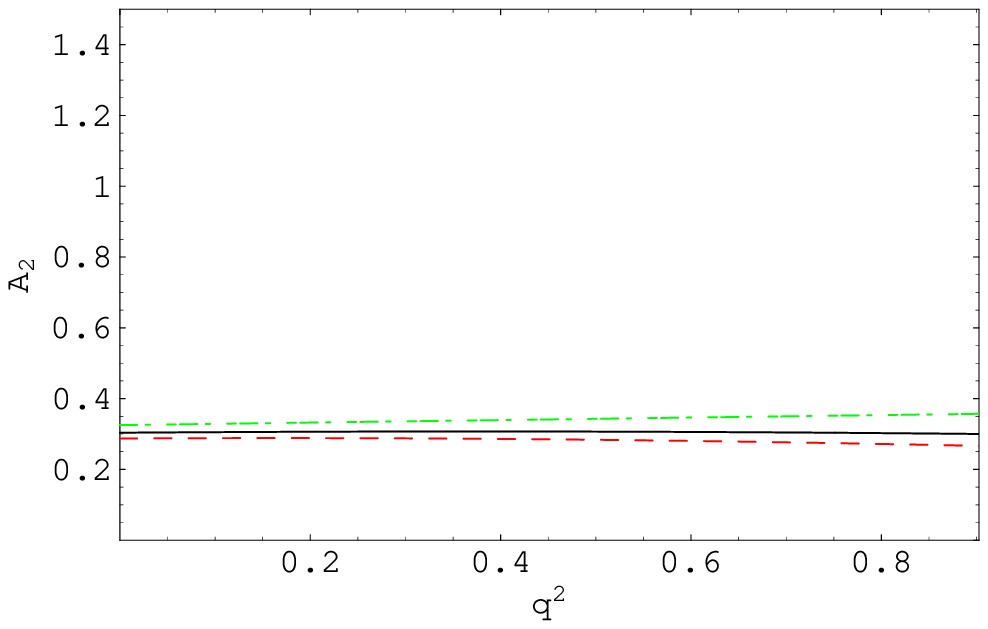}
\hspace{0.2in}
\includegraphics[width=2.2in]{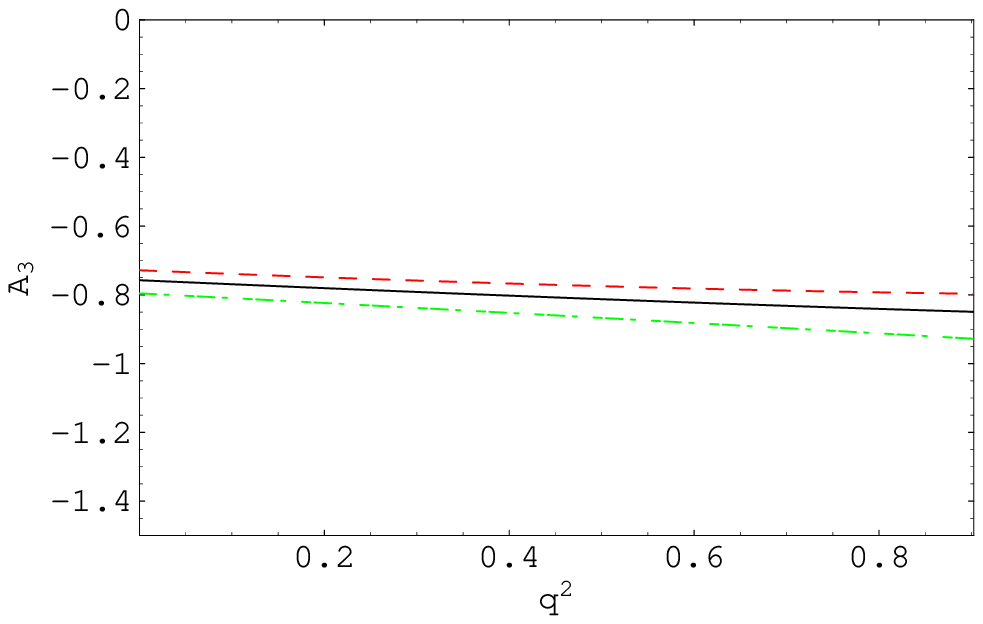}
\hspace{0.2in}
\includegraphics[width=2.2in]{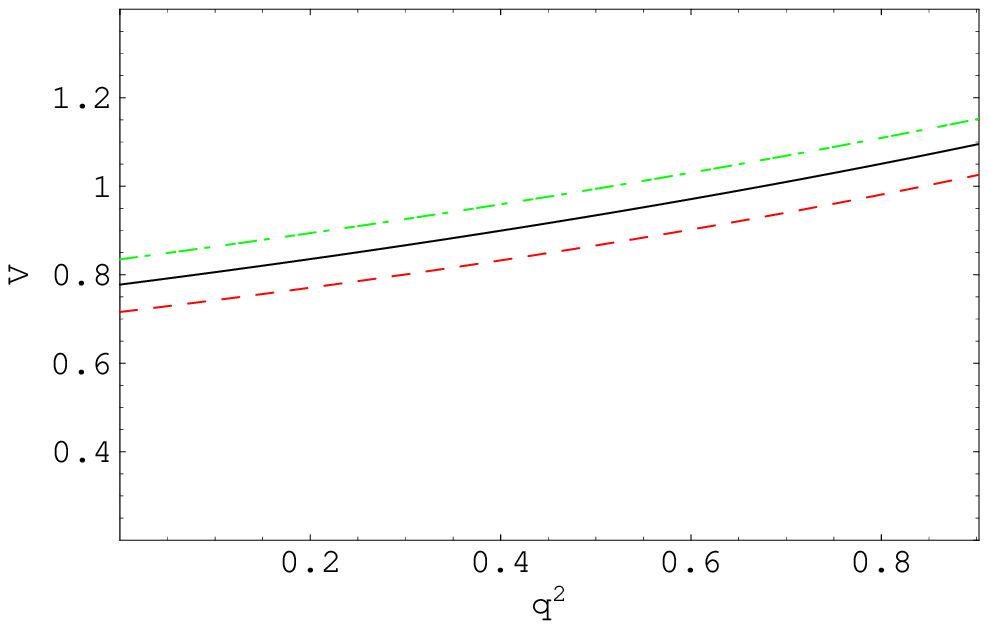}
\\
\flushleft{{\bf Fig.31}: Form factors of $D_s \rightarrow \phi$
decays as functions of $q^2$ obtained with considering the meson DAs
up to twist-4. The dashed, solid and dot dashed lines correspond to
$s_0 = 1.4$Gev, $T=1.6$Gev; $s_0=1.5$Gev, $T= 1.5$Gev and
$s_0=1.6$Gev, $T=1.4$GeV respectively, which reflect the possible
large uncertainties. }
\end{figure}

\begin{figure}
\centering
\includegraphics[width=2.2in]{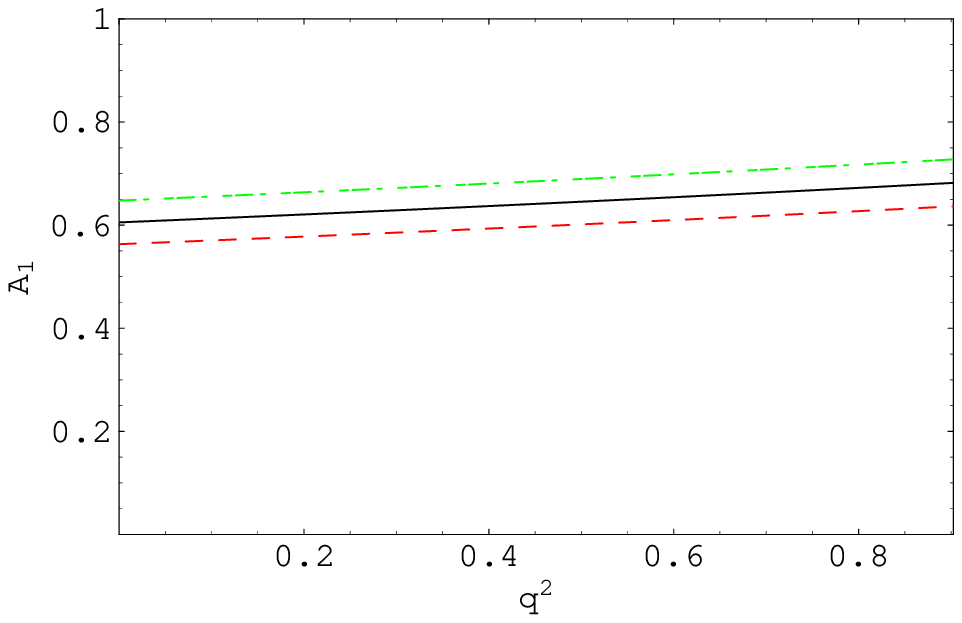}%
\hspace{0.2in}
\includegraphics[width=2.2in]{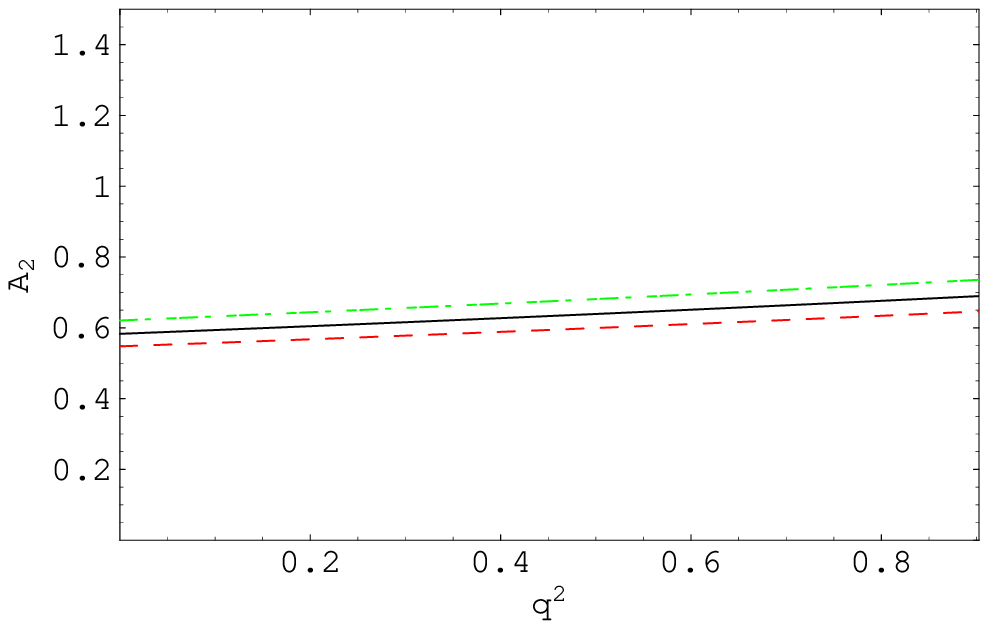}
\hspace{0.2in}
\includegraphics[width=2.2in]{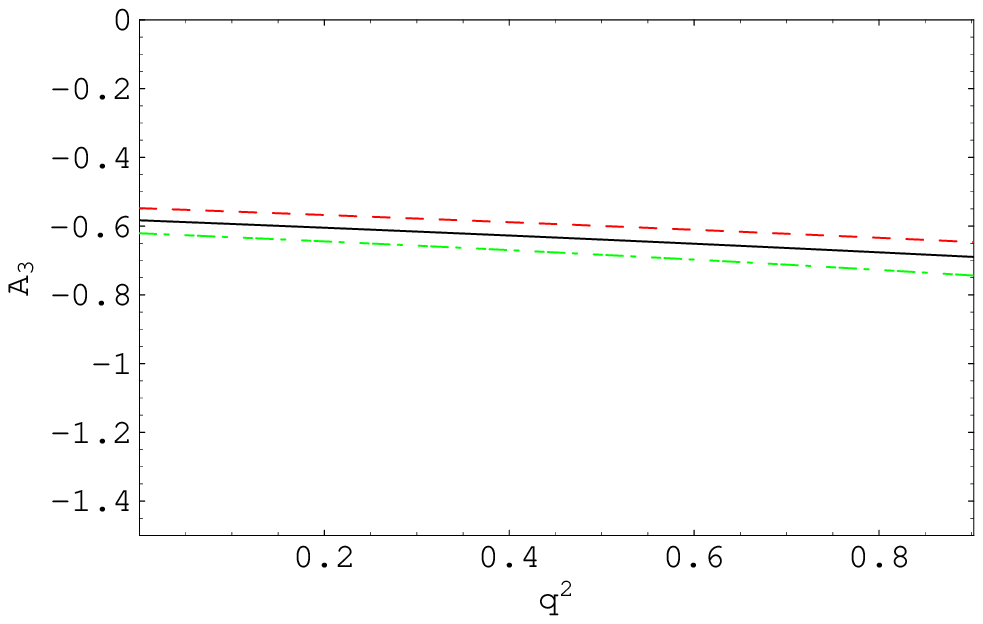}
\hspace{0.2in}
\includegraphics[width=2.2in]{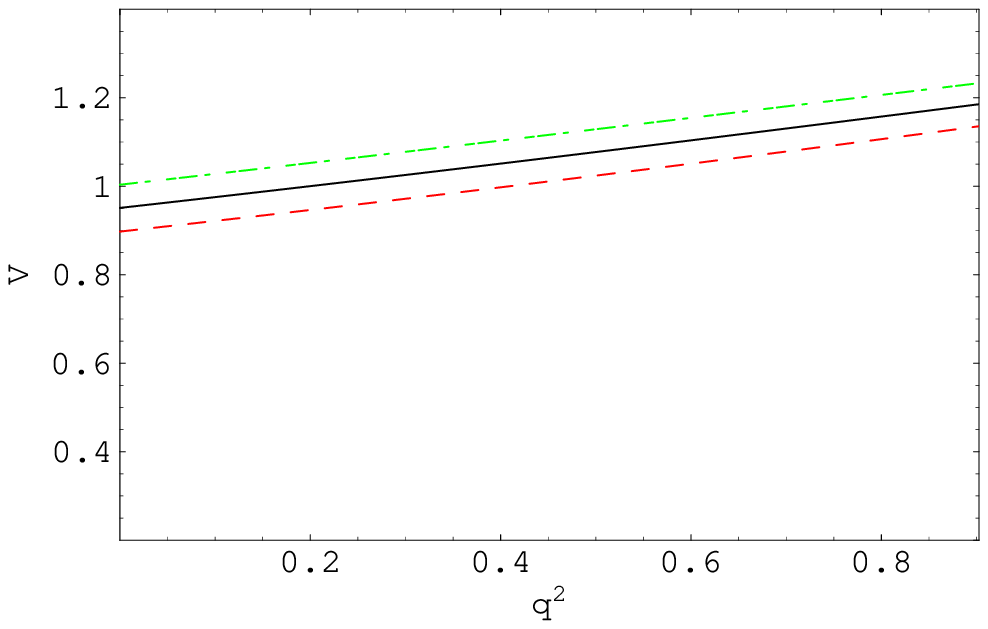}
\\
\flushleft{{\bf Fig.32}: Form factors of $D_s \rightarrow \phi$
decays as functions of $q^2$ obtained with only considering the
leading twist meson DAs. The dashed, solid and dot dashed lines
correspond to $s_0 = 1.4$Gev, $T=1.6$Gev; $s_0=1.5$Gev, $T= 1.5$Gev
and $s_0=1.6$Gev, $T=1.4$GeV respectively, which reflect the
possible large uncertainties. }
\end{figure}

\clearpage

\begin{table}
\centering \tabcolsep0.25in
\begin{tabular}{|c|c|c|c|c|}\hline\hline
& $\pi$ & $K$ & $\bar{K}$ & $\eta$ \\
\hline
$m_P (GeV)$ & 0.14 & 0.49 & 0.49 & 0.55 \\
\hline
$f_P (GeV)$ & 0.132 &  0.16 & 0.16 & 0.130 \\
\hline
$\rho_P^2$ &  0.0048 & 0.059 & 0.059 & 0.074 \\
\cline{2-5}
& 0.0063 & 0.078 & 0.078 & 0.098 \\
\hline
$a_1$ & 0 & 0.15 & -0.15 &  0 \\
\cline{2-5}
& 0 & 0.17 & -0.17 & 0 \\
\hline
$a_2$ & 0.35 & 0.16 & 0.16 & 0.16 \\
\cline{2-5}
& 0.41 & 0.21 & 0.21 & 0.21 \\
\hline
$a_3$ & 0 & 0.05 & -0.05 & 0 \\
\cline{2-5}
& 0 & 0.07 & -0.07 & 0 \\
\hline
$a_4$ & 0.18 & 0.06 & 0.06 & 0 \\
\cline{2-5}
& 0.23 & 0.08 & 0.08 & 0 \\
\hline\hline
\end{tabular}
\\
\flushleft{{\bf Tab.1}: Parameters relevant to specific pseudoscalar
mesons. For every $a_i$, the upper row corresponds to $\mu = \mu_b$
and the lower one to $\mu = \mu_c$.}
\end{table}

\begin{table}
\centering \tabcolsep0.25in
\begin{tabular}{|c|c|c|c|c|c|}\hline\hline
& $\rho$ & $K^*$ & $\bar{K}^*$ & $\omega$ & $\phi$ \\
\hline
$m_V(GeV)$ & 0.77 & 0.89 & 0.89 & 0.78 & 1.02 \\
\hline
$f_V(GeV)$ & 0.205 & 0.217 & 0.217 & 0.195 & 0.231 \\
\hline
$f_V^T(GeV) $ & 0.147 & 0.147 & 0.156 & 0.133 & 0.183 \\
\cline{2-6}
& 0.160 & 0.170 & 0.170 & 0.145 & 0.200 \\
\hline
$a^\parallel_1$ & 0 & 0.09 & -0.09 & 0 & 0 \\
\cline{2-6}
& 0 & 0.10 & -0.10 & 0 & 0 \\
\hline
$a^\parallel_2$ & 0.16 & 0.07 & 0.07 & 0.16 & 0 \\
\cline{2-6}
& 0.18 & 0.09 & 0.09 & 0.18 & 0 \\
\hline
$a^\perp_1$ & 0 & 0.09 & -0.09 & 0 & 0 \\
\cline{2-6}
& 0 & 0.10 & -0.10 & 0 & 0 \\
\hline
$a^\perp_2$ & 0.17 & 0.11 & 0.11 & 0.17 & 0 \\
\cline{2-6}
& 0.20 & 0.13 & 0.13 & 0.20 & 0 \\
\hline
$\delta_+$ & 0 & 0.22 & 0.22 & 0 & 0.41 \\
\cline{2-6}
& 0 & 0.24 & 0.24 & 0 & 0.46 \\
\hline
$ \delta_-$ & 0 & -0.22 & 0.22 & 0 & 0 \\
\cline{2-6}
& 0 & -0.24 & 0.24 & 0 & 0 \\
\hline
$\tilde{\delta}_+$ & 0 & 0.13 & 0.13 & 0 & 0.27 \\
\cline{2-6}
& 0 & 0.16 & 0.16 & 0 & 0.33 \\
\hline $\tilde{\delta}_-$ & 0 & -0.13 & 0.13 & 0 & 0 \\
\cline{2-6} & 0 & -0.16 & 0.16 & 0 & 0 \\
\hline\hline
\end{tabular}
\\
\flushleft{{\bf Tab.2}: Parameters relevant to specific vector
mesons. For every $a^{\parallel(\perp)}_i$, the upper row
corresponds to $\mu = \mu_b$ and the lower one to $\mu = \mu_c$.}
\end{table}

\begin{table}
\centering \tabcolsep0.10in
\begin{tabular}{|c|c|c|c|c|c|c|c|c|}\hline\hline
$C_1$ & $C_2$ & $C_3$ & $C_4$ & $C_5$ & $C_6$ & $C^{eff}_7$ & $C_9$ & $C_{10}$ \\
\hline
-0.360 & 1.169 & 0.017 & -0.036 & 0.010 & -0.048 & -0.355 & 4.42 & -4.398 \\
\hline\hline
\end{tabular}
\flushleft{{\bf Tab.3}: The SM Wilson coefficients used in the
numerical calculations of branching ratios for $B_{(s)}$ decays. }
\end{table}

\begin{table}
\centering \tabcolsep0.25in
\begin{tabular}{|c|c|c|c|c|}\hline\hline
Decay & \multicolumn{2}{|c|}{$F(0)$} & $a_F$ & $b_F$ \\
\hline
$B \rightarrow \pi$ & $f_+$ & $0.285^{+0.016}_{-0.015}$ & $1.31^{+0.02}_{-0.03}$ & $0.30^{+0.03}_{-0.03}$ \\
\cline{2-5}
& $f_0$ & $0.285^{+0.016}_{-0.015}$ & $0.33^{+0.02}_{-0.02}$ & $-0.61^{+0.02}_{-0.02}$ \\
\hline
$B \rightarrow K$ & $f_+$ & $0.345^{+0.021}_{-0.021}$ & $1.40^{-0.01}_{+0.01}$ & $0.36^{-0.01}_{+0.01}$ \\
\cline{2-5}
& $f_0$ & $0.345^{+0.021}_{-0.021}$ & $0.47^{+0.01}_{-0.01}$ & $-0.38^{+0.02}_{-0.04}$ \\
\hline
$B \rightarrow \eta$ & $f_+$ & $0.247^{+0.016}_{-0.016}$ & $1.52^{-0.02}_{+0.02}$ & $0.54^{-0.01}_{+0.02}$ \\
\cline{2-5}
& $f_0$ & $0.247^{+0.016}_{-0.016}$ & $0.63^{-0.01}_{+0.02}$ & $-0.21^{-0.01}_{+0.02}$ \\
\hline
$B_s \rightarrow K$ & $f_+$ & $0.296^{+0.018}_{-0.018}$ & $1.65^{-0.01}_{+0.01}$ & $0.75^{+0.01}_{-0.01}$ \\
\cline{2-5}
& $f_0$ & $0.296^{+0.018}_{-0.018}$ & $0.83^{-0.01}_{+0.01}$ & $-0.07^{-0.01}_{+0.01}$ \\
\hline
$B_s \rightarrow \eta$ & $f_+$ & $0.281^{+0.015}_{-0.016}$ & $1.48^{-0.01}_{+0.02}$ & $0.55^{-0.01}_{+0.01}$ \\
\cline{2-5}
& $f_0$ &  $0.281^{+0.015}_{-0.016}$ & $0.61^{-0.02}_{+0.02}$ & $-0.20^{-0.02}_{+0.04}$ \\
\hline\hline
\end{tabular}
\\
\flushleft{{\bf Tab.4}: Results of the form factors for $ B_{(s)}
\rightarrow P $ meson decays with $T = 2.0 \mp 0.2$Gev. For $B
\rightarrow P$, we choose $s_0 = 1.6 \pm 0.1 $Gev and for $B_s
\rightarrow P$, $s_0 = 1.9 \pm 0.1 $Gev. }
\end{table}

\begin{table}
\centering \tabcolsep0.25in
\begin{tabular}{|c|c|c|c|c|}\hline\hline
Decay & \multicolumn{2}{|c|}{$F(0)$} & $a_F$ & $b_F$ \\
\hline
$B \rightarrow K^*$ & $A_1$ & $0.266^{+0.016}_{-0.017}$ & $0.50^{+0.01}_{-0.01}$ & $-0.23^{+0.05}_{-0.05}$ \\
\cline{2-5}
& $A_2$ & $0.212^{+0.013}_{-0.013}$ & $1.00^{+0.01}_{-0.01}$ & $0.05^{+0.03}_{-0.04}$ \\
\cline{2-5}
& $A_3$ & $-0.256^{+0.016}_{-0.015}$ & $1.20^{+0.01}_{-0.01}$ & $0.20^{+0.01}_{-0.01}$ \\
\cline{2-5}
& $V$ & $0.331^{+0.019}_{-0.019}$ & $1.36^{-0.01}_{+0.02}$ & $0.42^{-0.01}_{+0.01}$ \\
\hline
$B \rightarrow \rho$ & $A_1$ & $0.232^{+0.013}_{-0.014}$ & $0.42^{+0.02}_{-0.01}$ & $-0.25^{+0.05}_{-0.03}$ \\
\cline{2-5}
& $A_2$ & $0.187^{+0.011}_{-0.012}$ & $0.98^{+0.01}_{-0.01}$ & $-0.03^{+0.02}_{-0.03}$ \\
\cline{2-5}
& $A_3$ & $-0.221^{+0.013}_{-0.013}$ & $1.16^{+0.01}_{-0.01}$ & $0.09^{+0.01}_{-0.01}$ \\
\cline{2-5}
& $V$ & $0.289^{+0.016}_{-0.016}$ & $1.32^{-0.01}_{+0.01}$ & $0.34^{-0.01}_{+0.01}$ \\
\hline
$B \rightarrow \omega$ & $A_1$ & $0.214^{+0.013}_{-0.012}$ & $0.43^{+0.01}_{-0.02}$ & $-0.26^{+0.05}_{-0.05}$ \\
\cline{2-5}
& $A_2$ & $0.170^{+0.010}_{-0.011}$ & $0.96^{+0.01}_{-0.01}$ & $-0.05^{+0.03}_{-0.04}$ \\
\cline{2-5}
& $A_3$ & $-0.202^{+0.012}_{-0.012}$ & $1.15^{+0.01}_{-0.01}$ & $0.07^{+0.01}_{-0.01}$ \\
\cline{2-5}
& $V$ & $0.268^{+0.014}_{-0.015}$ & $1.31^{-0.01}_{+0.01}$ & $0.31^{-0.01}_{+0.01}$ \\
\hline
$B_s \rightarrow K^*$ & $A_1$ & $0.227^{+0.010}_{-0.012}$ & $0.68^{+0.01}_{-0.01}$ & $-0.23^{+0.03}_{-0.02}$ \\
\cline{2-5}
& $A_2$ & $0.183^{+0.008}_{-0.010}$ & $1.20^{-0.01}_{+0.01}$ & $0.06^{-0.01}_{+0.01}$ \\
\cline{2-5}
& $A_3$ & $-0.222^{+0.011}_{-0.010}$ & $1.37^{+0.01}_{-0.01}$ & $0.23^{+0.01}_{-0.01}$ \\
\cline{2-5}
& $V$ & $0.285^{+0.013}_{-0.013}$ & $1.50^{-0.01}_{+0.01}$ & $0.46^{-0.01}_{+0.01}$ \\
\hline
$B_s \rightarrow \phi$ & $A_1$ & $0.271^{+0.014}_{-0.014}$ & $0.86^{-0.01}_{+0.02}$ & $-0.01^{-0.01}_{+0.02}$ \\
\cline{2-5}
& $A_2$ &  $0.212^{+0.011}_{-0.012}$ & $1.40^{-0.01}_{+0.01}$ & $0.33^{+0.01}_{-0.01}$ \\
\cline{2-5}
& $A_3$ & $-0.269^{+0.014}_{-0.014}$ & $1.56^{+0.02}_{-0.02}$ & $0.53^{+0.02}_{-0.03}$ \\
\cline{2-5}
& $V$ & $0.339^{+0.016}_{-0.017}$ & $1.63^{-0.02}_{+0.01}$ & $0.69^{-0.02}_{+0.01}$ \\
\hline\hline
\end{tabular}
\\
\flushleft{{\bf Tab.5a}: Results of the form factors for $ B_{(s)}
\rightarrow V $ meson decays obtained with considering the meson DAs
up to twist-4. For $B \rightarrow V$,
 we choose $s_0 = 1.7 \pm 0.1 $Gev, $T = 1.8 \mp 0.2 $Gev and for $B_s \rightarrow V$, $s_0 = 2.1 \pm 0.1 $Gev,
  $T = 2.0 \mp 0.2 $ Gev. }
\end{table}

\begin{table}
\centering \tabcolsep0.25in
\begin{tabular}{|c|c|c|c|c|}\hline\hline
Decay & \multicolumn{2}{|c|}{$F(0)$} & $a_F$ & $b_F$ \\
\hline
$B \rightarrow K^*$ & $A_1$ & $0.274^{+0.015}_{-0.016}$ & $0.52^{-0.01}_{+0.02}$ & $-0.03^{-0.01}_{+0.01}$ \\
\cline{2-5}
& $A_2$ & $0.268^{+0.014}_{-0.015}$ & $1.18^{-0.01}_{+0.01}$ & $0.45^{-0.01}_{+0.01}$ \\
\cline{2-5}
& $A_3$ & $-0.268^{+0.015}_{-0.014}$ & $1.18^{+0.01}_{-0.01}$ & $0.45^{+0.01}_{-0.01}$ \\
\cline{2-5}
& $V$ & $0.351^{+0.019}_{-0.019}$ & $1.46^{-0.02}_{+0.04}$ & $0.62^{-0.03}_{+0.05}$ \\
\hline
$B \rightarrow \rho$ & $A_1$ & $0.239^{+0.013}_{-0.014}$ & $0.46^{-0.01}_{+0.01}$ & $-0.09^{-0.01}_{+0.01}$ \\
\cline{2-5}
& $A_2$ & $0.230^{+0.013}_{-0.012}$ & $1.17^{-0.01}_{+0.01}$ & $0.35^{-0.01}_{+0.01}$ \\
\cline{2-5}
& $A_3$ & $-0.230^{+0.012}_{-0.013}$ & $1.17^{+0.01}_{-0.01}$ & $0.35^{+0.01}_{-0.01}$ \\
\cline{2-5}
& $V$ & $0.297^{+0.015}_{-0.016}$ & $1.40^{-0.02}_{+0.02}$ & $0.51^{-0.02}_{+0.03}$ \\
\hline
$B \rightarrow \omega$ & $A_1$ & $0.221^{+0.012}_{-0.013}$ & $0.46^{-0.01}_{+0.02}$ & $-0.08^{-0.01}_{+0.01}$ \\
\cline{2-5}
& $A_2$ & $0.211^{+0.011}_{-0.011}$ & $1.16^{-0.01}_{+0.01}$ & $0.35^{-0.01}_{+0.01}$ \\
\cline{2-5}
& $A_3$ & $-0.211^{+0.011}_{-0.011}$ & $1.16^{+0.01}_{-0.01}$ & $0.35^{+0.01}_{-0.01}$ \\
\cline{2-5}
& $V$ & $0.275^{+0.014}_{-0.015}$ & $1.40^{-0.02}_{+0.02}$ & $0.51^{-0.02}_{+0.04}$ \\
\hline
$B_s \rightarrow K^*$ & $A_1$ & $0.240^{+0.014}_{-0.014}$ & $0.58^{+0.01}_{-0.01}$ & $-0.15^{+0.03}_{-0.02}$ \\
\cline{2-5}
& $A_2$ & $0.240^{+0.013}_{-0.013}$ & $1.27^{-0.01}_{+0.01}$ & $0.46^{-0.01}_{+0.01}$ \\
\cline{2-5}
& $A_2$ & $-0.240^{+0.013}_{-0.013}$ & $1.27^{+0.01}_{-0.01}$ & $0.46^{+0.01}_{-0.01}$ \\
\cline{2-5}
& $V$ & $0.307^{+0.017}_{-0.017}$ & $1.52^{-0.01}_{+0.02}$ & $0.62^{-0.01}_{+0.01}$ \\
\hline
$B_s \rightarrow \phi$ & $A_1$ & $0.285^{+0.017}_{-0.018}$ & $0.73^{-0.01}_{+0.02}$ & $0.04^{-0.01}_{+0.01}$ \\
\cline{2-5}
& $A_2$ & $0.288^{+0.016}_{-0.016}$ & $1.37^{-0.01}_{+0.01}$ & $0.61^{-0.02}_{+0.01}$ \\
\cline{2-5}
& $A_3$ & $-0.288^{+0.016}_{-0.016}$ & $1.37^{+0.01}_{-0.01}$ & $0.61^{+0.01}_{-0.02}$ \\
\cline{2-5}
& $V$ & $0.376^{+0.021}_{-0.022}$ & $1.64^{-0.02}_{+0.02}$ & $0.81^{-0.02}_{+0.01}$ \\
\hline\hline
\end{tabular}
\flushleft{{\bf Tab.5b}: Results of the form factors for $ B_{(s)}
\rightarrow V $ meson decays obtained with only considering the
leading twist meson DAs. For $B \rightarrow V$, we choose $s_0 = 1.7
\pm 0.1 $Gev, $T = 2.0 \mp 0.2 $Gev and for $B_s \rightarrow V$,
$s_0 = 1.9 \pm 0.1 $Gev,
 $T = 2.0 \mp 0.2 $ Gev. }
\end{table}

\begin{table}
\centering \tabcolsep0.25in
\begin{tabular}{|c|c|c|c|c|}\hline\hline
Decay & \multicolumn{2}{|c|}{$F(0)$} & $a_F$ & $b_F$ \\
\hline
$D \rightarrow \pi$ & $f_+$ & $0.635^{+0.060}_{-0.057}$ & $1.01^{-0.12}_{+0.14}$ & $0.17^{-0.10}_{+0.13}$ \\
\cline{2-5}
& $f_0$ & $0.635^{+0.060}_{-0.057}$ & $0.64^{-0.01}_{+0.07}$ & $-0.20^{-0.04}_{+0.09}$ \\
\hline
$D \rightarrow K$ & $f_+$ & $0.661^{+0.067}_{-0.066}$ & $1.23^{-0.20}_{+0.22}$ & $0.69^{-0.15}_{+0.18}$ \\
\cline{2-5}
& $f_0$ & $0.661^{+0.067}_{-0.066}$ & $0.80^{-0.03}_{+0.05}$ & $-0.02^{-0.04}_{+0.06}$ \\
\hline
$D \rightarrow \eta$ & $f_+$ & $0.556^{+0.056}_{-0.053}$ & $1.25^{-0.04}_{+0.05}$ & $0.42^{-0.06}_{+0.05}$ \\
\cline{2-5}
& $f_0$ & $0.556^{+0.056}_{-0.053}$ & $0.65^{-0.01}_{+0.02}$ & $-0.22^{-0.03}_{+0.02}$ \\
\hline
$D_s \rightarrow K$ & $f_+$ & $0.820^{+0.080}_{-0.071}$ & $1.11^{-0.04}_{+0.07}$ & $0.49^{-0.05}_{+0.06}$ \\
\cline{2-5}
& $f_0$ & $0.820^{+0.080}_{-0.071}$ & $0.53^{-0.03}_{+0.04}$ & $-0.07^{-0.04}_{+0.04}$ \\
\hline
$D_s \rightarrow \eta$ & $f_+$ & $0.611^{+0.062}_{-0.054}$ & $1.20^{-0.02}_{+0.03}$ & $0.38^{-0.01}_{+0.01}$ \\
\cline{2-5}
& $f_0$ & $0.611^{+0.062}_{-0.054}$ & $0.64^{-0.01}_{+0.02}$ & $-0.18^{+0.04}_{-0.03}$ \\
\hline\hline
\end{tabular}
\flushleft{{\bf Tab.6}: Results of the form factors for $ D_{(s)}
\rightarrow P $ meson decays with $T = 1.5 \mp 0.2$Gev. For $D
\rightarrow P$, we choose $s_0 = 1.2 \pm 0.1 $Gev and for $D_s
\rightarrow P$, $s_0 = 1.4 \pm 0.1 $Gev. }
\end{table}

\begin{table}
\centering \tabcolsep0.25in
\begin{tabular}{|c|c|c|c|c|}\hline\hline
Decay & \multicolumn{2}{|c|}{$F(0)$} & $a_F$ & $b_F$ \\
\hline
$D \rightarrow K^*$ & $A_1$ & $0.571^{+0.020}_{-0.022}$ & $0.65^{-0.06}_{+0.10}$ & $0.66^{-0.18}_{+0.21}$ \\
\cline{2-5}
& $A_2$ & $0.345^{+0.034}_{-0.037}$ & $1.86^{+0.05}_{-0.22}$ & $-0.91^{+0.48}_{-0.97}$ \\
\cline{2-5}
& $A_3$ & $-0.723^{+0.065}_{-0.077}$ & $1.32^{+0.14}_{-0.09}$ & $1.28^{+0.22}_{-0.21}$ \\
\cline{2-5}
& $V$ & $0.791^{+0.024}_{-0.026}$ & $1.04^{-0.17}_{+0.25}$ & $2.21^{-0.12}_{+0.37}$ \\
\hline
$D \rightarrow \rho$ & $A_1$ & $0.599^{+0.035}_{-0.030}$ & $0.44^{-0.06}_{+0.10}$ & $0.58^{-0.04}_{+0.23}$ \\
\cline{2-5}
& $A_2$ & $0.372^{+0.026}_{-0.031}$ & $1.64^{-0.16}_{+0.10}$ & $0.56^{-0.28}_{+0.04}$ \\
\cline{2-5}
& $A_3$ & $-0.719^{+0.055}_{-0.066}$ & $1.05^{+0.15}_{-0.15}$ & $1.77^{-0.11}_{+0.20}$ \\
\cline{2-5}
& $V$ & $0.801^{+0.044}_{-0.036}$ & $0.78^{-0.20}_{+0.24}$ & $2.61^{+0.29}_{-0.04}$ \\
\hline
$D \rightarrow \omega$ & $A_1$ & $0.556^{+0.033}_{-0.028}$ & $0.45^{-0.05}_{+0.09}$ & $0.54^{-0.10}_{+0.17}$ \\
\cline{2-5}
& $A_2$ & $0.333^{+0.026}_{-0.030}$ & $1.67^{-0.15}_{+0.09}$ & $0.44^{-0.29}_{-0.05}$ \\
\cline{2-5}
& $A_3$ & $-0.657^{+0.053}_{-0.065}$ & $1.07^{+0.17}_{-0.14}$ & $1.77^{+0.14}_{-0.07}$ \\
\cline{2-5}
& $V$ & $0.742^{+0.041}_{-0.034}$ & $0.79^{-0.20}_{+0.22}$ & $2.52^{+0.28}_{-0.13}$ \\
\hline
$D_s \rightarrow K^*$ & $A_1$ & $0.589^{+0.040}_{-0.042}$ & $0.56^{-0.02}_{+0.02}$ & $-0.12^{+0.03}_{-0.02}$ \\
\cline{2-5}
& $A_2$ & $0.315^{+0.024}_{-0.018}$ & $0.15^{+0.22}_{-0.14}$ & $0.24^{-0.94}_{+0.83}$ \\
\cline{2-5}
& $A_3$ & $-0.675^{+0.027}_{-0.037}$ & $0.48^{-0.11}_{+0.13}$ & $-0.14^{+0.18}_{-0.17}$ \\
\cline{2-5}
& $V$ & $0.771^{+0.049}_{-0.049}$ & $1.08^{-0.02}_{+0.02}$ & $0.13^{+0.03}_{-0.02}$ \\
\hline
$D_s \rightarrow \phi$ & $A_1$ & $0.569^{+0.046}_{-0.049}$ & $0.84^{-0.05}_{+0.06}$ & $0.16^{-0.01}_{+0.01}$ \\
\cline{2-5}
& $A_2$ & $0.304^{+0.021}_{-0.017}$ & $0.24^{+0.18}_{-0.05}$ & $1.25^{-1.08}_{+1.02}$ \\
\cline{2-5}
& $A_3$ & $-0.757^{+0.029}_{-0.039}$ & $0.60^{-0.02}_{+0.07}$ & $0.60^{+0.31}_{-0.33}$ \\
\cline{2-5}
& $V$ & $0.778^{+0.057}_{-0.062}$ & $1.37^{-0.05}_{+0.04}$ & $0.52^{+0.04}_{-0.06}$ \\
\hline\hline
\end{tabular}
\flushleft{{\bf Tab.7a}: Results of the form factors for $ D_{(s)}
\rightarrow V $ meson decays obtained with considering the meson DAs
up to twist-4. For $D \rightarrow V$, we choose $s_0 = 1.9 \pm 0.1
$Gev, $T = 1.2 \mp 0.2 $Gev and for $D_s \rightarrow V$, $s_0 = 1.5
\pm 0.1 $Gev, $T = 1.5 \mp 0.1 $ Gev. }
\end{table}

\begin{table}
\centering \tabcolsep0.25in
\begin{tabular}{|c|c|c|c|c|}\hline\hline
Decay & \multicolumn{2}{|c|}{$F(0)$} & $a_F$ & $b_F$ \\
\hline
$D \rightarrow K^*$ & $A_1$ & $0.601^{+0.030}_{-0.029}$ & $0.51^{-0.02}_{+0.02}$ & $0.04^{+0.01}_{-0.01}$ \\
\cline{2-5}
& $A_2$ & $0.541^{+0.038}_{-0.033}$ & $0.91^{+0.05}_{-0.10}$ & $-0.68^{+0.12}_{-0.21}$ \\
\cline{2-5}
& $A_3$ & $-0.541^{+0.033}_{-0.038}$ & $0.91^{-0.10}_{+0.05}$ & $-0.68^{-0.21}_{+0.12}$ \\
\cline{2-5}
& $V$ & $0.796^{+0.032}_{-0.027}$ & $0.60^{-0.07}_{+0.13}$ & $1.53^{-0.13}_{+0.30}$ \\
\hline
$D \rightarrow \rho$ & $A_1$ & $0.590^{+0.031}_{-0.029}$ & $0.44^{-0.04}_{+0.05}$ & $0.20^{-0.03}_{+0.10}$ \\
\cline{2-5}
& $A_2$ & $0.528^{+0.036}_{-0.031}$ & $0.91^{+0.07}_{-0.13}$ & $-1.01^{+0.22}_{-0.41}$ \\
\cline{2-5}
& $A_3$ & $-0.528^{+0.031}_{-0.036}$ & $0.91^{-0.13}_{+0.07}$ & $-1.01^{-0.41}_{+0.22}$ \\
\cline{2-5}
& $V$ & $0.735^{+0.032}_{-0.025}$ & $0.48^{-0.11}_{+0.21}$ & $2.25^{-0.21}_{+0.61}$ \\
\hline
$D \rightarrow \omega$ & $A_1$ & $0.548^{+0.029}_{-0.027}$ & $0.45^{-0.03}_{+0.07}$ & $0.16^{-0.04}_{+0.14}$ \\
\cline{2-5}
& $A_2$ & $0.478^{+0.034}_{-0.029}$ & $0.91^{+0.07}_{-0.15}$ & $-1.12^{+0.23}_{-0.43}$ \\
\cline{2-5}
& $A_3$ & $-0.478^{+0.029}_{-0.034}$ & $0.91^{-0.15}_{+0.07}$ & $-1.12^{-0.43}_{+0.23}$ \\
\cline{2-5}
& $V$ & $0.679^{+0.030}_{-0.023}$ & $0.48^{-0.11}_{+0.20}$ & $2.20^{-0.21}_{+0.60}$ \\
\hline
$D_s \rightarrow K^*$ & $A_1$ & $0.582^{+0.037}_{-0.037}$ & $0.34^{+0.01}_{-0.01}$ & $-0.15^{+0.01}_{-0.04}$ \\
\cline{2-5}
& $A_2$ & $0.517^{+0.033}_{-0.030}$ & $0.60^{+0.03}_{-0.02}$ & $-0.06^{-0.01}_{+0.01}$ \\
\cline{2-5}
& $A_3$ & $-0.517^{+0.030}_{-0.033}$ & $0.60^{-0.02}_{+0.03}$ & $-0.06^{+0.01}_{-0.01}$ \\
\cline{2-5}
& $V$ & $0.854^{+0.041}_{-0.041}$ & $0.82^{-0.03}_{+0.04}$ & $0.42^{+0.01}_{-0.01}$ \\
\hline
$D_s \rightarrow \phi$ & $A_1$ & $0.605^{+0.043}_{-0.042}$ & $0.48^{-0.01}_{+0.02}$ & $-0.007^{-0.003}_{+0.003}$ \\
\cline{2-5}
& $A_2$ & $0.583^{+0.038}_{-0.036}$ & $0.70^{+0.01}_{-0.01}$ & $0.16^{+0.01}_{-0.03}$ \\
\cline{2-5}
& $A_3$ & $-0.583^{+0.036}_{-0.038}$ & $0.70^{-0.01}_{+0.01}$ & $0.16^{-0.03}_{+0.01}$ \\
\cline{2-5}
& $V$ & $0.951^{+0.053}_{-0.053}$ & $0.98^{-0.04}_{+0.05}$ & $0.57^{+0.02}_{-0.02}$ \\
\hline\hline
\end{tabular}
\flushleft{{\bf Tab.7b}: Results of the form factors for $ D_{(s)}
\rightarrow V $ meson decays obtained with only considering the
leading twist meson DAs. For $D \rightarrow V$, we choose $s_0 = 2.0
\pm 0.1 $Gev, $T = 1.5 \mp 0.2 $Gev and for $D_s \rightarrow V$,
$s_0 = 1.5 \pm 0.1 $Gev, $T = 1.5 \mp 0.1 $ Gev. }
\end{table}

\begin{table}
\centering \tabcolsep0.08in
\begin{tabular}{|c|c|c|c|c|c|}\hline\hline
Decays & Reference & $f_+(0)$ & $A_1(0)$ & $A_2(0)$ & $V(0)$ \\
\hline $B \rightarrow \pi(\rho) $ & This work &
$0.285^{+0.016}_{-0.015}$ & $0.232^{+0.013}_{-0.014}$ &
$0.187^{+0.011}_{-0.012}$ & $0.289^{+0.016}_{-0.016}$
\\
\cline{3-6} & & --- & $0.239^{+0.013}_{-0.014}$ &
$0.230^{+0.013}_{-0.012}$ &
$0.297^{+0.015}_{-0.016}$ \\
\cline{2-6}
& SR\cite{P1,P2} & $0.258 \pm 0.031$ & $0.242 \pm 0.024$ & $0.221 \pm 0.023$ & $0.323 \pm 0.029$ \\
\cline{2-6}
& QM\cite{BSW} & 0.333 & 0.283 & 0.283 & 0.329 \\
\hline $B \rightarrow K(K^*)$ & This work &
$0.345^{+0.021}_{-0.021}$ & $0.266^{+0.016}_{-0.017}$
& $0.212^{+0.013}_{-0.013}$ & $0.331^{+0.019}_{-0.019}$ \\
\cline{3-6} & &  --- & $0.274^{+0.015}_{-0.016}$ &
$0.268^{+0.014}_{-0.015}$
& $0.351^{+0.019}_{-0.019}$ \\
\cline{2-6}
& SR\cite{P1,P2} & $0.331 \pm 0.041$ & $0.292 \pm 0.028$ & $0.259 \pm 0.027$ & $0.411 \pm 0.033$ \\
\cline{2-6}
& QM\cite{BSW} & 0.379 & 0.328 & 0.331 & 0.369 \\
\hline $B \rightarrow \eta(\omega)$ & This work &
$0.247^{+0.016}_{-0.016}$ & $0.214^{+0.013}_{-0.012}$ &
$0.170^{+0.010}_{-0.011}$ & $0.268^{+0.014}_{-0.015}$ \\
\cline{3-6} & & --- & $0.221^{+0.012}_{-0.013}$ &
$0.211^{+0.011}_{-0.011}$ &
$0.275^{+0.014}_{-0.015}$ \\
\cline{2-6}
& SR\cite{P1,P2} & $0.275 \pm 0.036$ & $0.219 \pm 0.025$ & $0.198 \pm 0.022$ & $0.293 \pm 0.029$ \\
\cline{2-6}
& QM\cite{BSW} & 0.307 & 0.281 & 0.281 & 0.328 \\
\hline $B_s \rightarrow K(K^*)$ & This work &
$0.296^{+0.018}_{-0.018}$ & $0.227^{+0.010}_{-0.012}$ &
$0.183^{+0.008}_{-0.010}$ & $0.285^{+0.013}_{-0.013}$ \\
\cline{3-6} & & --- & $0.240^{+0.014}_{-0.014}$ &
$0.240^{+0.013}_{-0.013}$ &
$0.307^{+0.017}_{-0.017}$ \\
\cline{2-6}
& SR\cite{P1,P2} & --- & $0.233 \pm 0.023$ & $0.181 \pm 0.025$ & $0.311 \pm 0.026$ \\
\hline $B_s \rightarrow \eta(\phi)$ & This work &
$0.281^{+0.015}_{-0.016}$ & $0.271^{+0.014}_{-0.014}$ &
$0.212^{+0.011}_{-0.012}$ & $0.339^{+0.016}_{-0.017}$ \\
\cline{3-6} & & --- & $0.285^{+0.017}_{-0.018}$ &
$0.288^{+0.016}_{-0.016}$ &
$0.376^{+0.021}_{-0.022}$ \\
\cline{2-6}
& SR\cite{P1,P2} & --- & $0.311 \pm 0.029$ & $0.234 \pm 0.028$ & $0.434 \pm 0.035$ \\
\hline \hline
\end{tabular}
\flushleft{{\bf Tab.8}: Comparison of the form factors (at $q^2=0$)
given in this work with other groups for $B_{(s)}$ decays. The lower
row of this work corresponds to the results obtained with only
considering the leading twist meson DAs.}
\end{table}

\begin{table}
\centering \tabcolsep0.13in
\begin{tabular}{|c|c|c|c|c|c|}\hline\hline
Decays & Reference & $f_+(0)$ & $A_1(0)$ & $A_2(0)$ & $V(0)$ \\
\hline $D \rightarrow \pi(\rho)$ & This work &
$0.635^{+0.060}_{-0.057}$ & $0.599^{+0.035}_{-0.030}$ &
$0.372^{+0.026}_{-0.031}$ & $0.801^{+0.044}_{-0.036}$ \\
\cline{3-6} & & --- & $0.590^{+0.031}_{-0.029}$ &
$0.528^{+0.036}_{-0.031}$
& $0.735^{+0.032}_{-0.025}$ \\
\cline{2-6}
& QM\cite{BSW} & 0.692 & 0.775 & 0.923 & 1.225 \\
\hline $D \rightarrow K(K^*)$ & This work &
$0.661^{+0.067}_{-0.066}$ & $0.571^{+0.020}_{-0.022}$ &
$0.345^{+0.034}_{-0.037}$ & $0.791^{+0.024}_{-0.026}$ \\
\cline{3-6} & & --- & $0.601^{+0.030}_{-0.029}$ &
$0.541^{+0.038}_{-0.033}$
& $0.796^{+0.032}_{-0.027}$ \\
\cline{2-6}
& QM\cite{BSW} & 0.762 & 0.880 & 1.147 & 1.226 \\
\hline $D \rightarrow \eta(\omega)$ & This work &
$0.556^{+0.056}_{-0.053}$ & $0.556^{+0.033}_{-0.028}$ &
$0.333^{+0.026}_{-0.030}$ & $0.742^{+0.041}_{-0.034}$ \\
\cline{3-6} & & --- & $0.548^{+0.029}_{-0.027}$ &
$0.478^{+0.034}_{-0.029}$
& $0.679^{+0.030}_{-0.023}$ \\
\cline{2-6}
& QM\cite{BSW} & 0.681 & 0.772 & 0.920 & 1.236 \\
\hline $D_s \rightarrow K(K^*)$ & This work &
$0.820^{+0.080}_{-0.071}$ & $0.589^{+0.040}_{-0.042}$ &
$0.315^{+0.024}_{-0.018}$ & $0.771^{+0.049}_{-0.049}$\\
\cline{3-6} & & --- & $0.582^{+0.037}_{-0.037}$ &
$0.517^{+0.033}_{-0.030}$
& $0.854^{+0.041}_{-0.041}$ \\
\cline{2-6}
& QM\cite{BSW} & 0.643 & 0.717 & 0.853 & 1.250 \\
\hline $D_s \rightarrow \eta(\phi)$ & This work &
$0.611^{+0.062}_{-0.054}$ & $0.569^{+0.046}_{-0.049}$ &
$0.304^{+0.021}_{-0.017}$ & $0.778^{+0.057}_{-0.062}$ \\
\cline{3-6} & & --- & $0.605^{+0.043}_{-0.042}$ &
$0.583^{+0.038}_{-0.036}$
& $0.951^{+0.053}_{-0.053}$ \\
\cline{2-6}
& QM\cite{BSW} & 0.723 & 0.820 & 1.076 & 1.319 \\
\hline\hline
\end{tabular}
\flushleft{{\bf Tab.9}: Comparison of the form factors (at $q^2=0$)
given in this work with other groups for $D_{(s)}$ decays. The lower
row of this work corresponds to the results obtained with only
considering the leading twist meson DAs.}
\end{table}

\begin{table}
\centering \tabcolsep0.15in
\begin{tabular}{|c|c|c|c|}\hline\hline
Decays & Reference & $R_V$ & $R_2$ \\
\hline
$D \rightarrow K^*$ & This work & $1.39^{+0.09}_{-0.10}$ & $0.60^{+0.09}_{-0.08}$  \\
\cline{3-4} & &  $1.32^{+0.13}_{-0.10}$ & $0.90^{+0.11}_{-0.09}$ \\
\cline{2-4}
& FOCUS\cite{ML} & $1.504 \pm 0.057 \pm 0.039$ & $0.875 \pm 0.049 \pm 0.064$  \\
\cline{2-4}
& BEATRICE\cite{MA} & $1.45 \pm 0.23 \pm 0.07$ & $1.00 \pm 0.15 \pm 0.03$ \\
\cline{2-4}
& E791\cite{EM} & $1.87 \pm 0.08 \pm 0.07$ & $0.73 \pm 0.06 \pm 0.08$ \\
\hline $D_s \rightarrow \phi$ & This work & $1.37^{+0.24}_{-0.21}$ & $0.53^{+0.10}_{-0.06}$ \\
\cline{3-4} & & $1.57^{+0.21}_{-0.18}$
& $0.96^{+0.14}_{-0.12}$ \\
\cline{2-4}
& FOCUS\cite{JML} & $1.549 \pm 0.250 \pm 0.145$ & $0.713 \pm 0.202 \pm 0.266$ \\
\cline{2-4}
& E791\cite{EMA} & $2.27 \pm 0.35 \pm 0.22$ & $1.570 \pm 0.250 \pm 0.190$ \\
\hline\hline
\end{tabular}
\flushleft{{\bf Tab.10}: Comparison of measurements and theoretical
predictions for the form factor ratios $R_V$, $R_2$. The lower row
of this work corresponds to the results obtained with only
considering the leading twist meson DAs.}
\end{table}

\begin{table}
\centering \tabcolsep0.18in
\begin{tabular}{|c|c|c|c|c|}\hline\hline
Decays  & Reference & $f_T(0)$ & $T_1(0)=T_2(0) $ & $T_3(0)$ \\
\hline $B \rightarrow \pi(\rho) $ & This work &
$0.267^{+0.015}_{-0.014}$ & $0.256^{+0.015}_{-0.015}$ &
$0.175^{+0.010}_{-0.010}$
\\
\cline{3-5} & & --- & $0.264^{+0.014}_{-0.015}$ &
$0.198^{+0.010}_{-0.010}$  \\
\cline{2-5}
& SR\cite{P1,P2} & $0.253 \pm 0.028$ & $0.267 \pm 0.021$ & $0.176 \pm 0.016$  \\
\hline $B \rightarrow K(K^*)$ & This work &
$0.347^{+0.021}_{-0.021}$ & $0.293^{+0.017}_{-0.018}$
& $0.193^{+0.011}_{-0.011}$  \\
\cline{3-5} & & --- & $0.306^{+0.017}_{-0.018}$ &
$0.226^{+0.011}_{-0.012}$  \\
\cline{2-5}
& SR\cite{P1,P2} & $0.358 \pm 0.037$ & $0.333 \pm 0.028$ & $0.202 \pm 0.018$  \\
\hline $B \rightarrow \eta(\omega)$ & This work &
$0.248^{+0.017}_{-0.016}$ & $0.237^{+0.013}_{-0.014}$ &
$0.160^{+0.009}_{-0.009}$  \\
\cline{3-5} & & --- & $0.244^{+0.013}_{-0.014}$ &
$0.181^{+0.009}_{-0.010}$  \\
\cline{2-5}
& SR\cite{P1,P2} & $0.285 \pm 0.029$ & $0.242 \pm 0.022$ & $0.155 \pm 0.015$  \\
\hline $B_s \rightarrow K(K^*)$ & This work &
$0.288^{+0.018}_{-0.017}$ & $0.251^{+0.012}_{-0.012}$ &
$0.169^{+0.008}_{-0.008}$  \\
\cline{3-5} & & --- & $0.268^{+0.015}_{-0.015}$ &
$0.203^{+0.010}_{-0.011}$  \\
\cline{2-5}
& SR\cite{P1,P2} & --- & $0.260 \pm 0.024$ & $0.136 \pm 0.016$  \\
\hline $B_s \rightarrow \eta(\phi)$ & This work &
$0.282^{+0.015}_{-0.016}$ & $0.299^{+0.015}_{-0.016}$ &
$0.191^{+0.010}_{-0.010}$ \\
\cline{3-5} & & --- & $0.321^{+0.019}_{-0.019}$ &
$0.239^{+0.013}_{-0.013}$  \\
\cline{2-5}
& SR\cite{P1,P2} & --- & $0.349 \pm 0.033$ & $0.175 \pm 0.018$  \\
\hline $D \rightarrow \pi(\rho)$ & This work &
$0.520^{+0.040}_{-0.038}$ &  $0.658^{+0.038}_{-0.031}$ &
$0.326^{+0.026}_{-0.023}$  \\
\cline{2-5} & & --- & $0.633^{+0.031}_{-0.028}$ &
$0.287^{+0.016}_{-0.012}$  \\
\hline $D \rightarrow K(K^*)$  & This work &
$0.633^{+0.051}_{-0.052}$ & $0.629^{+0.021}_{-0.023}$ &
$0.308^{+0.028}_{-0.025}$ \\
\cline{2-5}& & --- & $0.652^{+0.030}_{-0.029}$ &
$0.293^{+0.016}_{-0.011}$ \\
\hline $D \rightarrow \eta(\omega)$  & This work &
$0.559^{+0.045}_{-0.043}$ & $0.610^{+0.036}_{-0.030}$ &
$0.293^{+0.025}_{-0.022}$  \\
\cline{2-5} & & --- & $0.586^{+0.029}_{-0.026}$ &
$0.255^{+0.016}_{-0.010}$ \\
\hline $D_s \rightarrow K(K^*)$ & This work &
$0.792^{+0.068}_{-0.057}$ & $0.639^{+0.042}_{-0.044}$ &
$0.272^{+0.015}_{-0.009}$ \\
\cline{2-5} &  & --- & $0.656^{+0.039}_{-0.037}$ &
$0.335^{+0.015}_{-0.012}$ \\
\hline $D_s \rightarrow \eta(\phi)$  & This work &
$0.595^{+0.053}_{-0.043}$ & $0.620^{+0.048}_{-0.053}$ &
$0.286^{+0.014}_{-0.011}$  \\
\cline{2-5} &  & --- & $0.689^{+0.044}_{-0.045}$ &
$0.387^{+0.017}_{-0.017}$ \\
\hline\hline
\end{tabular}
\flushleft{{\bf Tab.11}: The penguin type form factors at $q^2=0$
given via the relations of Eqs.(27-30). For comparison, the values
calculated by other groups are also listed. The lower row of this
work corresponds to the results obtained with only considering the
leading twist meson DAs.}
\end{table}

\FloatBarrier

\begin{table}[t]
\centering \tabcolsep0.23in
\renewcommand\arraystretch{0.75}
\begin{tabular}{|c|c|c|c|}\hline\hline
Decays & $e^+e^-$ & $\mu^+\mu^-$ & $\tau^+\tau^-$ \\
\hline $B^0 \rightarrow K^0 \ell^+ \ell^-$ & $56.3^{+7.0}_{-6.6}$ &
$56.2^{+7.0}_{-6.5}$
& $16.0^{+1.8}_{-1.8}$ \\
\cline{2-4}
& $4.5^{+9.0}_{-8.0}$\cite{HFAG}  & $61.8^{+18.5}_{-15.5}$ \cite{HFAG} & --- \\
\hline $B^+ \rightarrow K^+ \ell^+ \ell^-$ & $61.1^{+7.6}_{-7.2}$ &
$60.9^{+7.6}_{-7.1} $
& $17.4^{+2.0}_{-1.9} $ \\
\cline{2-4}
& $52.2^{+9.4}_{-9.0 }$\cite{HFAG} & $51.8 \pm 9.2 $ \cite{HFAG} & ---\\
\hline $B^0 \rightarrow \pi^0 \ell^+ \ell^-$ & $0.87^{+0.11}_{-0.10}
$ & $0.87^{+0.11}_{-0.10} $
& $0.41^{+0.06}_{-0.06} $ \\
\cline{2-4}
& --- & --- & --- \\
\hline $B^+ \rightarrow \pi^+ \ell^+ \ell^-$ & $1.89^{+0.23}_{-0.22}
$ & $1.88^{+0.24}_{-0.21} $
 & $0.90^{+0.13}_{-0.12} $ \\
\cline{2-4}
& $<3.9 \times 10^5$ \cite{PDG} & $<9.1 \times 10^5$ \cite{PDG}& --- \\
\hline $B^0 \rightarrow \eta \ell^+ \ell^-$ & $0.41^{+0.05}_{-0.04}
$ & $0.41^{+0.05}_{-0.04} $
& $0.11^{+0.01}_{-0.01} $ \\
\cline{2-4}
& --- & --- & --- \\
\hline $B_s \rightarrow \bar{K^0} \ell^+ \ell^-$ &
$1.99^{+0.21}_{-0.20}$ & $1.99^{+0.21}_{-0.20} $
& $0.74^{+0.07}_{-0.07} $ \\
\cline{2-4}
& --- & --- & --- \\
\hline $B_s \rightarrow \eta \ell^+ \ell^-$ & $12.0^{+1.2}_{-1.2}$ &
$12.0^{+1.2}_{-1.2} $
& $3.4^{+0.4}_{-0.4} $ \\
\cline{2-4}
& --- & --- & --- \\
\hline\hline
\end{tabular}
\flushleft{{\bf Tab.12}: The branching ratios of $ B \rightarrow P $
semileptonic rare decays, where the corresponding experimental
values are also given for comparison, unit: $10^{-8}$. }
\end{table}

\begin{table}
 \renewcommand\arraystretch{0.75}
 \tabcolsep0.10in
\begin{tabular}{|c|c|c|c|c|}\hline\hline
Decays & $e^+e^-$ & $\mu^+\mu^-$ & $\tau^+\tau^-$ & $\gamma$ \\
\hline $B^0 \rightarrow {K^*}^0 \ell^+ \ell^- (\gamma )$ &
$10.6^{+1.2}_{-1.1} $
& $10.2^{+1.1}_{-1.1} $ & $0.99^{+0.08}_{-0.09} $ & $387^{+46}_{-44} $ \\
\cline{2-5} & $12.8^{+3.0}_{-2.9}$ \cite{HFAG}  & $14.8 \pm 2.6$
 \cite{HFAG}
& --- & $430 \pm 40$ \cite{PDG} \\
\hline $B^+ \rightarrow {K^*}^+ \ell^+ \ell^- (\gamma)$ &
$11.5^{+1.3}_{-1.2} $ & $11.1^{+1.2}_{-1.2}$ &
$1.07^{+0.09}_{-0.10}$
& $420^{+50}_{-48} $ \\
\cline{2-5} & $12.2^{+7.3}_{-6.5}$ \cite{HFAG} &
$14.4^{+5.4}_{-4.5}$ \cite{HFAG} & --- &
$403 \pm 26 $ \cite{HFAG} \\
\hline $B^0 \rightarrow \rho^0 \ell^+ \ell^- (\gamma )$ &
$0.19^{+0.02}_{-0.02} $
& $0.18^{+0.02}_{-0.02}$ & $0.020^{+0.002}_{-0.002}$ & $6.4^{+0.7}_{-0.7} $ \\
\cline{2-5}
& --- & --- & --- & $3.8 \pm 1.8 $ \cite{HFAG} \\
\hline $B^+ \rightarrow \rho^+ \ell^+ \ell^- (\gamma)$ &
$0.40^{+0.04}_{-0.04} $
& $0.39^{+0.04}_{-0.04} $ & $0.040^{+0.004}_{-0.004} $ & $13.8^{+1.6}_{-1.5} $ \\
\cline{2-5}
& --- & --- & --- & $ 6.8^{+3.6}_{-3.1}$  \cite{HFAG} \\
\hline $B^0 \rightarrow \omega \ell^+ \ell^- (\gamma )$ &
$0.16^{+0.02}_{-0.02}$
& $0.16^{+0.02}_{-0.02} $ & $0.020^{+0.002}_{-0.002} $ & $5.5^{+0.6}_{-0.6} $ \\
\cline{2-5}
& --- & --- & --- & $5.4^{+2.3}_{-2.1}$ \cite{HFAG}  \\
\hline $B_s \rightarrow \bar{K^ {* 0}} \ell^+ \ell^- (\gamma )$ &
$0.40^{+0.04}_{-0.04} $ & $0.38^{+0.03}_{-0.03} $ &
$0.050^{+0.004}_{-0.004} $
& $12.0^{+1.1}_{-1.2} $ \\
\cline{2-5}
& --- & --- & --- & --- \\
\hline $B_s \rightarrow \phi \ell^+ \ell^- (\gamma)$ &
$12.3^{+1.1}_{-1.2} $
& $11.8^{+1.1}_{-1.1}$ & $1.23^{+0.10}_{-0.12} $ & $391^{+40}_{-41} $ \\
\cline{2-5}
& --- & $<32 $ \cite{HFAG} & --- & --- \\
\hline\hline
\end{tabular}
\flushleft{{\bf Tab.13a}: The branching ratios of $ B \rightarrow V
$ semileptonic and radiative rare decays obtained with considering
the meson DAs up to twist-4, where the corresponding experimental
values are also given for comparison, unit: $10^{-7}$. }
\end{table}

\begin{table}
\centering
\renewcommand\arraystretch{0.9}
 \tabcolsep0.10in
\begin{tabular}{|c|c|c|c|c|}\hline\hline
Decays & $e^+e^-$ & $\mu^+\mu^-$ & $\tau^+\tau^-$ & $\gamma$ \\
\hline $B^0 \rightarrow {K^*}^0 \ell^+ \ell^- (\gamma )$ &
$8.7^{+0.9}_{-0.9}$
& $8.2^{+0.8}_{-0.9}$ & $0.75^{+0.07}_{-0.07} $ & $421^{+48}_{-47}$ \\
\cline{2-5}
& $12.8^{+3.0}_{-2.9}$\cite{HFAG}  & $14.8 \pm 2.6$ \cite{HFAG} & --- & $430 \pm 40$\cite{PDG} \\
\hline $B^+ \rightarrow {K^*}^+ \ell^+ \ell^- (\gamma)$ &
$9.4^{+1.0}_{-1.0}$ & $8.9^{+0.9}_{-1.0}$ & $0.81^{+0.08}_{-0.08}$
& $457^{+52}_{-51}$ \\
\cline{2-5} & $12.2^{+7.3}_{-6.5}$\cite{HFAG} &
$14.4^{+5.4}_{-4.5}$\cite{HFAG} & --- &
$403 \pm 26$\cite{HFAG} \\
\hline $B^0 \rightarrow \rho^0 \ell^+ \ell^- (\gamma )$ &
$0.15^{+0.02}_{-0.02}$
& $0.14^{+0.02}_{-0.02}$ & $0.016^{+0.002}_{-0.002}$ & $6.7^{+0.7}_{-0.7}$ \\
\cline{2-5}
& --- & --- & --- & $3.8 \pm 1.8$ \cite{HFAG} \\
\hline $B^+ \rightarrow \rho^+ \ell^+ \ell^- (\gamma)$ &
$0.32^{+0.04}_{-0.03}$
& $0.31^{+0.03}_{-0.03}$ & $0.034^{+0.004}_{-0.003}$ & $(14.6^{+1.6}_{-1.5}$ \\
\cline{2-5}
& --- & --- & --- & $6.8^{+3.6}_{-3.1}$ \cite{HFAG} \\
\hline $B^0 \rightarrow \omega \ell^+ \ell^- (\gamma )$ &
$0.13^{+0.01}_{-0.01}$
& $0.12^{+0.01}_{-0.01}$ & $0.013^{+0.001}_{-0.001}$ & $(5.8^{+0.6}_{-0.7}$ \\
\cline{2-5}
& --- & --- & --- & $5.4^{+2.3}_{-2.1}$\cite{HFAG}  \\
\hline $B_s \rightarrow \bar{K^ {* 0}} \ell^+ \ell^- (\gamma )$ &
$0.32^{+0.04}_{-0.03}$ & $0.30^{+0.03}_{-0.03}$ &
$0.037^{+0.004}_{-0.004}$
& $13.6^{+1.6}_{-1.4}$ \\
\cline{2-5}
& --- & --- & --- & --- \\
\hline $B_s \rightarrow \phi \ell^+ \ell^- (\gamma)$ &
$9.7^{+1.2}_{-1.1}$
& $9.2^{+1.1}_{-1.0}$ & $0.89^{+0.10}_{-0.09} $ & $453^{+55}_{-53}$ \\
\cline{2-5}
& --- & $<32$ \cite{HFAG} & --- & --- \\
\hline\hline
\end{tabular}
\flushleft{{\bf Tab.13b}: The branching ratios of $ B \rightarrow V
$ semileptonic and radiative rare decays obtained with only
considering the leading twist meson DAs, where the corresponding
experimental values are also given for comparison, unit: $10^{-7}$.}
\end{table}

\begin{table}
\centering \renewcommand\arraystretch{0.9}
 \tabcolsep0.25in
\begin{tabular}{|c|c|c|c|}\hline\hline
Decays & $e \nu_e$ & $\mu \nu_\mu$ & $\tau \nu_\tau$ \\
\hline $B^0 \rightarrow \pi^- \ell^+ \nu_\ell$ &
$1.28^{+0.16}_{-0.15} $ & $1.28^{+0.16}_{-0.15}$
& $0.93^{+0.12}_{-0.11}$ \\
\cline{2-4}
& $1.33 \pm 0.22$\cite{PDG} &  $1.33 \pm 0.22$ \cite{PDG} & --- \\
\hline $B^+ \rightarrow \pi^0 \ell^+ \nu_\ell$ &
$0.70^{+0.09}_{-0.08}$ & $0.70^{+0.09}_{-0.08}$
 & $0.50^{+0.07}_{-0.06}$ \\
\cline{2-4}
& $0.90 \pm 0.28$ \cite{PDG} & --- & --- \\
\hline $B^+ \rightarrow \eta \ell^+ \nu_\ell$ &
$0.33^{+0.04}_{-0.04}$ & $0.33^{+0.04}_{-0.04}$
 & $0.21^{+0.02}_{-0.02}$ \\
\cline{2-4}
& $0.8 \pm 0.4$ \cite{PDG} & $0.8 \pm 0.4$ \cite{PDG} & --- \\
\hline $B_s \rightarrow K^- \ell^+ \nu_\ell$ & $1.47^{+0.15}_{-0.15}
$ & $1.46^{+0.16}_{-0.14} $
& $1.02^{+0.10}_{-0.10}$ \\
\cline{2-4}
& --- & --- & --- \\
\hline\hline
\end{tabular}
\flushleft{{\bf Tab.14}: The branching ratios of $ B \rightarrow P $
semileptonic decays, where the corresponding experimental values are
also given for comparison, unit: $10^{-4}$. }
\end{table}

\clearpage

\begin{table}[t]
\centering \centering
 \tabcolsep0.25in
\begin{tabular}{|c|c|c|c|}\hline\hline
Decays & $e \nu_e$ & $\mu \nu_\mu$ & $\tau \nu_\tau$ \\
\hline $B^0 \rightarrow \rho^- \ell^+ \nu_\ell$ &
$2.69^{+0.28}_{-0.28}$
& $2.69^{+0.28}_{-0.28}$ & $1.38^{+0.15}_{-0.15}$ \\
\cline{2-4}
& $2.6 \pm 0.7$\cite{PDG} &  $2.6 \pm 0.7$\cite{PDG} & --- \\
\hline $B^+ \rightarrow \rho^0 \ell^+ \nu_\ell$ &
$1.46^{+0.15}_{-0.15}$
& $1.46^{+0.15}_{-0.15}$ & $0.75^{+0.08}_{-0.08}$ \\
\cline{2-4}
& $1.34^{+0.32}_{-0.35}$\cite{PDG} & $1.34^{+0.32}_{-0.35}$\cite{PDG} & --- \\
\hline $B^+ \rightarrow \omega \ell^+ \nu_\ell$ &
$1.27^{+0.13}_{-0.13}$
& $1.26^{+0.14}_{-0.12}$ & $0.65^{+0.07}_{-0.07}$ \\
\cline{2-4}
& $<2.1$ \cite{PDG} & $<2.1$ \cite{PDG}  & --- \\
\hline $B_s \rightarrow K^{* -} \ell^+ \nu_\ell$ &
$2.91^{+0.26}_{-0.26}$
& $2.91^{+0.25}_{-0.26}$ & $1.58^{+0.13}_{-0.13}$ \\
\cline{2-4}
& --- & --- & --- \\
\hline\hline
\end{tabular}
\flushleft{{\bf Tab.15a}: The branching ratios of $ B \rightarrow V
$ semileptonic decays obtained with considering the meson DAs up to
twist-4, where the corresponding experimental values are also given
for comparison, unit: $10^{-4}$. }
\end{table}

\begin{table}
\centering  \tabcolsep0.25in
\begin{tabular}{|c|c|c|c|}\hline\hline
Decays & $e \nu_e$ & $\mu \nu_\mu$ & $\tau \nu_\tau$ \\
\hline $B^0 \rightarrow \rho^- \ell^+ \nu_\ell$ &
$2.47^{+0.26}_{-0.26}$
& $2.47^{+0.26}_{-0.26}$ & $1.27^{+0.13}_{-0.13}$ \\
\cline{2-4}
& $2.6 \pm 0.7$\cite{PDG} &  $2.6 \pm 0.7$\cite{PDG} & --- \\
\hline $B^+ \rightarrow \rho^0 \ell^+ \nu_\ell$ &
$1.34^{+0.14}_{-0.14}$
& $1.34^{+0.14}_{-0.14}$ & $0.69^{+0.07}_{-0.07}$ \\
\cline{2-4}
& $1.34^{+0.32}_{-0.35}$\cite{PDG} & $1.34^{+0.32}_{-0.35}$\cite{PDG} & --- \\
\hline $B^+ \rightarrow \omega \ell^+ \nu_\ell$ &
$1.16^{+0.12}_{-0.12}$
& $1.16^{+0.12}_{-0.12}$ & $0.60^{+0.06}_{-0.06}$ \\
\cline{2-4}
& $<2.1$ \cite{PDG} & $<2.1$\cite{PDG} & --- \\
\hline $B_s \rightarrow K^{* -} \ell^+ \nu_\ell$ &
$2.63^{+0.30}_{-0.28}$
& $2.63^{+0.29}_{-0.29}$ & $1.41^{+0.15}_{-0.15}$ \\
\cline{2-4}
& --- & --- & --- \\
\hline\hline
\end{tabular}
\flushleft{{\bf Tab.15b}: The branching ratios of $ B \rightarrow V
$ semileptonic decays obtained with only considering the leading
twist meson DAs, where the corresponding experimental values are
also given for comparison, unit: $10^{-4}$.}
\end{table}

\FloatBarrier

\begin{table}[t]
\centering \tabcolsep0.46in
\begin{tabular}{|c|c|c|}\hline\hline
Decays & $e^+e^-$ & $\mu^+\mu^-$  \\
\hline
$D^0 \rightarrow \pi^0 \ell^+ \ell^-$ & $0.92^{+0.17}_{-0.12}$ & $0.89^{+0.11}_{-0.10}$ \\
\cline{2-3}
& $<45$\cite{PDG}  & $<180$ \cite{PDG} \\
\hline
$D^+ \rightarrow \pi^+ \ell^+ \ell^-$ & $4.68^{+0.88}_{-0.59} $ & $4.49^{+0.57}_{-0.49}$ \\
\cline{2-3}
& $<52$\cite{PDG} & $<8.8$ \cite{PDG} \\
\hline
$D^0 \rightarrow \eta \ell^+ \ell^-$ & $0.24^{+0.05}_{-0.03}$ & $0.24^{+0.05}_{-0.04}$ \\
\cline{2-3}
& $<110 $\cite{PDG} & $<530$ \cite{PDG}  \\
\hline
$D_s^+ \rightarrow K^+ \ell^+ \ell^-$ & $2.80^{+0.52}_{-0.39}$ & $2.78^{+0.51}_{-0.39}$ \\
\cline{2-3}
& $<1600$\cite{PDG}  & $<36$\cite{PDG} \\
\hline\hline
\end{tabular}
\flushleft{{\bf Tab.16}: The branching ratios of $ D \rightarrow P $
semileptonic rare decays, where the corresponding experimental
values are also given for comparison, unit: $10^{-6}$. }
\end{table}

\begin{table}
\centering \tabcolsep0.25in
\begin{tabular}{|c|c|c|c|}\hline\hline
Decays & $e^+e^-$ & $\mu^+\mu^-$ & $\gamma$ \\
\hline $D^0 \rightarrow \rho^0 \ell^+ \ell^- (\gamma )$ &
$0.85^{+0.07}_{-0.07}$
& $0.83^{+0.07}_{-0.06}$ & $2.3^{+0.3}_{-0.2}$ \\
\cline{2-4}
& $<100$ \cite{PDG} & $<22$\cite{PDG}  & $<240$\cite{PDG}  \\
\hline $D^+ \rightarrow \rho^+ \ell^+ \ell^- (\gamma)$ &
$4.33^{+0.38}_{-0.34}$
& $4.19^{+0.37}_{-0.32}$ & $11.5^{+1.4}_{-1.1}$ \\
\cline{2-4}
& --- & $<560$ \cite{PDG}  & --- \\
\hline $D^0 \rightarrow \omega \ell^+ \ell^- ( \gamma )$ &
$0.71^{+0.07}_{-0.06}$
& $0.68^{+0.07}_{-0.05}$ & $1.9^{+0.3}_{-0.2}$ \\
\cline{2-4}
& $<180 $ \cite{PDG} & $<830$ \cite{PDG} & $<240$ \cite{PDG} \\
\hline $D_s^+ \rightarrow K^{* +} \ell^+ \ell^- (\gamma )$ &
$2.24^{+0.26}_{-0.28}$
& $2.17^{+0.26}_{-0.26}$ & $5.26^{+0.73}_{-0.70}$ \\
\cline{2-4}
& --- & $<1400$\cite{PDG}  & --- \\
\hline\hline
\end{tabular}
\flushleft{{\bf Tab.17a}: The branching ratios of $ D \rightarrow V
$ semileptonic and radiative rare decays obtained with considering
the meson DAs up to twist-4, where the corresponding experimental
values are also given for comparison, unit: $10^{-6}$. }
\end{table}

\begin{table}
\centering \centering
\renewcommand\arraystretch{0.9}
 \tabcolsep0.25in
\begin{tabular}{|c|c|c|c|}\hline\hline
Decays & $e^+e^-$ & $\mu^+\mu^-$ & $\gamma$ \\
\hline $D^0 \rightarrow \rho^0 \ell^+ \ell^- (\gamma )$ &
$0.84^{+0.07}_{-0.07}$
& $0.82^{+0.07}_{-0.07}$ & $2.1^{+0.2}_{-0.2}$ \\
\cline{2-4}
& $<100$ \cite{PDG} & $<22$ \cite{PDG} & $<240$ \cite{PDG} \\
\hline $D^+ \rightarrow \rho^+ \ell^+ \ell^- (\gamma)$ &
$4.26^{+0.36}_{-0.33}$
& $4.14^{+0.34}_{-0.33}$ & $10.7^{+1.0}_{-1.0}$ \\
\cline{2-4}
& --- & $<560$\cite{PDG} & --- \\
\hline $D^0 \rightarrow \omega \ell^+ \ell^- ( \gamma )$ &
$0.70^{+0.06}_{-0.06}$
& $0.68^{+0.06}_{-0.06}$ & $1.8^{+0.2}_{-0.2}$ \\
\cline{2-4}
& $<180$\cite{PDG} & $<830 $\cite{PDG} & $<240 $\cite{PDG} \\
\hline $D_s^+ \rightarrow K^{* +} \ell^+ \ell^- (\gamma )$ &
$1.89^{+0.25}_{-0.23}$
& $1.82^{+0.25}_{-0.21}$ & $5.56^{+0.67}_{-0.62}$ \\
\cline{2-4}
& --- & $<1400 $\cite{PDG} & --- \\
\hline\hline
\end{tabular}
\flushleft{{\bf Tab.17b}: The branching ratios of $ D \rightarrow V
$ semileptonic and radiative rare decays obtained with only
considering the leading twist meson DAs, where the corresponding
experimental values are also given for comparison, unit: $10^{-6}$.}
\end{table}

\begin{table}
\centering \renewcommand\arraystretch{0.9}
 \tabcolsep0.40in
\begin{tabular}{|c|c|c|}\hline\hline
Decays & $e \nu_e$ & $\mu \nu_\mu$ \\
\hline
$D^0 \rightarrow K^- \ell^+ \nu_\ell$ & $32^{+4.7}_{-4.3}$ & $31.5^{+4.6}_{-4.2}$ \\
\cline{2-3}
& $35.8 \pm 1.8$\cite{PDG} & $31.9 \pm 1.7$\cite{PDG} \\
\hline
$D^+ \rightarrow \bar{K}^0 \ell^+ \nu_\ell$ & $81.2^{+11.9}_{-10.8}$ & $79.8^{+11.6}_{-10.6}$ \\
\cline{2-3}
& $67 \pm 9 $\cite{PDG} & $70^{+30}_{-20}$\cite{PDG} \\
\hline
$D^0 \rightarrow \pi^- \ell^+ \nu_\ell$ & $2.78^{+0.35}_{-0.30}$ & $2.75^{+0.35}_{-0.30}$ \\
\cline{2-3}
& $3.6 \pm 0.6$\cite{PDG} & --- \\
\hline
$D^+ \rightarrow \pi^0 \ell^+ \nu_\ell$ & $3.52^{+0.45}_{-0.38}$ & $3.49^{+0.45}_{-0.38}$ \\
\cline{2-3}
& $3.1 \pm 1.5$\cite{PDG} & $3.1 \pm 1.5$\cite{PDG} \\
\hline
$D^+ \rightarrow \eta \ell^+ \nu_\ell$ & $0.86^{+0.16}_{-0.15}$ & $0.84^{+0.16}_{-0.14}$ \\
\cline{2-3}
& $<5 $\cite{PDG} & $<5 $\cite{PDG} \\
\hline
$D_s^+ \rightarrow K^0 \ell^+ \nu_\ell$ & $3.90^{+0.74}_{-0.57}$ & $3.83^{+0.72}_{-0.56}$ \\
\cline{2-3}
& --- & --- \\
\hline
$D_s^+ \rightarrow \eta \ell^+ \nu_\ell$ & $12.7^{+2.6}_{-2.0}$ & $12.5^{+2.5}_{-2.0}$ \\
\cline{2-3}
& $25 \pm 7$\cite{PDG} & $25 \pm 7$\cite{PDG} \\
\hline\hline
\end{tabular}
\flushleft{{\bf Tab.18}: The branching ratios of $ D \rightarrow P $
semileptonic decays, where the corresponding experimental value are
also given for comparison, unit: $10^{-3}$. }
\end{table}

\begin{table}[t]
\centering \renewcommand\arraystretch{0.75}
 \tabcolsep0.40in
\begin{tabular}{|c|c|c|}\hline\hline
Decays & $e \nu_e$ & $\mu \nu_\mu$ \\
\hline
$D^0 \rightarrow K^{* -} \ell^+ \nu_\ell$ & $21.2^{+0.9}_{-0.9}$ & $20.1^{+0.9}_{-0.8}$ \\
\cline{2-3}
& $21.5 \pm 3.5$\cite{PDG} & --- \\
\hline
$ D^+ \rightarrow \bar{K}^{* 0} \ell^+ \nu_\ell$ & $53.7^{+2.4}_{-2.3}$ & $51.0^{+2.3}_{-2.1}$ \\
\cline{2-3}
& $55 \pm 7 $\cite{PDG} & $55 \pm 4$\cite{PDG} \\
\hline
$D^0 \rightarrow \rho^- \ell^+ \nu_\ell$ & $1.81^{+0.18}_{-0.13} $ & $1.73^{+0.17}_{-0.13}$ \\
\cline{2-3}
& --- & --- \\
\hline
$D^+ \rightarrow \rho^0 \ell^+ \nu_\ell$ & $2.29^{+0.23}_{-0.16}$ & $2.20^{+0.21}_{-0.16}$ \\
\cline{2-3}
& $2.5 \pm 1.0$\cite{PDG} & $3.4 \pm 0.8$\cite{PDG} \\
\hline
$D^+ \rightarrow \omega \ell^+ \nu_\ell$ & $1.93^{+0.20}_{-0.14}$ & $1.85^{+0.19}_{-0.13}$ \\
\cline{2-3}
& --- & --- \\
\hline
$D_s^+ \rightarrow K^{* 0} \ell^+ \nu_\ell$ & $2.33^{+0.29}_{-0.30}$ & $2.24^{+0.27}_{-0.29}$ \\
\cline{2-3}
& --- & --- \\
\hline
$D_s^+ \rightarrow \phi \ell^+ \nu_\ell$ & $25.3^{+3.7}_{-4.0}$ & $24.0^{+3.5}_{-3.7} $ \\
\cline{2-3}
& $20 \pm 5 $\cite{PDG} & $20 \pm 5$\cite{PDG} \\
\hline\hline
\end{tabular}
\flushleft{{\bf Tab.19a}: The branching ratios of $ D \rightarrow V
$ semileptonic decays obtained with considering the meson DAs up to
twist-4, where the corresponding experimental values are also given
for comparison, unit: $10^{-3}$. }
\end{table}

\begin{table}
\centering \renewcommand\arraystretch{0.75} \tabcolsep0.40in
\begin{tabular}{|c|c|c|}\hline\hline
Decays & $e \nu_e$ & $\mu \nu_\mu$ \\
\hline
$D^0 \rightarrow K^{* -} \ell^+ \nu_\ell$ & $21.2^{+1.8}_{-1.8}$ & $20.2^{+1.7}_{-1.7}$ \\
\cline{2-3}
& $21.5 \pm 3.5$\cite{PDG} & --- \\
\hline
$ D^+ \rightarrow \bar{K}^{* 0} \ell^+ \nu_\ell$ & $53.9^{+4.6}_{-4.5} $ & $51.3^{+4.4}_{-4.2}$ \\
\cline{2-3}
& $55 \pm 7 $\cite{PDG} & $55 \pm 4$\cite{PDG} \\
\hline
$D^0 \rightarrow \rho^- \ell^+ \nu_\ell$ & $1.61^{+0.13}_{-0.13}$ & $1.55^{+0.13}_{-0.12}$ \\
\cline{2-3}
& --- & --- \\
\hline
$D^+ \rightarrow \rho^0 \ell^+ \nu_\ell$ & $2.04^{+0.16}_{-0.16}$ & $1.96^{+0.16}_{-0.15}$ \\
\cline{2-3}
& $2.5 \pm 1.0$\cite{PDG} & $3.4 \pm 0.8$\cite{PDG} \\
\hline
$D^+ \rightarrow \omega \ell^+ \nu_\ell$ & $1.72^{+0.15}_{-0.14}$ & $1.65^{+0.14}_{-0.13}$ \\
\cline{2-3}
& --- & --- \\
\hline
$D_s^+ \rightarrow K^{* 0} \ell^+ \nu_\ell$ & $1.87^{+0.24}_{-0.22}$ & $1.79^{+0.24}_{-0.21}$ \\
\cline{2-3}
& --- & --- \\
\hline
$D_s^+ \rightarrow \phi \ell^+ \nu_\ell$ & $23.3^{+3.3}_{-3.1}$ & $22.2^{+3.1}_{-3.0} $ \\
\cline{2-3}
& $20 \pm 5 $\cite{PDG} & $20 \pm 5$\cite{PDG} \\
\hline\hline
\end{tabular}
\flushleft{{\bf Tab.19b}: The branching ratios of $ D \rightarrow V
$ semileptonic decays obtained with only considering the leading
twist meson DAs, where the corresponding experimental values are
also given for comparison, unit: $10^{-3}$.}
\end{table}

\end{document}